\documentclass[fleqn,usenatbib]{mnras}

\usepackage{newtxtext,newtxmath}

\usepackage[T1]{fontenc}

\DeclareRobustCommand{\VAN}[3]{#2}
\let\VANthebibliography\thebibliography
\def\thebibliography{\DeclareRobustCommand{\VAN}[3]{##3}\VANthebibliography}

\newcommand{\sbcrit}[1]{$\mu_{0}(g)\,\gtrsim\,#1\,$mag/arcsec$^{2}$}
\newcommand{\surfbright}[1]{#1\,mag/arcsec$^{2}$}
\newcommand{\radcrit}[1]{$r_e\,\gtrsim\,#1\,$kpc}

\usepackage{graphicx}	
\usepackage{amsmath}	
\usepackage{orcidlink}

\title[NUDGEs in the Subaru Wide-field Clusters]{\centering Ultra-diffuse Galaxy Analogues in the Subaru\\Hyper-Suprime Cam Wide-field Clusters}

\author[N. A. Makda et al.]{
N. A. Makda,$^{1,2}\orcidlink{0000-0002-2611-0043}$
, S.-L. Blyth$^{1}\orcidlink{0000-0002-5777-0036}$
\& R. E. Skelton,$^{1,2}\orcidlink{0000-0001-7393-3336}$
\\
$^{1}$Department of Astronomy, University of Cape Town, Private Bag X3, Rondebosch 7701, South Africa\\
$^{2}$South African Astronomical Observatory, PO Box 9, Observatory, 7935, South Africa\\
}

\date{Accepted XXX. Received YYY; in original form ZZZ}

\pubyear{2022}

\begin{document}
\label{firstpage}
\pagerange{\pageref{firstpage}--\pageref{lastpage}}
\maketitle

\begin{abstract}
We perform a systematic statistical study of ultra-diffuse galaxy analogues (NUDGEs) in a large sample of galaxy clusters to investigate their properties with respect to the host clusters. We used data from the Hyper Suprime-Cam Subaru Strategic Program wide field survey and find a total of 5057 NUDGEs exceeding the background counts in 51 out of 66 galaxy clusters.  The clusters span the redshift range 0.08$\,<\,$z$\,<\,$0.15 and they have a mass range of $0.95\times10^{14}\,\text{M}_\odot - 8.34\times10^{14}\,\text{M}_\odot$. The properties of these NUDGEs are found to be similar to UDGs studied in previous works and reaffirm that they are an extension of a continuous galaxy distribution. The number of NUDGEs as a function of cluster halo mass for our sample follows the power law: $N\propto M_{200}^{0.78\pm\,0.28} $. This fit is consistent with previous UDG studies and, together with our NUDGE sizes distributions, matches well with the simulations of UDGs in cored dark matter haloes formed by tidal stripping. The NUDGE density distribution with respect to clustercentric radius of our sample is flatter than previous UDG studies, although the red NUDGEs in this sample show a statistically significant decrease in density with respect to clustercentric radius, indicating that red UDGs may be more affected by their environment than blue UDGs. 

\end{abstract}




\section{Introduction}

\subsection{Ultra-diffuse galaxies}

Ultra-diffuse galaxies (UDGs) are a subclass of low surface brightness galaxies (LSBs) which classify the largest and faintest LSB sources. The term ultra-diffuse galaxy was coined by \cite{vandokkum} after finding an abundance of these sources in the Coma cluster. \cite{vandokkum} initially set the criteria for a galaxy to be classified as a UDG, as requiring a central $g$-band surface brightness of \sbcrit{24} and an effective radius of \radcrit{1.5}. The criteria were established not necessarily to find a new type of galaxy but rather to identify the largest and faintest galaxies and  study this extreme branch of the galaxy continuum. There have been many studies since and the criteria have been modified due to the differences in instrumentation and analysis strategies of various authors/groups.\\

Previous investigations observed similar galaxies in smaller numbers in various environments \citep{sandage1984, impey1988, Schwartzenberg1995, Conselice2003}, however their abundances in clusters were only relatively recently discovered. The potentially large dark matter content and interest in galaxy formation processes have prompted a renewed interest in these extreme LSBs \citep{vandokkum}. Due to the recent advances in large telescopes, detectors, and deep survey programs, the study of extremely faint and extended sources can be performed over large areas relatively quickly and systematically. Further studies of UDGs have since confirmed their abundance in clusters (e.g \citealt{koda2015,venhola2017,janssens2019,mancera2019,iodice2020,lamarca2022}), as well as lower density groups (e.g \citealt{trujillo2017_feb,cohen2018,gannon2021}) and the field (e.g \citealt{greco2018,prole2019,barbosa2020}). These studies also show that UDGs are found to have stellar masses comparable to dwarf galaxies while having sizes akin to large spiral galaxies.\\ 

UDGs have been observed in a range of different environments and possess a range of different properties. Several studies have shown that a variety of different UDG formation mechanisms are necessary to account for their diverse environments, observed morphologies, metallicities, and dark matter fractions, as supported by both observational research \citep{papstergis2017,bennet2018,vandokkum2022,boselli2022} and simulations \citep{amorisco2016,baushev2018,shin2020}. These UDG formation mechanisms can be separated into two main subsets: the so-called "internal" processes and "external" processes. UDGs observed to be isolated in the field are thought to be formed from internal processes whereas UDGs in clusters or dense environments are expected to be affected by external processes.

\subsection{UDG formation mechanisms} 

\subsubsection{Internal processes}
Internal formation processes are expected to be the main contributing mechanism resulting in isolated and field UDGs, due to the lack of external environmental factors like the tidal forces from neighbouring galaxies and harsh conditions within a dense cluster environment. \cite{amorisco2016} used the standard model of disc formation to demonstrate that UDGs can be formed from dwarf galaxies with high spin haloes. The high angular momentum of the halo would cause the dwarf to expand and the surface brightness to decrease. \cite{rong2017} came to a similar hypothesis that UDG progenitors are dwarf galaxies and their spatially extended sizes result from a late formation time and a high spin halo, observed in the Millenium II cosmological simulation \citep{boylan2009} and Phoenix simulation \citep{gao2012}. Cosmological zoom-in hydrodynamical simulations from the NIHAO project \citep{wang2015} were used by \cite{dicintio2017} to determine that UDGs can be formed in haloes similar to dwarf galaxies, concluding that extended star formation in dwarfs at earlier times may result in strong gas outflows which result in passively evolving UDGs. \cite{papstergis2017} observed several isolated UDGs; one was observed to be gas-rich and three others gas-poor, suggesting multiple mechanisms for UDG formation. The range of different formation scenarios and the need for multiple scenarios to explain their various properties indeed imply a few different formation processes are needed to produce the observed UDG sample. More observational constraints on the properties of isolated field UDGs will help to distinguish which of these possible scenarios has a larger impact on UDG formation.\\

The study of isolated UDGs in the field is particularly difficult, as this requires a deep, blind photometric search of the field to identify candidates and subsequent spectroscopic or \textsc{Hi} follow-up to measure their distances, which can then be used to characterise their properties. Illustratively, \cite{vanDokkum2015} required 1.5 hours with the Low Resolution Imaging Spectrometer \citep{oke1995} on the Keck I telescope to determine the radial velocity of the brightest UDG in the Coma cluster (Dragonfly 44). To measure the spatially resolved stellar kinematics of Dragonfly 44 the Keck Cosmic Web Imager \citep{Morrissey2012,Morrissey2018} on the Keck II telescope required a total observation time of 25.3 h \citep{vandokkum2019}. Recently, \cite{gault2021} used the  Karl G. Jansky Very Large Array to study the \textsc{Hi} properties of 12 UDGs; approximately 6h was spent on each source. In recent times significant improvements in technology have allowed deep photometric surveys to scan large portions of the sky and advances in software have enabled improvements in data reduction techniques to detect to fainter surface brightnesses. The improvements in spectroscopic and radio instrumentation as well as data processing methods have also improved the ability to characterise the distances of these faint galaxies \citep{gault2021,iodice2023}. The benefits of these technological advances are now beginning to bear fruit in the low surface brightness regime.

\subsubsection{External processes}

External formation processes are expected to have the most impact in dense environments, either through a galaxy's interaction with the dense intracluster gas or cluster tidal fields, or through galaxy-galaxy interactions which are more common in the dense environments of clusters.\\

Ram-pressure stripping is a process whereby galaxies falling into the cluster environment experience the force of passing through the cluster's hot intracluster medium (ICM). If the pressure experienced by galaxies passing through the ICM is greater than the galaxies' gravitational restoring force then the interstellar medium (ISM) of these galaxies may be stripped \citep{gunn1972,boselli2022}. The stripping of ISM in galaxies may initially result in dust lanes and trailing star formation as the galaxy passes through the cluster \citep{kenney1999,kenney2004,crowl2005,hess2022}. The stripped ISM in these galaxies is expected to be the atomic gas reservoir supplying star formation, which may subsequently cause quenching in these galaxies \citep{boselli2022}.\\

\cite{safarzadeh2017} studied the in-fall of satellite galaxies into clusters and found that the change in gravitational potential due to ram-pressure stripping causes these satellite galaxies to become puffy and appear as UDGs. \cite{carleton2019} propose a similar proposition: tidally stripped dwarf satellite galaxies within cored dark matter haloes may represent the population of observed UDGs in clusters. The radial distribution of UDGs observed in the Coma cluster paired with the UDG number vs cluster halo mass relation and the reproduce-ability of UDGs in the Illustris-dark simulation motivate this proposition. \cite{tremmel2020} used the RomulusC cosmological simulations \citep{tremmel2019} to study the origin of 122 cluster UDGs. The selected UDGs in the simulation fit the \cite{vandokkum} UDG criteria and are shown to agree with the properties (absolute magnitude, effective radius and sersic index) of UDGs observed in clusters \citep{vandokkum, vanderburg2016,mancera2018}, although some inconsistencies with metallicity do occur. \cite{tremmel2020} show that these UDGs form from accreting dwarf galaxies experiencing quenching due to ram pressure stripping at early infall times.\\

The diffuse and extended nature of UDGs suggests that their survival in clusters is unlikely and moreover, their abundance in the cluster environment is quite unexpected. It is expected that UDGs cannot survive in the central core regions of clusters due to the strong forces caused by the cluster potential. Several studies have observed fewer UDGs in the centres of clusters \citep{vanderburg2016,wittman2017,janssens2019}, which may indicate the strong fields there and the effect of the cluster potential. \cite{venhola2017} found in the Fornax cluster that the large UDGs were distorted and elongated, which potentially implies that they formed from tidally disturbed galaxies. The abundance of UDGs in the Coma cluster had suggested a possible large dark matter content in these galaxies to enable them to survive the harsh cluster environment. \cite{vandokkum} suggested that due to the expected large dark matter content, their large sizes, and the red stellar populations, they were likely formed from M33-like haloes that lost their gas after the initial first generation of star formation due to tidal forces \citep{peng2016,danieli2022,gannon2023,buzzo2024}.\\

Tidal harassment is one of the processes which impact galaxy formation due to nearby neighbouring galaxies, especially in dense clusters, through close encounters. The interaction is a process by which the gravitational pulls of neighbouring galaxies cause the stripping of stars and gas, and may result in star-bursts and morphological evolution \citep{moore1996,moore1998}. \cite{karunakaran2023} studied satellite UDGs around Milky Way-like hosts and found a subsample ($\sim$$20\%$) of UDGs which show tidal features or extended morphologies consistent with having experienced recent or ongoing tidal interactions. Tidal harassment may transform dwarf galaxies into UDGs by expanding them or they may be formed from the collapse of the stripped tidal debris, although in general UDGs do not look like tidal debris but rather like disrupted dwarfs \citep{bennet2018}. Collisions of galaxies in the dense cluster environment are rare. However, the galaxy-galaxy interactions of gas-rich galaxies may result in the ejection of the gas and the suppression of star formation \citep{baushev2018}. Some evidence shows tidal stripping to be an unlikely cause for UDG formation due to the lack of visual signs which are expected in tidal stripping systems \citep{mowla2017}. Also, studies of UDGs in galaxy clusters like Coma and Virgo show high globular cluster richness, suggesting these systems experience minimal tidal stripping interactions \citep{forbes2020}. UDGs that experience tidal stripping are expected to have fewer globular clusters due to disruptions caused during tidal stripping. \cite{buzzo2024} studied UDGs in the MATLAS survey and observed their sample of UDGs to have intermediate-to-old ages and to be metal poor. Furthermore, they distinguish two classes of UDGs from their sample: the first is consistent with UDGs forming from “puffed-up dwarfs” due to internal mechanisms. The second class is consistent with a “failed galaxy” scenario at earlier times and quiescent evolution. They suggest that tidal stripping is therefore an unlikely dominant UDG formation route \citep{buzzo2024}.\\ 

\cite{duc2014} studied a tidal dwarf galaxy (TDG) in the tidal tail of an elliptical galaxy and observed properties comparable to those of UDGs: a low central surface brightness and an effective radius larger than typical dwarfs. These TDGs are proposed to form through galaxy merger events and their similarities to UDGs indicate that they may share the same merger formation mechanism. \cite{lelli2015} studied the gas dynamics of TDGs around interacting systems and observed that these galaxies are likely to be dark matter deficient. The tidal effects of interacting systems and the lack of dark matter in TDGs imply that observed UDGs in similar environments may have formed from the collisional debris of interacting and merging galaxies. \cite{bennet2018} identify two UDGs with a clear association with a host galaxy halo and associated tidal material, suggesting UDGs may form from galaxy interactions expanding normal dwarf galaxies or possibly forming from tidal material. \cite{shin2020} used idealized resolution simulations and the IllustrisTNG simulation to to show that UDGs can form from high-velocity galaxy collisions. \cite{vandokkum2022} studied the line-of-sight distances and radial velocities of two UDGs in the NGC 1052 group. The observed measurements are consistent with the end-result of a single bullet-dwarf collision as well as a trail of low-luminosity objects likely originating from the same event.\\

Some studies predicted a sample of UDGs to originate from low surface brightness galaxies which later become quenched to form UDGs through multiple cluster processes. \cite{roman2017b} studied UDGs in three nearby groups and suggest that old extended low surface brightness dwarf galaxies form in the field and are then drawn into groups and form UDGs and are subsequently gathered by clusters during group accretion. In the dense cluster environment both the cluster potential well in addition to interactions with cluster galaxies impact the formation of UDGs. \cite{ferremateu2018} showed that the star formation histories for several spectroscopically observed UDGs in the Coma cluster are consistent with predictions for a dwarf-like origin. Subsequently this suggests many UDGs are field dwarfs that become quenched after cluster accretion. However, this only explained the properties for a sub-sample of UDGs and therefore it is expected that different formation mechanisms are expected to reproduce the scatter in the properties of the rest of the sample. \cite{benavides2021} used hydrodynamical simulations to show that isolated UDGs in the field could be formed as backsplash galaxies which have over time become distant from their host cluster or group. They suggest UDGs may be observed in large numbers in filaments and voids as extended galaxies stripped of gas and dark matter compared to dwarfs of comparable stellar mass. \cite{sales2020} studied the properties of simulated cluster UDGs in the IllustrisTNG simulation and suggest two distinct populations of UDGs, `born' UDGs and `tidal' UDGs. They found that `born' UDGs are similar to low surface brightness galaxies in the field, which lose their gas during cluster accretion. The `tidal' UDGs are shown to have formed from a variety of more massive galaxies and become UDGs upon cluster accretion through cluster tidal effects. The sample of `tidal' UDGs entered the cluster at earlier times and were found to be more common near the centre of cluster than the `born' UDG population.\\

\subsection{Aims of our work}

The wide range of densities and properties of each cluster implies the environment and its effects vary not only between clusters but also within each cluster. UDG counts vs cluster halo mass can indicate how the number of UDGs changes depending on cluster mass and, by extension, cluster environment. \cite{vanderburg2017} studied this relationship for galaxy clusters and groups, and determined that the UDG counts and cluster halo mass follow a power law relation: $N\propto~M^{1.11\pm\,0.07}$. Subsequent studies of UDGs in nearby groups and clusters, including \cite{roman2017a,roman2017b}, \cite{mancera2019}, \cite{forbes2020} and in the Fornax cluster \citep{venhola2021} are consistent with this relation. The number of UDGs increases with respect to cluster halo mass, possibly indicating dense environments with harsher conditions have more UDGs. External UDG formation processes in clusters directly affect the UDG counts vs cluster halo mass relation due to the harsher environments in more massive clusters. Galaxy interactions with the dense intracluster gas and cluster tidal fields may cause stripping of gas resulting in quenching and increased numbers of UDGs forming. Galaxy-galaxy interactions are expected to result in stripped tidal debris which may form UDGs. Also, more massive clusters accrete more galaxies and more UDGs which may have formed in the field due to their larger gravitational pull and sizes. Galaxies in the more dense environments of massive clusters are expected to experience these external processes with a higher frequency and intensity than in smaller galaxy groups which results in the observed increase in the number of UDGs with increasing cluster halo mass. However, the variability within clusters and spatial distribution of the large galaxies, which we expect to have the greatest tidal effects, need to be investigated further to fully understand how density within these clusters and groups effects UDG formation and evolution.\\

Through observations and simulations there are many plausible formation mechanisms which may be responsible for the formation and evolution of UDGs and multiple mechanisms are likely in the dense cluster environment. We studied the characteristics of UDG-analogues (NUDGEs, defined in section \ref{sec:udg_select}) in a large sample of clusters and their associated properties. This study focuses on examining these clusters using a systematic method and increasing the sample of UDGs studied and the effect of the cluster environment on the formation of UDGs. To try to shed light on formation mechanisms we study the spatial distribution of NUDGEs in the clusters and compare their locations and density to the cluster radius, as well as their size distributions. The effect of the different cluster processes are expected to affect the properties of UDGs and show variations in their properties, therefore we compare the characteristics of the blue and red NUDGEs in our sample. This paper is ordered in the following way. Section 2 describes the dataset we used and the cluster catalogue employed. Section 3 details the method we implemented to identify and classify NUDGEs in the selected clusters and the techniques to measure completeness. Section 4 describes the results of the identified NUDGEs and section 5 investigates the effect of the cluster environment on NUDGEs. Section 6 summarizes our results and conclusions. We used the following cosmological parameters throughout: $H_0~=~67.7~\text{km}\,\text{Mpc}^{-1}\,\text{s}^{-1}, \Omega_m~=~0.3, \Omega_\Lambda~=~0.7$ \citep{planck2016}.\\

\section{Data}

We used data from the publicly available third data release of the Hyper Suprime-Cam Subaru Strategic Program (HSC-SSP; \citealt{miyazaki2018, aihara2022}), a wide-field imaging survey in the $g,\,r,\,i,\,z,\,y$ optical bands. The redMaPPer cluster catalogue \citep{Redmapper} was used to identify galaxy clusters in the regions covered by the Subaru wide field. The HSC-SSP data and redMaPPer catalogue are detailed in the following sections. 

\subsection{Imaging Data}

The main scientific goals of the HSC-SSP are to study the distribution of dark matter in the universe through observations of weak gravitational lensing and to investigate galaxy evolution \citep{aihara2018}. The survey utilized the Hyper Suprime-Cam (HSC) 870 megapixel prime focus optical imaging camera on the 8.2\,m Subaru telescope on Maunakea in Hawaii. The camera consists of 116 CCD chips and each is 2048$\,\times\,$4096 pixels in size, equivalent to 5.73\arcmin$\times$ 11.47\arcmin\ with gaps of either 12\arcsec\ or 53\arcsec\ between them. The 3rd public data release (\citealt{bosch2018}; PDR3) wide layer of the HSC-SSP covers an area of$~\sim$\,670 deg$^2$ at full image depth of $\sim$\,26\,mag at 5 sigma for point sources in the $g$-band. The large area coverage and depth of the image allow for statistical studies of faint sources. The images have a pixel scale of 0.168\arcsec/pixel and the median seeing is$~\sim$\,0.79\arcsec\ , 0.75\arcsec\  and 0.61\arcsec\ in the \textit{g, r, i} bands respectively.\\

\begin{figure}
\centering
\includegraphics[width=0.5\textwidth,trim={1.8cm 0.8cm 2cm 2.4cm},clip]{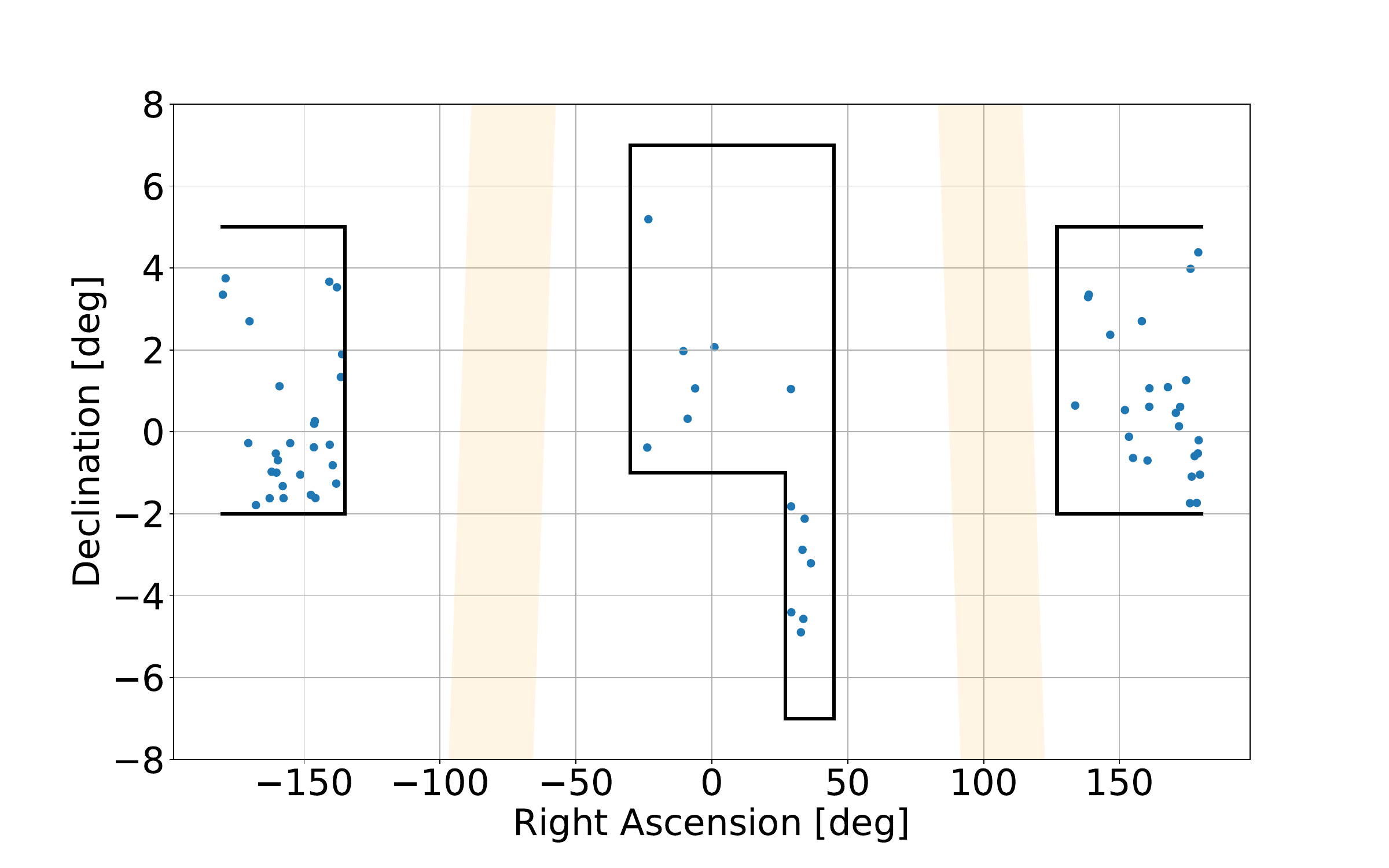}
\caption{The spatial positions of the cluster sample are indicated with blue dots as well as the overall coverage of the two wide field survey regions used in our study (black outlined regions). The Galactic plane is highlighted in faint orange shading.}\label{fig:subaru_world}
\end{figure}

The survey used a dither pattern to offset the telescope between successive exposures to ensure the data collected was uniform and the gaps between the CCDs were filled. The reduced images were coadded with a total exposure time of$~\sim$\,10, 10 and 20 mins in the \textit{g, r, i} respectively. The \textit{hscpipe8} pipeline was used to reduce, sky-subtract and co-add the PDR3 data. The processed data were split into tracts covering$~\sim$\,1.7~square degree regions, which are further divided into patches covering$~\sim$\,12~square arcmin sub-regions. There are three main regions which make up the wide field survey. In our study we used the two main regions at declination 0$^{\circ}$ in the publicly available archival PDR3 wide layer of the HSC-SSP\footnote[1]{\label{footnote1}Available from the HSC-SSP website, see \url{https://hsc-release.mtk.nao.ac.jp/doc/}} ($g,\,r,\,i$ bands) and only the patches with coverage of the identified clusters were extracted (the 2 main regions used are shown in Figure \ref{fig:subaru_world}). As an example, the patches can be seen in Figure \ref{fig:cluster9723} as the black dotted lines on the Subaru image of cluster 9723. We used the PSF model$^{\ref{footnote1}}$ generated at the centre of each patch for source fitting within each patch.\\ 

\color{black}
\subsection{redMaPPer Cluster Catalogue}\label{sec:remapper}

The redMaPPer cluster catalogue \citep{Redmapper} is the output of a galaxy cluster survey based on the SDSS DR8 dataset \citep{aihara2011} spanning $\,\sim10\,000\,$deg$^2$ which reports $\,\sim25\,000$ galaxy clusters over the redshift range z~$\in$~[0.08, 0.55]. The redMaPPer cluster catalogue was created using the \text{redMaPPer} algorithm, which is a red sequence cluster finder that determines the cluster membership of galaxies by utilizing colour and proximity. The coverage of the Subaru Wide field dataset is $\,\sim670\,$deg$^2$, and a total of 2462 redMaPPer clusters lie in this region. We selected all the redMaPPer clusters in the two main regions of the Subaru wide field in the redshift range z~$\in$~[0.08, 0.15] as our cluster sample for this study, resulting in a total of 66 clusters.\\

We use the cluster richness estimate ($\lambda$) provided by the redMaPPer catalogue to calculate each cluster's mass ($M_{200}$) \citep{simet2018}:
\begin{equation}\label{eqn:m200}
    \indent \langle M_{200}|\lambda \rangle = 10^{14.369\pm0.021\,\mathrm{stat}.\,\pm0.023\,\mathrm{sys}.} \Bigg(\frac{\lambda}{40}\Bigg)^{1.30^{+0.10}_{-0.09}}
\end{equation}

The radius of each cluster is calculated as  \citep{navarro1996, coe2010}:
\begin{equation}\label{eqn:r200}
    \indent R_{200} = \Bigg( \frac{G\,M_{200}}{100\,H_0^2}\Bigg)^{\tfrac{1}{3}}
\end{equation}

where \textit{G} is the gravitational constant. We used the $R_{200}$ radius to perform the initial selection of potential cluster members. Figure \ref{fig:cluster9723} shows the Subaru image of cluster 9723 with the cluster's $R_{200}$ and the redMaPPer galaxies indicated. 

\section{ Method}

In this paper we study NUDGEs (defined in section \ref{sec:udg_select}) in low redshift (\textit{z} < 0.15) clusters identified in the redMapper cluster catalogue with the HSC-SSP wide field data. We used \textsc{SExtractor} to detect NUDGEs in the Subaru data and \textsc{galfit} to model and measure their properties. In the following sections we describe the  pipeline created to automate the detection and characterisation of NUDGEs in clusters. \\

UDGs are difficult to detect and characterise due to their low surface brightnesses, and in most cases UDGs do not have spectroscopic redshifts due to the long observation times required to obtain them. This results in samples of cluster UDGs contaminated with foreground and background sources. In our study we quantified the true number of NUDGEs in each cluster by selecting a `background' comparison sample annulus around the cluster with an area of equal size to each respective cluster ($1.5\,R_{200}\,<\,r\,<\,1.8\,R_{200}$). The region of the background sample is near enough to the cluster such that the effects of cosmic variance are insignificant and far enough from the cluster such that we do not expect significant numbers of cluster members in the background sample. The methods implemented in this pipeline on the clusters were identically applied to the background annuli to provide an unbiased and quantitative measurement of NUDGEs in the foreground/background. Comparing the cluster NUDGEs to the background sample allowed us to estimate the true number of NUDGEs in the cluster. 

\subsection{\label{sourcedetection}Source detection}

We used the Source Extractor (\textsc{SExtractor}; \citealt{sextractor1996}) software to identify all galaxies in each cluster up to the $R_{200}$ radius and in the background annulus. The clusters in our sample span a range of sizes and exist in environments with different local densities, which results in a large variation in the number of detections in each cluster. The close proximity of the background annulus to the cluster implies that the density of foreground and background sources should be similar in both regions and the variation in the local environment density can be well attributed.\\

We obtained all the Subaru images with coverage of the clusters and background regions. We used \textsc{SExtractor} primarily as a detection tool on the $g$-band images with follow up characterisation in the $r$ and $i$ bands. Through testing different regions we found that \textsc{SExtractor} consistently identified more extended sources in the $g$-band than the $r$ and $i$ bands, and therefore we used the $g$-band for detection. We aimed to have a high completeness in source detection, therefore the \textsc{SExtractor} configuration parameters were set to be sensitive to faint objects. A Gaussian filter with a FWHM of 4 pixels was used, and detected sources were required to be above a 1.0$\,\sigma$ detection threshold and to have a size of at least 4 pixels to remove compact sources. To remove point sources and sources smaller than our expected NUDGEs, we applied a radius criterion to the $g$-band \textsc{SExtractor} detection of $r_e$~>~2.5 pixels, where $r_e$ is the SExtractor "flux radius" (equivalent to effective radius). 

\begin{figure*}
\centering
\includegraphics[width=1.2\textwidth,trim={3.6cm 2.8cm 1.5cm 7cm},clip]{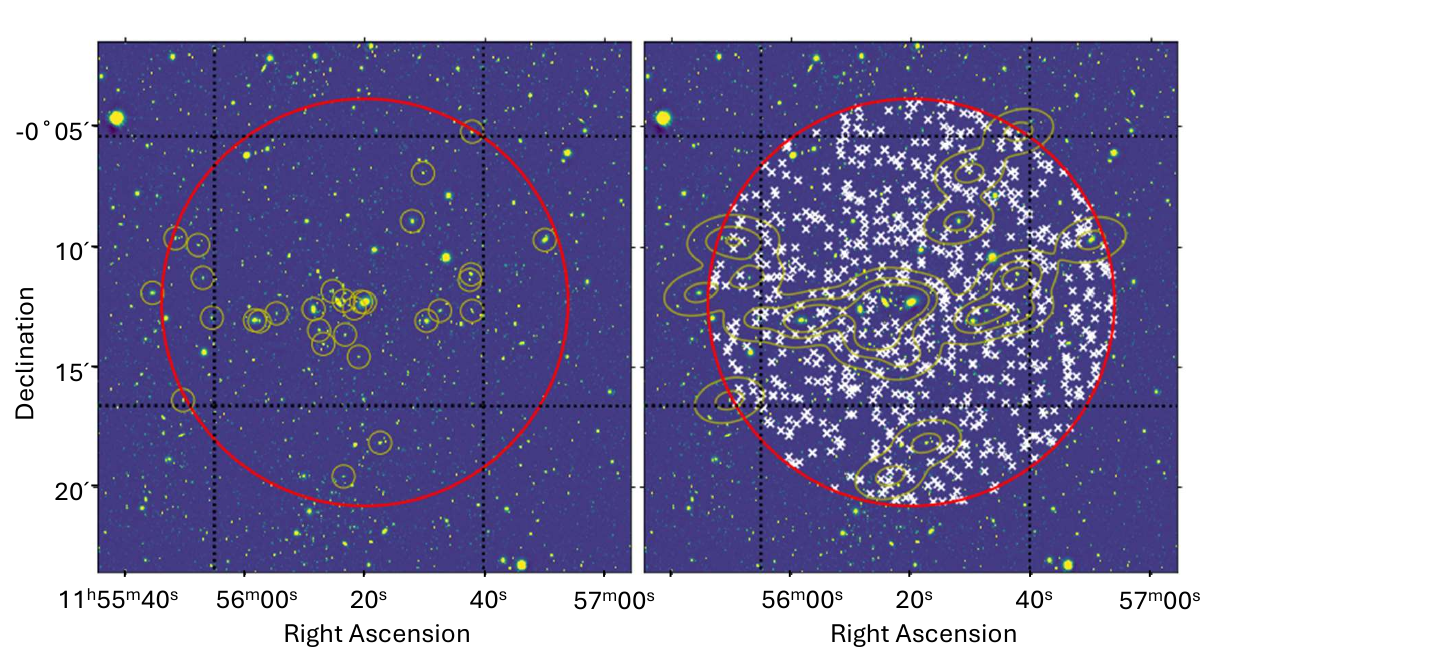}
\caption{\label{fig:cluster9723} Subaru image of cluster 9723, both panels show the same image. The red circle indicates the $R_{200}$ cluster radius. In the left panel the positions redMaPPer galaxies are identified by the yellow circles. In the right panel the white crosses identify the positions of NUDGE candidates and the yellow contours indicate the density of the redMaPPer galaxies. The black lines show the individual patches of the Subaru data with respect to the cluster. Each patch was investigated independently.}
\end{figure*}

\subsection{\label{fitting}Surface brightness fitting}

The \textsc{galfit} software \citep{peng2002} was used to measure the total magnitude, effective radius, S\'ersic index and axis ratio of each of the sources identified with \textsc{SExtractor} in the \textit{g, r} and \textit{i} band images. \textsc{galfit} was run independently in each band using initial input parameters (effective radius, total magnitude and axis ratio) measured by \textsc{SExtractor} for the respective bands, which were then allowed to vary. Cluster UDGs have been observed to be dwarf-like, diffuse and elliptical, which can be approximated well by a single S\'ersic function. Previous studies (eg. \citealt{vandokkum, koda2015}) show that a S\'ersic index of $n\sim\,1$ estimates the light distribution well. \cite{smudges2022} found a similar S\'ersic index $n\sim\,0.8$. Therefore we set the input S\'ersic index to 1 initially and allowed it to vary during \textsc{galfit} modelling.\\

We measured the background noise in 100 regions in the Subaru images and found a background level $\sim0.005$ ADUs. We set this value as the initial background and allowed it to vary. For each source we used the PSF at the centre of each image patch as measured and provided by the HSC-SSP. A convolution box of 300$\,\times\,$300 pixels in size was used to fit each source. The magnitudes were corrected for Galactic extinction \citep{schlafly2011}. We corrected all the sources within each cluster by the same amount, i.e. the extinction value at the centre of the cluster. To calculate the mean effective surface brightness from the measured \textsc{galfit} properties we used \citep{graham2005}:
\begin{equation}\label{eqn:meanSB}
    \indent<\mu_{e}> = m_{t} + 2.5\, \mathrm{log}_{10}(2\,\times\,q\,\pi\,r_{e}^{2})
\end{equation}

where $m_t$ is total magnitude, $q$ is the axis ratio and $r_e$ is the semi-major axis half-light radius. 

\subsection{\label{sec:completeness}Completeness Estimation}

Completeness testing was used to understand and quantify the number of sources we missed during the detection and classification of NUDGEs in our cluster sample. The number of sources detected in images is less than the true number that exist in that region, due to detection limitations of the instruments and software. Additionally, NUDGEs have very low surface brightnesses, therefore, background noise further impacts their detection in the data. We need accurate estimates of the numbers of NUDGEs in clusters to gain insight into their formation mechanisms. Briefly, we performed the completeness test by creating models and inserting them into the data. We then used \textsc{SExtractor} and \textsc{galfit} to find and characterise them (as described previously). We then quantified their recovery fractions and the quality of the measured properties. The results of the completeness test enable us to statistically study the effect of background image noise on NUDGE detection and provide reliable estimates of the true number of NUDGEs in the data. The completeness test is described in more detail below. 

\subsubsection{\label{sec:galsim}Model creation }

We created models of extended sources using \textsc{galsim} \citep{galsim2015}, spanning a wide range of morphologies. The properties of these models uniformly cover the following range: effective radii~$r_e~=~0.5~-~7.0\,$arcsec with~0.5~arcsec intervals, effective surface brightness~$\mu_e~=~20.5~-~27.0\,$mag/arcsec$^2$ with~0.5~mag/arcsec$^2$ intervals, S\'ersic index~$n~=~0.5~-~5.0$ with 0.5~intervals, and axis ratios~$q~=~0.2~-~1.0$ with 0.1 intervals. To consistently create well modelled faint sources with a smooth profile we multiplied the \textsc{galsim} input total flux (calculated from the effective surface brightness and effective radius) by a factor of 10 000. The resulting models were then re-normalized by dividing the flux in each pixel across all the models by 10 000. A total of 30\,240 unique models were created, covering a broad range of morphologies and properties similar to those expected for UDGs and NUDGEs.\\

Following the method we used to measure the properties of the cluster NUDGEs, we characterised the \textsc{galsim} models by fitting them with \textsc{galfit}. The \textsc{galsim} properties were used as initial input values in the \textsc{galfit} templates and the models were fitted with single S\'ersic profiles. Altogether, \textsc{galsim} was used to create the initial models covering a wide range of morphologies, and \textsc{galfit} was used to properly characterise and remodel them. The completeness test uses the models created by \textsc{galfit} with well characterised properties and any further reference to models unless otherwise stated refers to these \textsc{galfit} models.\\

The models selected for the completeness test were chosen to cover a broad range of effective radii and mean effective surface brightnesses. We randomly selected 10 unique models in each grid square. For squares with less than 10 unique models we randomly repeated existing models.

\subsubsection{\label{sec:insertmodels}Insertion and recovery of models}

\begin{figure}
\centering
\includegraphics[width=0.95\linewidth,trim={9.0cm 3.0cm 9.0cm 6.50cm},clip]{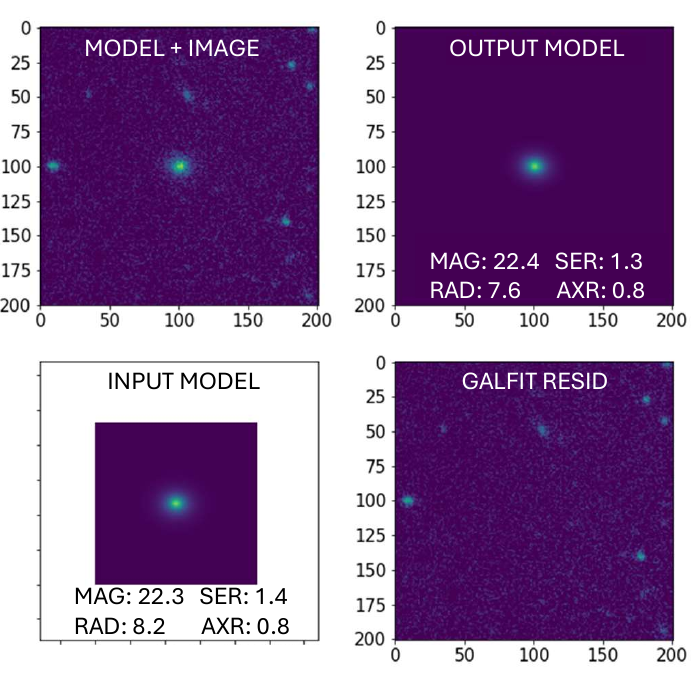}
\caption{\label{fig:model_measure} Measuring the model properties during completeness testing. Top-left: Subaru image cutout with our model added. Top-right: Model recovered by \textsc{galfit}. Bottom-left: Initial \textsc{galfit} input model that was inserted into the Subaru image. Bottom-right: Residual image created by \textsc{galfit} by subtracting the recovered model from the image. "MAG", "SER", "RAD" and "AXR" refers to the apparent magnitude, S\'ersic index, effective radius in arcsecs and axis ratio respectively.}
\end{figure}

A single HSC-SSP image in each of the $g,r,i$-bands was used to perform the completeness test and each band was tested independently. The image was chosen to contain no redMaPPer clusters. We used \textsc{SExtractor} to create a segmentation map of the image and identified one hundred 80 x 80 pixel non-overlapping regions with no source detections. The sample of NUDGE models were then added to the HSC-SSP image and each model was inserted 10 times over in the identified one hundred regions. We used the identification and characterisation pipeline, described in sections \ref{sourcedetection} and \ref{fitting}, to recover the models. Figure~\ref{fig:model_measure} displays an example comparison between the NUDGE model inserted into the HSC-SSP data, \textsc{galfit} refitted model, input model and the residual (top left, top right, bottom left and bottom right panels respectively). In this example the properties of the input and recovered models are similar and well recovered. This was repeated for each model in the $g,r,i$ bands.\\

\begin{figure*}
\centering

\begin{minipage}{.99\textwidth}
  \centering
  \includegraphics[width=1\linewidth,trim={4.0cm 1.0cm 3.5cm 1.0cm},clip]{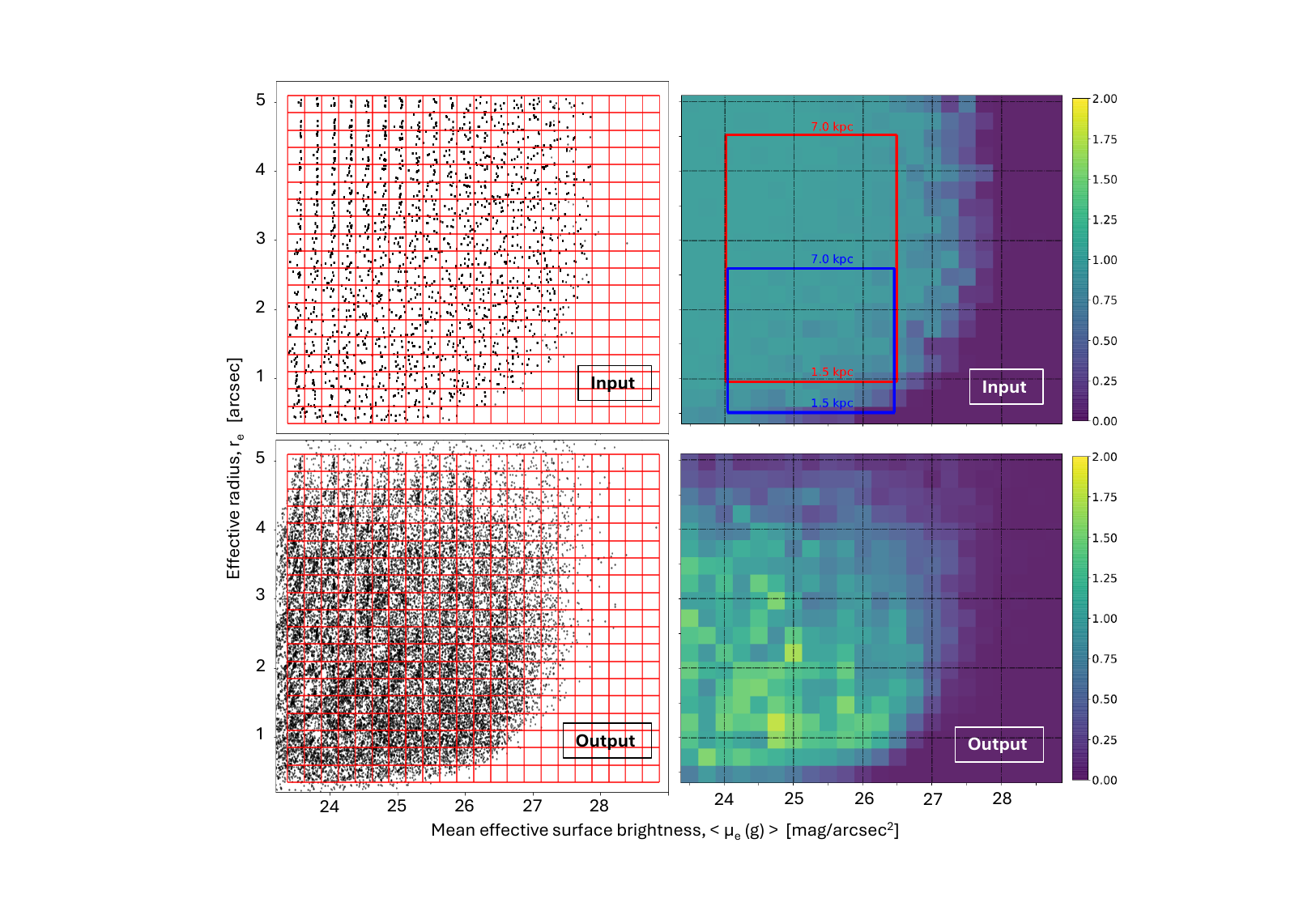}
  \label{fig:detect_comp}
\end{minipage}
\caption{\label{fig:completeness_det_char} Plots showing the effective radius vs Mean effective surface brightness in $g$-band for the sample of recovered models from the completeness test. Top-left panel: The input distribution of of our recovered models (each point constitutes up-to 10 models). Bottom-left panel: The distribution of the output properties of the recovered models. The output properties of some models are recovered outside of the completeness grid, particularly toward brighter mean effective surface brightnesses. Top-right panel: Detection plot for the sample of recovered models (input properties). The completeness fraction shown by the colourbar is measured against the total number of models inserted in each bin with respect to their input model properties. Each cluster has a completeness correction with respect to its redshift, which is between the maximum and minimum regions shown (the red and blue rectangles, lowest and highest redshift respectively). Bottom-right panel: Completeness measured with respect to the output properties. The number of models in these bins increase beyond 1, as typically the models are recovered to smaller sizes and brighter surface brightnesses, implying lower recovery fractions in the top-right of this plot and higher fractions toward the bottom left.}
\end{figure*}

The positions of the detected sources were required to be within 1~arcsec of the input positions to be counted as identified. By repeating each model at least 10 times and inserting them into 10 different positions in the image we could improve the statistics and minimize the effect of erroneous characterisations. The top-left panel of Figure~\ref{fig:completeness_det_char} displays the input values of radius and surface brightness for all the models that were recovered. This plot can be directly compared to the bottom-left panel of Figure~\ref{fig:completeness_det_char}, where the difference between the plots indicates the incompleteness. We can see that the faintest models were not recovered. The recovered radius - surface brightness distribution is shown in the bottom-left panel and by comparing these two panels we can see the difference between the input and recovered properties. There is a dissimilarity between the input and recovered properties and generally models are recovered to be brighter and smaller than their input values. This difference can be attributed to several factors, including the faint nature of the models, the effect of background noise and the limitations of the software.

\subsubsection{\label{sec:completenessanalysis}Completeness analysis}

\begin{figure}
\centering
\includegraphics[width=0.475\textwidth,trim={0.6cm 0.6cm 2.5cm 2.0cm},clip]{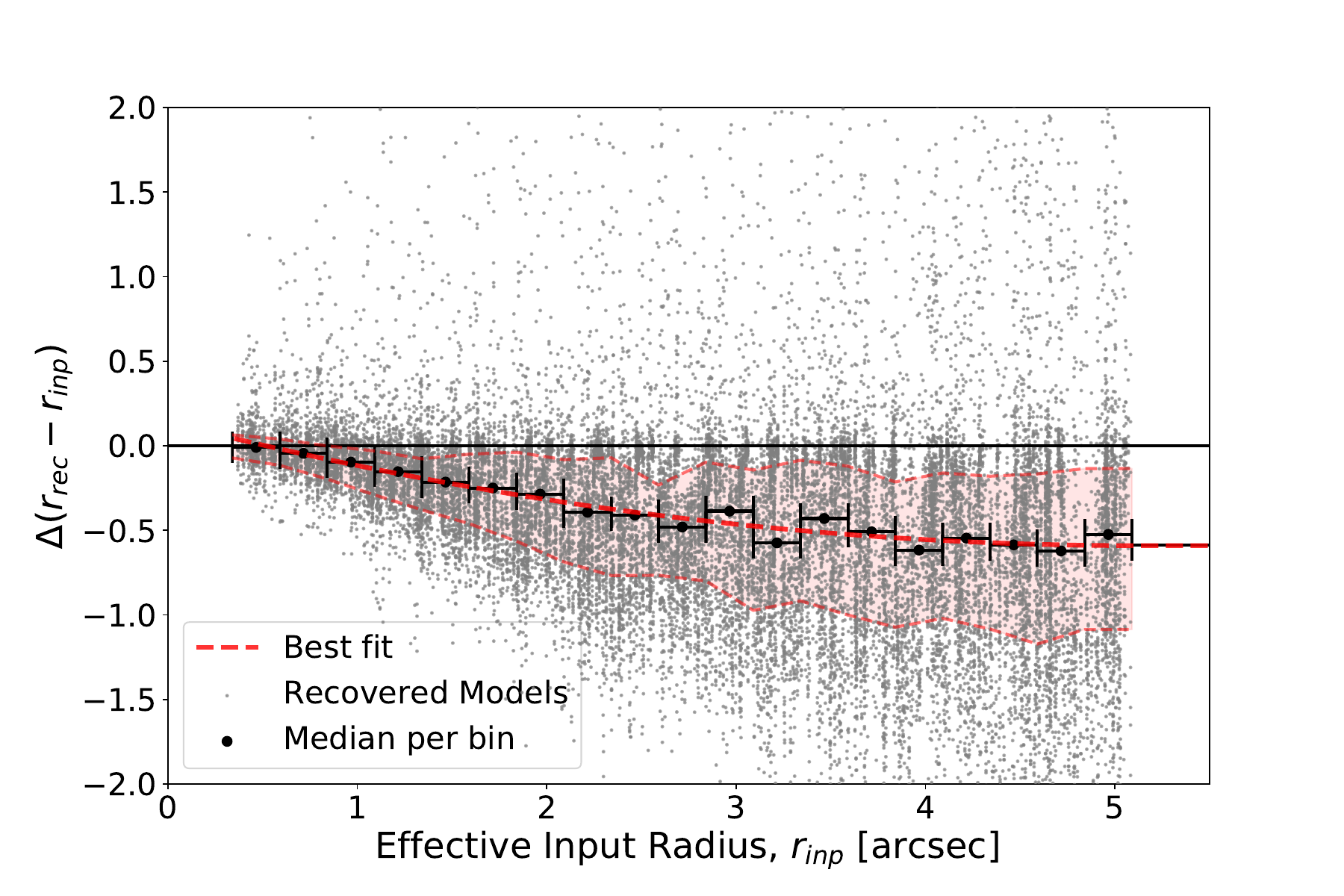}
\caption{\label{fig:radius_offset} The difference in radius between the recovered \textsc{galfit} radius and the input model radius as a function of the input model radius. The black markers indicate the median value in each bin and the red curve indicates the best-fit second-order polynomial to the median values. The black error-bars indicate the binning of the data, which is identical to the detection and completeness plots (Figure~\ref{fig:completeness_det_char}), and  the red shading indicates the interquartile range. The difference between the measured and input model radii increases as the radius increases, as does the scatter.}
\end{figure}

\begin{figure}
\centering
\includegraphics[width=0.48\textwidth,trim={0.6cm 0.6cm 2.5cm 2.0cm},clip]{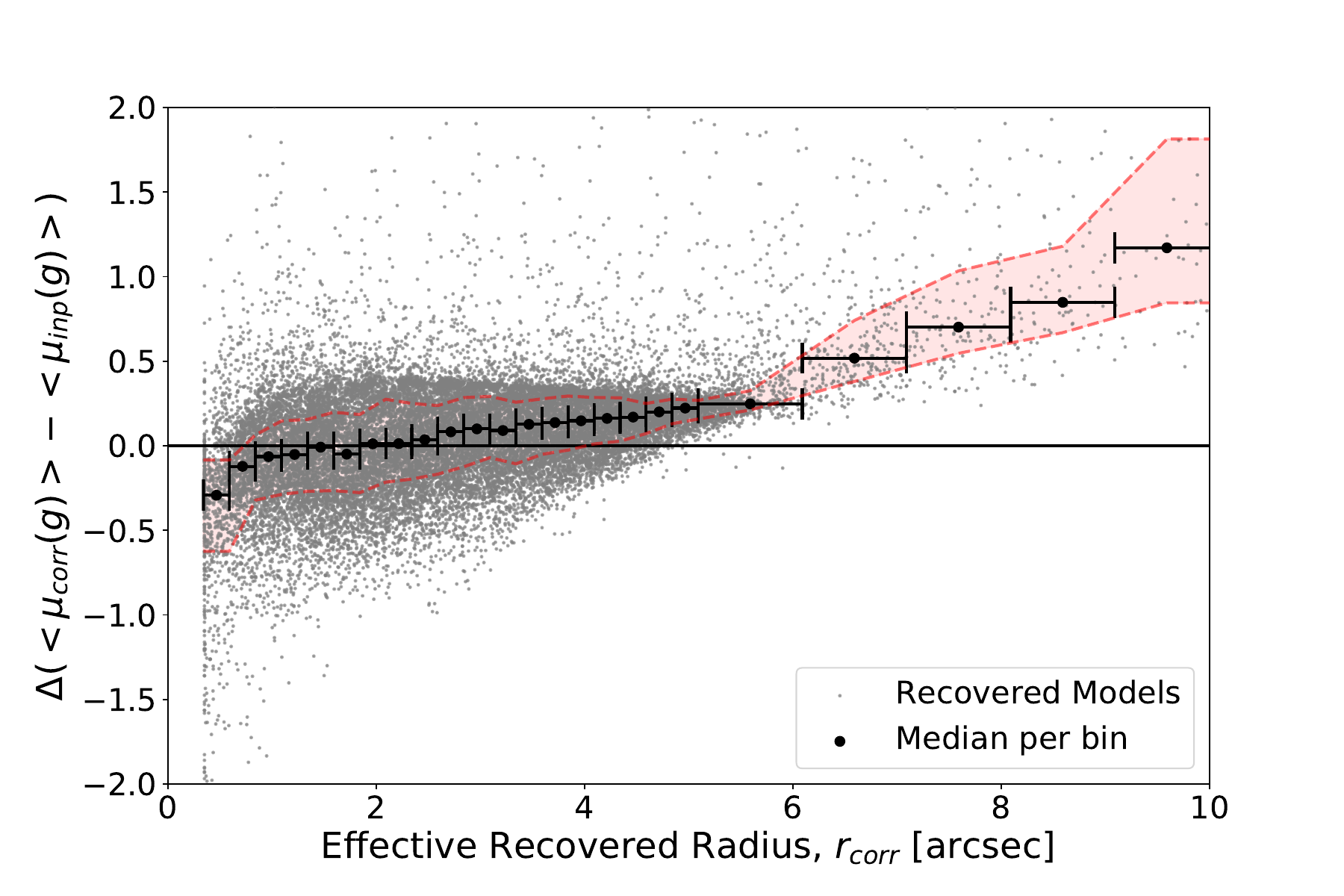}
\caption{\label{fig:sb_offset} The difference in surface brightness between the calculated surface brightness and the input model value as a function of the median corrected radius. The black dots indicate the median value for each bin and the black error-bars indicate the binning of the data, which is identical to the detection and completeness plots (Figure \ref{fig:completeness_det_char}). The red shading indicates the interquartile range. The recovered surface brightness shows large scatter at lowest radii (input is fainter than recovered), which converges as the radius increases toward 2 arcsec. The recovered surface brightness becomes fainter than the input as the radius increases beyond 2 arcsec.}
\end{figure}

The input properties of our models and their recovered values show a roughly consistent difference, which we quantify and correct using the following steps. The difference in radius, input vs recovered (Figure~\ref{fig:radius_offset}), shows a systematic and quantifiable trend with respect to the input effective radius. We binned the input effective radius and fitted a 2$^{\text{nd}}$ order polynomial (best fit parameters: $y=0.03x^2-0.28x+0.14$, $x<5$) to the median values in each bin. Figure~\ref{fig:radius_offset} shows that the recovered radii are smaller than the input radii and the difference increases as the input effective radius increases up to $\sim\,$4 arcsec at which point it becomes approximately constant. The scatter also increases with respect to the input effective radius, indicating that larger input effective radii are more difficult to recover accurately. The best fit polynomial function maps each recovered radius to an input radius (i.e. the expected "corrected" radius). The surface brightness is calculated using radius, therefore any error in radius propagates through to the estimation of surface brightness. The corrected radius was used to calculate the corrected surface brightness. We compare the corrected surface brightness to the input surface brightness vs corrected effective radius (Figure~\ref{fig:sb_offset}) and we see that for radii < 5" the surface brightnesses roughly agree. At larger input values the differences in radii and surface brightnesses increase rapidly and are not well recovered, however, these models fall outside our NUDGE criteria (section \ref{result_char}) and completeness correction. Therefore, we observe that this correction does not have an effect on the number of NUDGEs detected, instead, it corrects the measured effective radius and surface brightness of NUDGEs. The completeness test also established a meaningful measure of the error on the recovered properties which were recorded for each source. These corrected values more accurately represent the values of the detected and characterized sources. As a final correction we implemented a K-correction across all bands ($g, r, i$) for the NUDGEs and NUDGE-like sources in the cluster and background annuli (described in \citealt{chilingarian2010, chilingarian2012}).\\

\subsubsection{\label{sec:completenesscorrection}Completeness correction}

There are two important aspects of completeness testing: the successful detection of the models; and the accuracy of the recovered properties. Firstly, to determine the recovery fraction of the detected models we measure the number of models recovered in each bin and compare to the number of models inserted. Figure \ref{fig:completeness_det_char} (top right panel) shows the detection completeness; the fraction of sources recovered, based on the properties of the input models. We have good detection ($\gtrsim$50\%) up to a surface brightness of 27.5 mag/arcsec$^2$ in the $g$-band. The lower-right corner of this plot is incomplete and difficult to recover due to the models being both small and faint. The red and blue rectangles indicate the regions used to determine the completeness correction fraction for the lowest and highest redshift clusters respectively (the completeness correction for each cluster is determined based on redshift and our NUDGE radius and surface brightness criteria, see section \ref{sec:udg_select}). The sizes of NUDGEs in arcsec are dependent on cluster distance, therefore the correction fraction we applied is dependent on cluster redshift. The bottom right panel shows the fraction of sources recovered now binned by their recovered properties. The recovery fraction colourbar is measured from 0 - 2, to indicate which bins now include more than the 100 input sources. These plots roughly show that models with large-faint input values are recovered to have smaller-brighter values, bins observed to have recovery fractions over 1 are typically found toward the bottom-left.\\

\begin{figure}
\noindent
\includegraphics[width=0.48\textwidth,trim={4.5cm 9.0cm 5.4cm 59.4cm},clip]{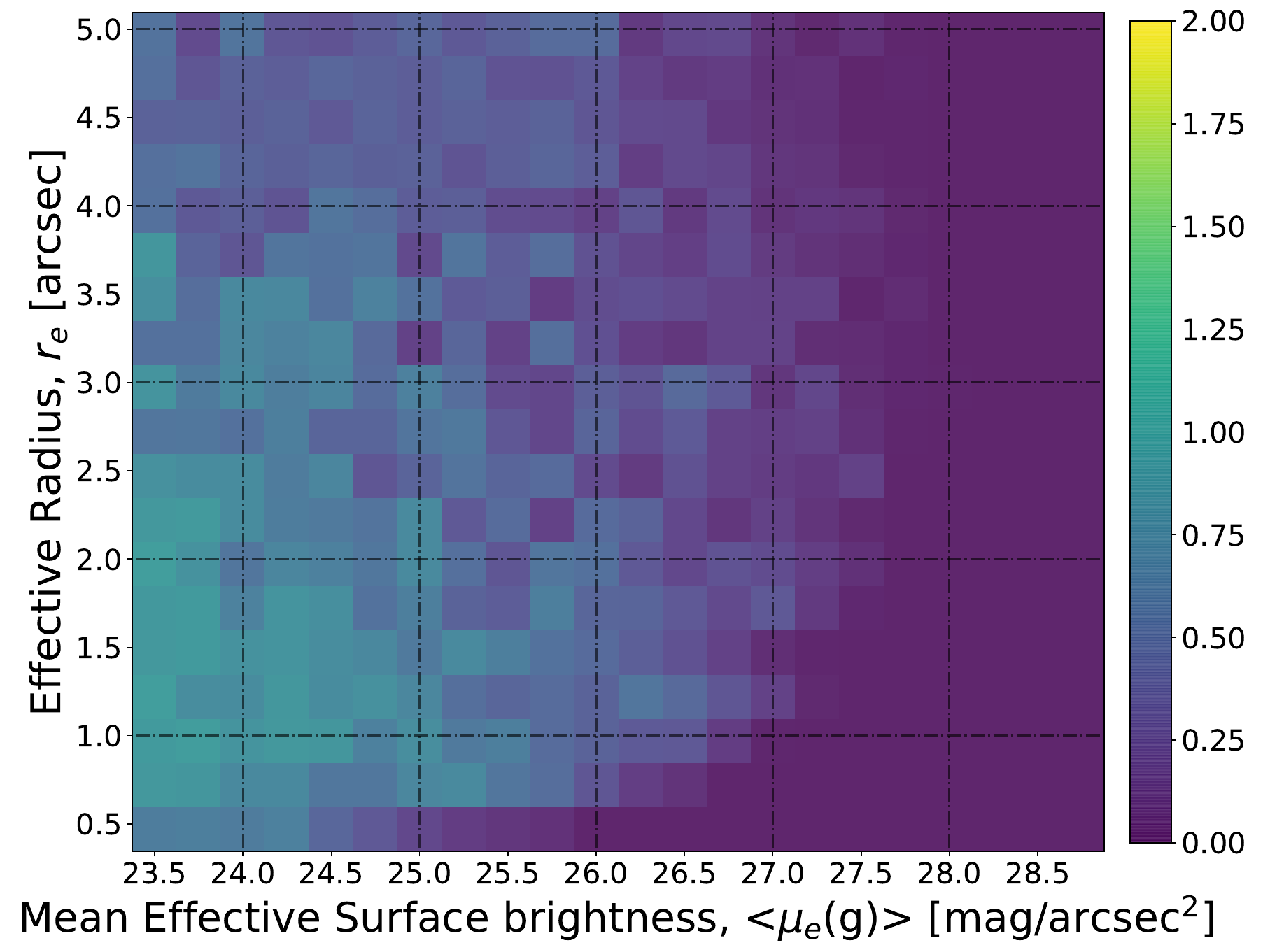}
\vspace{-2.5ex}
\caption{\label{fig:completeness_diff} Model accuracy plot comparing the input and recovered properties in the completeness test (Figure \ref{fig:completeness_det_char}). Models with less than 0.25 dex difference between input model properties and recovered properties after median property correction are shown. We see good accuracy at small radii and brighter surface brightnesses, which decreases toward larger and fainter values.}
\end{figure}

The second aspect of the completeness test is to compare the recovered properties to the input values. Recovered and input properties with similar values indicate the models were well fitted, whereas large discrepancies indicate poorly recovered models. Figure \ref{fig:completeness_diff} shows the accuracy of successfully recovered models, where successfully recovered models are defined to have an effective radius within~0.25~arcsec and surface brightness within~0.25~mag/arcsec$^2$ of the input values. There is a large difference between the detection and model accuracy plots, showing that while we detected a good portion of the models, their properties are not always well fitted. The smallest and brightest sources are well recovered, however, most of the regions with $\geq\,$90\% detection only reach an accuracy of $\sim\,$50\%. \\

The completeness study shows that we have a satisfactory detection fraction of $\geq\,$90\% for sources with a mean effective surface brightnesses <$\mu_e$(g)>~$\leq$~27.5~mag/arcsec$^2$ as modelled by \textsc{galfit}. The recovery fraction determined in the detection plot is given in Table \ref{tab:cl_table1} as the completeness percentage and is the last step to get our final NUDGE counts.

\subsection{NUDGE selection \label{sec:udg_select}}

\cite{vannest2022} have shown that the selection criteria of UDGs in low density environments impact their resulting measured properties and subsequently their predicted formation pathways, noting that this effect decreases with increasing density. Therefore, to avoid introducing any biases we only apply criteria in effective radius and mean effective surface brightness similar to \cite{vanderburg2016} (~24.0~$\leq\, \langle\mu_e(g)\rangle\, \leq$~\surfbright{26.5} and 1.5~$\leq\,r_e\,\leq$~7.0$\,$kpc). These criteria were used to directly compare to the work by \cite{vanderburg2016,vanderburg2017} on UDGs in clusters. However, this range includes slightly brighter sources than the \cite{vandokkum} criteria, therefore we identify our sample as UDG-analogues (NUDGEs, previously defined by \cite{forbes2024} as UDGs that do not strictly pass the \cite{vandokkum} criteria). The same selection criteria were applied to the background annulus. Note that during the completeness estimation the properties of the models used extend beyond these criteria.

\subsection{Removing sources of contamination}

In the Subaru images there are bright cluster galaxies and background/foreground sources which may obscure the detection and characterisation of NUDGEs due to light extending from them and illuminating the surrounding regions in the images. Furthermore, some artefacts appear in the images which may have been caused during data reduction, possibly due to over-subtraction of the background, residual artefacts caused by bright sources, or the dithering patterns and CCD gaps. Another issue is that intracluster light (ICL) illuminates the regions within clusters around galaxies. Typically the effect is prominent in central regions and decreases with increasing radius. The implication of this is that the detection of low surface brightness galaxies is more incomplete near the centre compared to outer regions. Some studies have used a correction with respect to distance from the cluster centre to correct for UDGs not detected due to ICL (e.g.\citealt{vanderburg2016}). However, the relationship between UDG density and the central regions of clusters is not well understood and applying any ICL correction may introduce biases into the UDG sample which we want to avoid.\\

We masked the brightest sources that showed clear background over-subtraction (dips) around them, artefacts, as well as the central $0.1~R/R_{200}$ for all clusters to avoid incompleteness due to light from bright central galaxies, and to maintain statistical uniformity throughout our sample. The authors of this paper independently examined all of the images identifying all potentially bright sources and artefacts. These regions were manually masked using the \textsc{DS9} software and region files, and all sources identified in these regions were removed from the results.\\

Additionally, we divided the clusters and background regions in our sample into 8 equal sections and masked sections which overlapped with background clusters beyond our sample in the redMaPPer cluster catalogue (z<0.32). This removes possible contamination from NUDGEs in background clusters.\\

We correct for the missing NUDGEs in these regions in two different ways: the first is to calculate an overall fraction of missing NUDGEs using the total number counted and the fraction of missing area, assuming that the masked regions have the same NUDGE density as the average density of the rest of the cluster. The second method involves adding a fractional weighting to each remaining NUDGE equivalent to the missing area. This weighting is useful for statistically studying the collective NUDGE properties of each cluster within the $R_{200}$.\\

\subsection{\label{bgcontamination}Background contamination}

Generally photometric redshifts and comparison to the cluster red sequences are used to determine the membership of UDGs in clusters due to their low surface brightnesses and the limits of spectroscopic redshift observations. However, using photometric redshifts may introduce contamination from background and foreground sources with similar colours to the cluster red sequence. Also, implementing a red sequence association results in the bluer cluster members being excluded from the cluster sample. To quantify the number of background and foreground sources while avoiding introducing a colour bias to our cluster sample, we did the same analysis on background annuli around our clusters. We used these background samples of NUDGE-like sources to estimate the contamination for each cluster in our sample and to determine the true number of NUDGEs in these clusters.\\

\begin{figure}
\centering
\begin{minipage}{0.5\textwidth}
\includegraphics[width=1\textwidth,trim={2cm 1.0cm 3cm 2cm},clip]{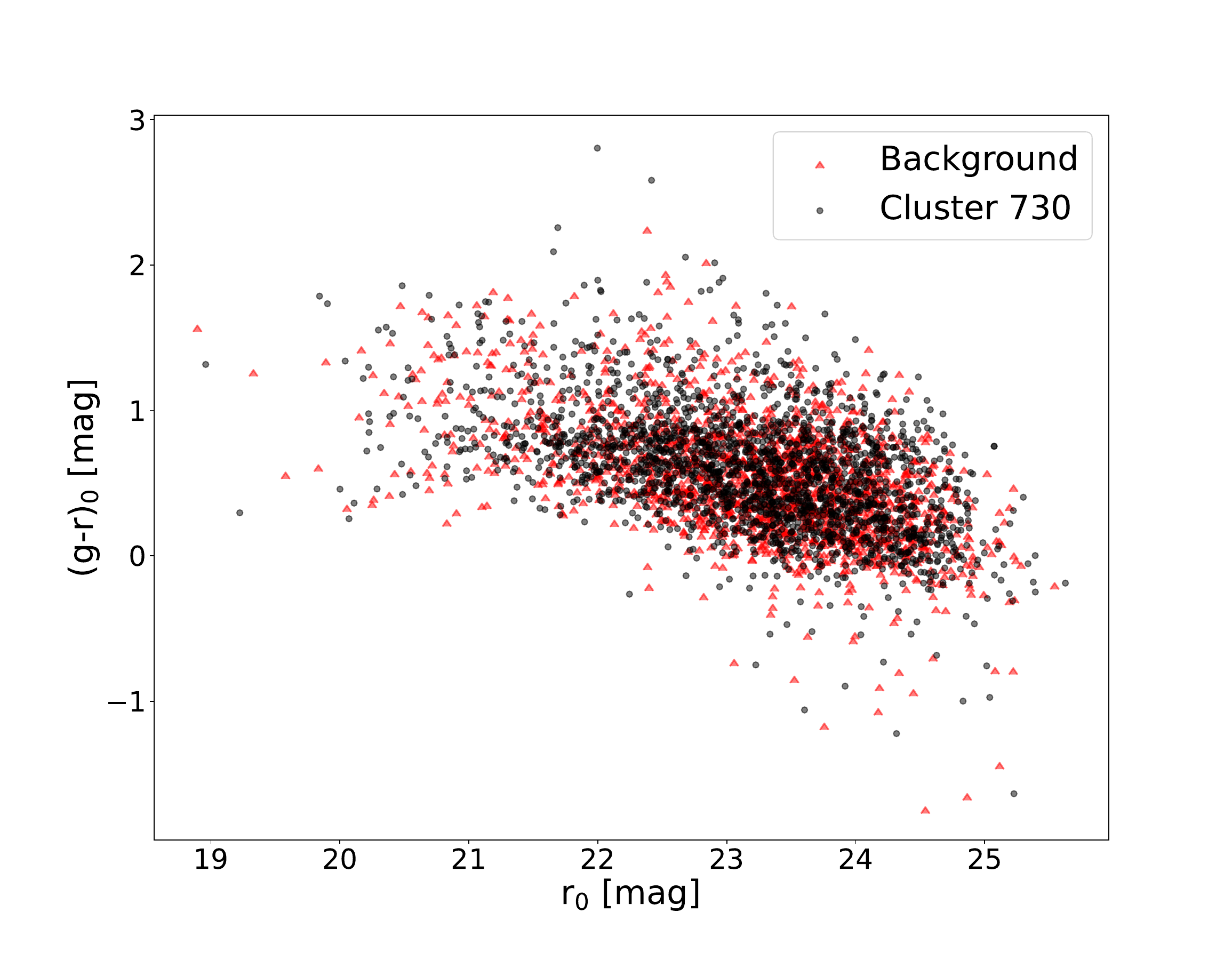}
\caption{\label{fig:bg_cont1} Comparison of the properties of the cluster NUDGEs (black points) and background NUDGE-like sources (red points), for an example cluster (ID: 730). The distributions in the colour-magnitude plane overlap, suggesting the sample of cluster NUDGE candidates and background NUDGE-like sources are quite similar and indistinguishable in the colour-magnitude plane. }
\centering
\includegraphics[width=1\textwidth,trim={2cm 1.0cm 3cm 5cm},clip]{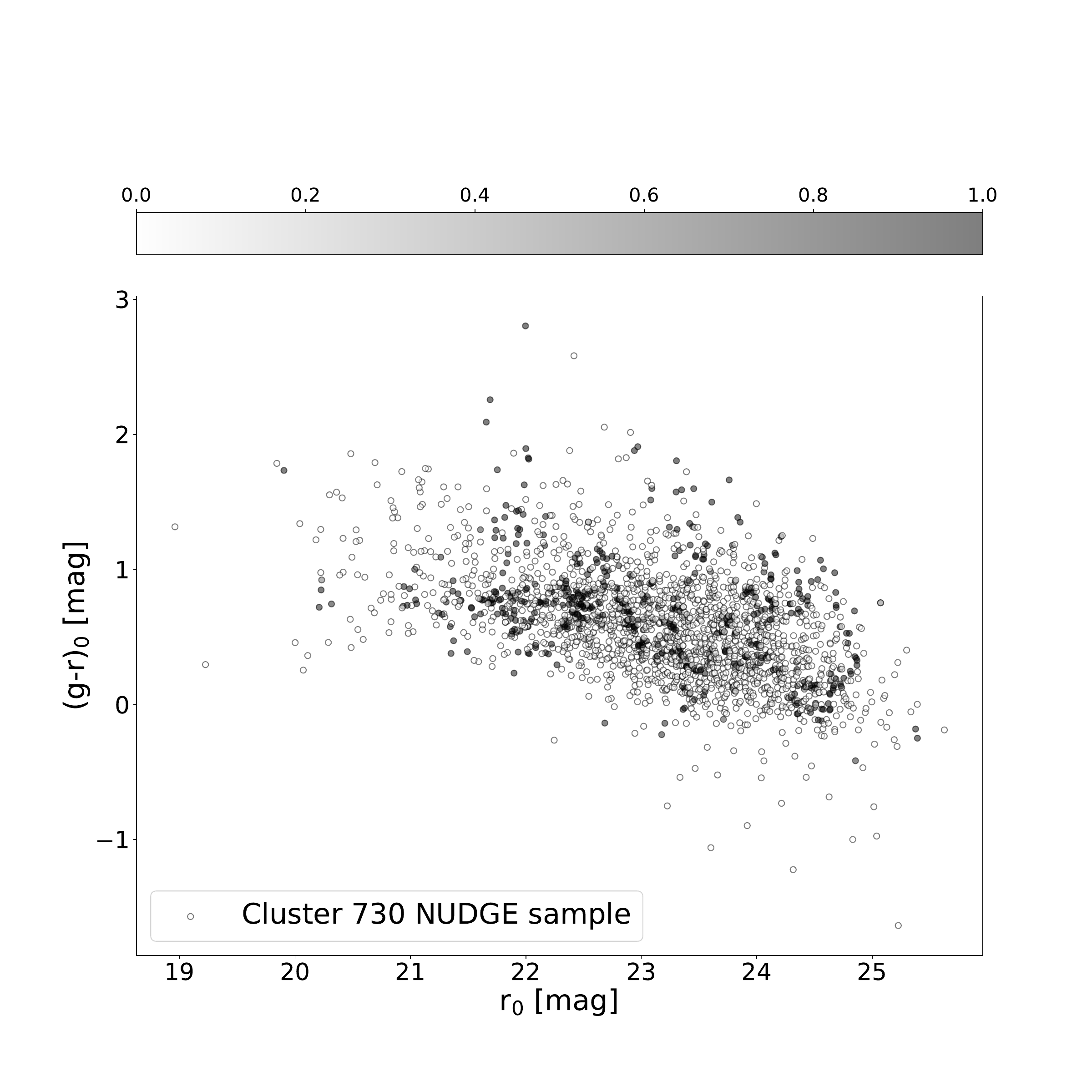}
\caption{\label{fig:bg_cont2} The resulting colour vs magnitude weighted cluster NUDGE sample for cluster 730. The grey scale shows how likely a galaxy is to remain after the background subtraction iterations, thus darker points represent an excess of NUDGEs in the cluster over the background. }
\end{minipage}
\end{figure}

After masking the bright sources and artefacts in the clusters and background annuli their areas were often no longer exactly equivalent. To effectively measure the background contamination we weighted the cluster and background sources, scaling the weighting value to the $R_{200}$ area of each cluster. We then compared the NUDGE candidates from each cluster and background annulus region in the $(g\,-\,r)$~-~$r$ magnitude plane and systematically subtracted NUDGE candidates (using their weightings) from the list of cluster members based on their proximity to background sources in the colour-magnitude plane, as illustrated in Figures~\ref{fig:bg_cont1}~and~\ref{fig:bg_cont2}. We also investigated the $(g\,-\,r)\,-\,(r\,-\,i)$ plane and found that the samples overlap similarly to the $(g\,-\,r)\,-\,r$ plane, with no significant difference in colour between the background and cluster sample distributions. To prevent introducing any bias caused by the ordering of source subtraction, we repeated the removal process for each cluster 20 times, reordering the background sources randomly. We also computed the properties of the background subtracted samples after 5, 10 and 20 iterations, to determine the variation in their properties and found the samples were near identical with little change in the properties (this is demonstrated with an example cluster in Figure~\ref{fig:iteration}). This implies that the same galaxies (or galaxies with similar properties) are removed regardless of the order with which the cluster list is compared to the background list objects. The NUDGE candidates for cluster 9723 are shown in the right panel of Figure \ref{fig:cluster9723}.\\

\begin{table*}
    \centering
\caption{The sample of clusters studied in the Subaru wide field survey, identified in the redMaPPer cluster catalogue (sorted by $M_{200}$).}\label{tab:cl_table1}
    \begin{tabular}{l|c|c|c|c|c|c|c|c|c|c|c|c|l} \hline 
          ID&Cluster&  RA &  Dec&  z$_{\text{phot}\  \lambda}$&  $M_{200}$&  $R_{200}$&  NUDGE$_{cl}$& NUDGE$_{bg}$ &NUDGE$_{f}$  & Comp &NUDGE $_{c}$ \\
  & & J2000 & J2000 & &  [$10^{14}\,M_{\odot}$]&  Mpc&  &  &    & \%& \\ \hline
  1& 15889& 214.18& -1.62& 0.12& 0.95& 0.96& 513.6& 506.1& 7.5 $\pm\,{31.9}$& 90.9&8.2 $\pm\,{33.4}$\\\hline
  2& 18640& 174.46& 1.26& 0.12& 0.95& 0.96& 776.0& 727.7& 48.3 $\pm\,{38.8}$& 90.9&52.7 $\pm\,{40.5}$\\\hline
  3& 9033& 213.95& 0.26& 0.12& 0.96& 0.97& 448.2& 465.3& -17.1 $\pm\,{30.2}$& 90.4&0 \\\hline
  4& 17332& 349.52& 1.97& 0.12& 0.97& 0.97& 379.2& 396.6& -17.4 $\pm\,{27.9}$& 90.4&0 \\\hline
  5& 15735& 219.28& 3.67& 0.15& 0.99& 0.97& 266.7& 252.5& 14.2 $\pm\,{22.7}$& 84.4&16.4 $\pm\,{24.5}$\\\hline
  6& 14697& 32.76& -4.89& 0.13& 1.02& 0.98& 565.0& 457.8& 107.2 $\pm\,{32.0}$& 90.4&117.5 $\pm\,{33.5}$\\\hline 
  7& 12112& 212.5& -1.54& 0.11& 1.03& 0.99& 450.5& 391.7& 58.8 $\pm\,{29.0}$& 91.4&63.8 $\pm\,{30.2}$\\\hline
  8& 13527& 221.83& -1.26& 0.15& 1.04& 0.99& 668.7& 609.0& 59.7 $\pm\,{35.7}$& 84.4&69.0 $\pm\,{38.4}$\\\hline
  9& 16596& 158.17& 2.7& 0.13& 1.04& 0.99& 385.9& 417.5& -31.6 $\pm\,{28.3}$& 90.4&0 \\\hline
  10& 10986& 176.53& -1.09& 0.12& 1.06& 1.0& 1066.6& 938.1& 128.5 $\pm\,{44.8}$& 90.9&140.2 $\pm\,{46.8}$\\\hline
  11& 11853& 172.28& 0.61& 0.12& 1.06& 1.0& 304.9& 305.9& -1.0 $\pm\,{24.7}$& 90.9&0 \\\hline
  12& 18042& 200.36& -0.69& 0.11& 1.08& 1.01& 826.0& 779.4& 46.6 $\pm\,{40.1}$& 91.4&50.7 $\pm\,{41.8}$\\\hline
  13& 6242& 192.28& -1.79& 0.09& 1.08& 1.01& 371.5& 357.9& 13.5 $\pm\,{27.0}$& 93.9&14.4 $\pm\,{27.8}$\\\hline
  14& 16983& 336.23& -0.38& 0.14& 1.09& 1.01& 493.1& 431.6& 61.5 $\pm\,{30.4}$& 84.4&71.1 $\pm\,{32.7}$\\\hline
  15& 11469& 220.52& -0.82& 0.14& 1.09& 1.01& 1003.9& 963.9& 40.0 $\pm\,{44.4}$& 89.7&44.2 $\pm\,{46.6}$\\\hline
  16& 10769& 178.94& 4.38& 0.15& 1.1& 1.01& 805.4& 718.6& 86.8 $\pm\,{39.0}$& 84.4&100.4 $\pm\,{42.0}$\\\hline
  17& 18362& 179.55& -1.05& 0.13& 1.11& 1.01& 581.4& 554.3& 27.0 $\pm\,{33.7}$& 90.4&29.7 $\pm\,{35.3}$\\\hline
  18& 17416& 160.91& 0.61& 0.12& 1.11& 1.01& 1126.5& 1147.3& -20.8 $\pm\,{47.7}$& 90.9&0 \\\hline
  19& 13670& 202.14& -1.33& 0.14& 1.14& 1.02& 533.3& 441.9& 91.4$\pm\,{31.2}$& 84.4&105.7 $\pm\,{33.6}$\\\hline
  20& 15600& 178.81& -0.53& 0.13& 1.16& 1.03& 501.2& 440.2& 61.0 $\pm\,{30.7}$& 90.4&66.9 $\pm\,{32.1}$\\\hline
  21& 12317& 202.43& -1.62& 0.14& 1.17& 1.03& 503.9& 454.8& 49.1 $\pm\,{31.0}$& 89.7&54.1 $\pm\,{32.5}$\\\hline
  22& 9723& 179.08& -0.21& 0.11& 1.19& 1.04& 571.6& 529.1& 42.5 $\pm\,{33.2}$& 91.4&46.2 $\pm\,{34.6}$\\\hline
  23& 12741& 189.93& 2.7& 0.15& 1.24& 1.05& 991.6& 1064.6& -73.0 $\pm\,{45.3}$& 84.4&0 \\\hline
  24& 11537& 351.09& 0.32& 0.15& 1.26& 1.06& 200.2& 154.0& 46.2 $\pm\,{18.8}$& 84.4&53.4 $\pm\,{20.2}$\\\hline
  25& 9128& 154.93& -0.64& 0.1& 1.27& 1.06& 1290.6& 1091.6& 199.0 $\pm\,{48.8}$& 91.8&215.4 $\pm\,{50.8}$\\\hline
  26& 11934& 353.86& 1.06& 0.1& 1.29& 1.06& 651.1& 539.7& 111.4 $\pm\,{34.5}$& 91.8&120.5 $\pm\,{35.9}$\\\hline
  27& 6419& 197.33& -1.62& 0.08& 1.3& 1.07& 1064.5& 972.3& 92.3 $\pm\,{45.1}$& 93.9&97.9 $\pm\,{46.5}$\\\hline
  28& 8868& 199.6& -0.53& 0.11& 1.3& 1.07& 687.3& 670.4& 16.9 $\pm\,{36.8}$& 90.9&18.5 $\pm\,{38.5}$\\\hline
  29& 6922& 336.65& 5.19& 0.11& 1.3& 1.07& 330.2& 314.1& 16.1 $\pm\,{25.4}$& 91.4&17.4 $\pm\,{26.5}$\\\hline
  30& 6545& 170.69& 0.46& 0.09& 1.31& 1.07& 466.7& 428.3& 38.4 $\pm\,{29.9}$& 92.2&41.4 $\pm\,{31.1}$\\\hline
  31& 6437& 223.97& 1.89& 0.13& 1.32& 1.07& 581.2& 568.7& 12.5 $\pm\,{33.9}$& 89.7&13.8 $\pm\,{35.6}$\\\hline
  32& 15059& 223.51& 1.34& 0.15& 1.33& 1.08& 249.6& 325.8& -76.2 $\pm\,{24.0}$& 84.4&0\\\hline
  33& 7629& 29.16& -1.82& 0.15& 1.35& 1.08& 364.6& 381.1& -16.5 $\pm\,{27.3}$& 84.4&0 \\\hline\hline

    \end{tabular}
    
    \begin{flushleft}
        {Notes. Column 1: Identification number; Column 2: Cluster name as referenced in \cite{Redmapper}; Column 3,4: RA and DEC in degrees; Column 5: redshift determined in redMaPPer cluster catalogue using cluster galaxies photometric redshifts; Column 6: $M_{200}$ calculated using redMaPPer cluster richness (equation \ref{eqn:m200}); Column 7: $R_{200}$ calculated using equation \ref{eqn:r200}; Column 8: Number of NUDGE candidates in our clusters after applying similar cuts to \cite{vanderburg2016} and before background subtraction; Column 9: Number of NUDGE candidates in the background annulus after applying similar cuts to \cite{vanderburg2016}; Column 10: Number of NUDGE candidates after background subtraction and their Poisson errors; Column 11: Completeness percentage determined from redshift (section \ref{sec:completenesscorrection}); Column 12: Number of NUDGE candidates after performing completeness correction and their Poisson errors (Note that clusters that show no excess above background have 0 NUDGEs).}
    \end{flushleft}

\end{table*}

\begin{table*}
    \centering        
    \begin{tabular}{l|c|c|c|c|c|c|c|c|c|c|c|c|l} \hline 
          ID&Cluster&  RA &  Dec&  z$_{\text{phot}\  \lambda}$&  $M_{200}$&  $R_{200}$&  NUDGE$_{cl}$& NUDGE$_{bg}$ &NUDGE$_{f}$ & Comp &NUDGE $_{c}$ \\
  & & J2000 & J2000 & &  [$10^{14}\,M_{\odot}$]&  Mpc&  &  &   & \%& &\\ \hline
  34& 6356& 204.89& -0.28& 0.15& 1.42& 1.1& 980.9& 874.0& 106.8 $\pm\,{43.1}$& 84.4&123.5 $\pm\,{46.3}$\\\hline
  35& 9906& 178.38& -1.73& 0.13& 1.44& 1.11& 487.0& 411.2& 75.9 $\pm\,{30.0}$& 90.4&83.2 $\pm\,{31.4}$\\\hline
  36& 10447& 138.69& 3.35& 0.14& 1.45& 1.11& 598.4& 640.0& -41.6 $\pm\,{35.2}$& 84.4&0 \\\hline
  37& 4995& 167.77& 1.09& 0.1& 1.46& 1.11& 399.2& 503.4& -104.2 $\pm\,{30.0}$& 91.8&0 \\\hline
  38& 10225& 29.23& -4.41& 0.14& 1.47& 1.11& 725.9& 597.6& 128.3 $\pm\,{36.4}$& 89.7&141.6 $\pm\,{38.2}$\\\hline
  39& 5569& 213.72& 0.2& 0.12& 1.48& 1.11& 587.6& 496.1& 91.5 $\pm\,{32.9}$& 90.4&100.3 $\pm\,{34.4}$\\\hline
  40& 6836& 222.08& 3.53& 0.12& 1.52& 1.12& 709.0& 647.0& 62.0 $\pm\,{36.8}$& 90.9&67.6 $\pm\,{38.5}$\\\hline
  41& 2236& 181.11& 3.75& 0.15& 1.55& 1.13& 239.7& 220.4& 19.3 $\pm\,{21.5}$& 84.4&22.4 $\pm\,{23.1}$\\\hline
  42& 9754& 33.34& -2.88& 0.14& 1.58& 1.14& 814.0& 578.3& 235.7 $\pm\,{37.3}$& 84.4&272.4 $\pm\,{40.1}$\\\hline
  43& 3873& 146.55& 2.37& 0.13& 1.58& 1.14& 793.2& 851.3& -58.1 $\pm\,{40.6}$& 90.4&0 \\\hline
  44& 4312& 133.65& 0.64& 0.12& 1.65& 1.16& 1160.6& 1106.1& 54.5 $\pm\,{47.6}$& 90.9&59.4 $\pm\,{49.7}$\\\hline
  45& 6934& 176.1& 3.98& 0.13& 1.66& 1.16& 877.5& 935.4& -57.9 $\pm\,{42.6}$& 90.4&0 \\\hline
  46& 3783& 198.06& -0.97& 0.09& 1.79& 1.19& 1001.8& 915.2& 86.6 $\pm\,{43.8}$& 93.9&91.9 $\pm\,{45.1}$\\\hline
  47& 4875& 200.95& 1.11& 0.11& 1.82& 1.19& 1576.1& 1396.1& 171.0 $\pm\,{54.4}$& 91.4&185.8 $\pm\,{56.7}$\\\hline
  48& 1961& 171.84& 0.14& 0.13& 1.83& 1.2& 586.2& 543.6& 42.6 $\pm\,{33.6}$& 90.4&46.7 $\pm\,{35.2}$\\\hline
  49& 4322& 36.44& -3.21& 0.14& 1.86& 1.2& 790.0& 827.1& -37.1 $\pm\,{40.2}$& 84.4&0 \\\hline
  50& 4198& 138.41& 3.29& 0.14& 1.86& 1.2& 768.2& 717.9& 50.4 $\pm\,{38.6}$& 84.4&58.2 $\pm\,{41.4}$\\\hline
  51& 5365& 160.97& 1.06& 0.11& 1.87& 1.2& 1814.1& 1780.0& 34.1 $\pm\,{60.0}$& 91.4&37.0 $\pm\,{62.5}$\\\hline
  52& 3374& 199.82& -1.0& 0.08& 1.92& 1.22& 1485.0& 1317.5& 167.5 $\pm\,{52.9}$& 93.9&177.7 $\pm\,{54.5}$\\\hline
  53& 3682& 177.59& -0.59& 0.14& 2.06& 1.24& 869.7& 727.4& 142.3 $\pm\,{40.0}$& 89.7&157.0 $\pm\,{42.0}$\\\hline
  54& 2185& 160.25& -0.7& 0.13& 2.11& 1.26& 1338.5& 1361.9& -23.4 $\pm\,{52.0}$& 90.4&0 \\\hline
  55& 3365& 0.96& 2.07& 0.1& 2.29& 1.29& 1139.8& 1061.4& 78.4 $\pm\,{46.9}$& 91.4&85.2 $\pm\,{48.9}$\\\hline
  56& 1799& 153.44& -0.12& 0.09& 2.29& 1.29& 1132.1& 965.1& 167.0 $\pm\,{45.8}$& 92.2&180.1 $\pm\,{47.6}$\\\hline
  57& 1505& 29.09& 1.04& 0.08& 2.49& 1.33& 1412.3& 1196.2& 216.2 $\pm\,{51.1}$& 94.1&228.9 $\pm\,{52.6}$\\\hline
  58& 1443& 175.87& -1.74& 0.11& 2.6& 1.35& 1200.7& 1092.7& 108.0 $\pm\,{47.9}$& 91.4&117.3 $\pm\,{59.9}$\\\hline
  59& 1808& 189.48& -0.27& 0.14& 2.67& 1.36& 1271.1& 1122.6& 148.5 $\pm\,{48.9}$& 89.7&163.8 $\pm\,{51.4}$\\\hline
  60& 1454& 208.6& -1.05& 0.14& 2.73& 1.37& 832.1& 775.6& 56.5 $\pm\,{40.1}$& 84.4&65.3 $\pm\,{43.1}$\\\hline
  61& 1079& 151.96& 0.53& 0.09& 3.02& 1.41& 1496.3& 1277.0& 219.2 $\pm\,{52.7}$& 92.2&236.4 $\pm\,{54.7}$\\\hline
  62& 800& 219.43& -0.32& 0.13& 3.74& 1.52& 1676.8& 1619.9& 56.9 $\pm\,{57.4}$& 90.4&62.4 $\pm\,{60.1}$\\\hline
  63& 962& 33.67& -4.57& 0.15& 3.79& 1.53& 1321.5& 1388.3& -66.8 $\pm\,{52.1}$& 84.4&0 \\\hline
  64& 730& 34.15& -2.12& 0.14& 3.79& 1.53& 1958.2& 1566.1& 392.1 $\pm\,{59.4}$& 84.4&453.2 $\pm\,{63.8}$\\\hline
  65& 39& 180.11& 3.35& 0.14& 7.96& 1.95& 988.8& 913.3& 75.5 $\pm\,{43.6}$& 89.7&83.2 $\pm\,{45.8}$\\\hline
  66& 46& 213.6& -0.38& 0.14& 8.34& 1.98& 2000.9& 1887.4& 115.6 $\pm\,{62.4}$& 89.7&127.5 $\pm\,{65.5}$\\\hline\hline

    \end{tabular}
    
    \begin{flushleft}
        {Notes. Table continues.}
    \end{flushleft}
\end{table*}

\section{Results : NUDGE properties}

\begin{figure*}
\centering
\centering
\includegraphics[width=0.85\textwidth,trim={4.5cm 3.5cm 4.5cm 3.5cm},clip]{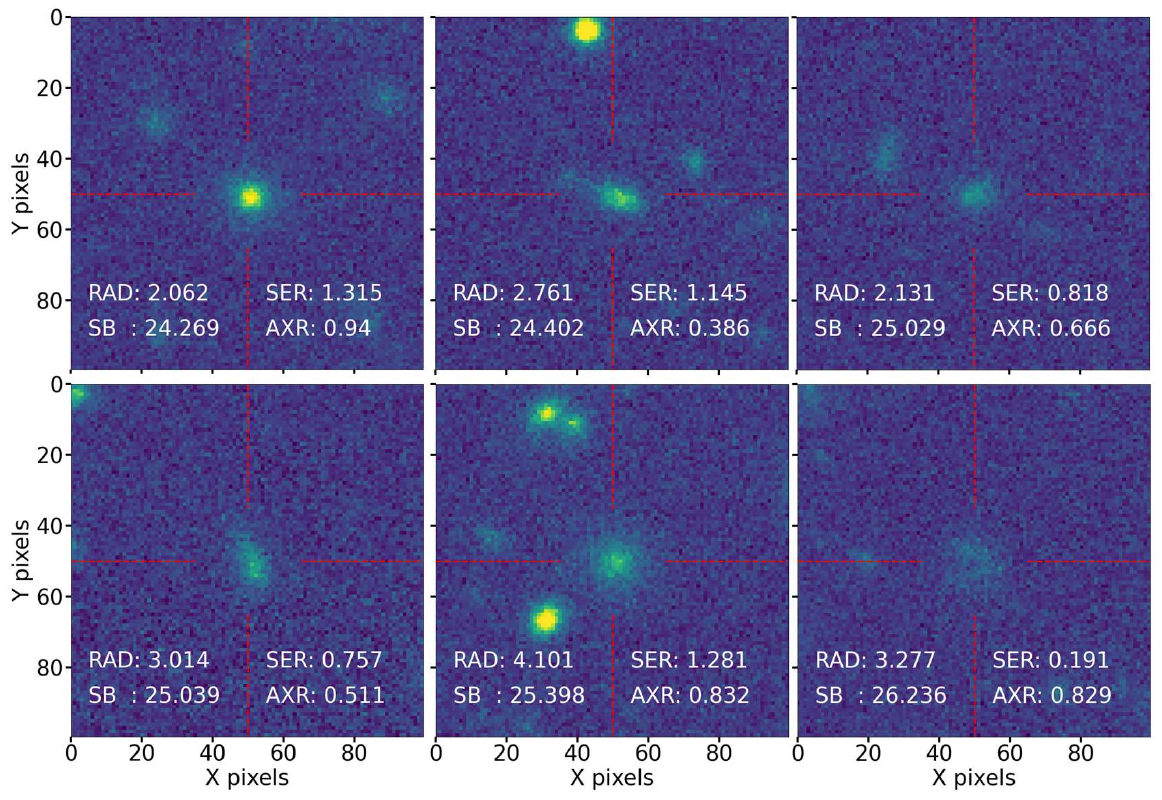}
\caption{\label{fig:UDG_images}Examples of NUDGEs identified in this study, shown in the Subaru g-band. The properties shown were determined with \textsc{galfit}, the radius ("RAD") is in kpc, the mean effective surface brightness ("SB") is in mag/arcsec$^2$. "SER" and "AXR" refer to the S\'ersic index and axis ratio respectively.}
\end{figure*}

The NUDGE candidates in our sample of redMaPPer clusters comprises all the identified and characterised NUDGEs after background subtraction (section~\ref{bgcontamination}); a total of 5057 NUDGEs in 51 clusters. The number of NUDGEs represented here is calculated based on the weighted galaxy-by-galaxy subtraction method. For comparison, in the coma cluster ($\sim 10^{15}\,M_{\odot}$) \cite{vandokkum} found 47 UDGs. \cite{koda2015} followed up, identifying a total of 332 UDGs in Coma using Subaru archival data and \cite{vanderburg2016} found an average of 104 UDGs in 8 nearby galaxy clusters. A sample of 6 NUDGEs from our sample is shown in Figure \ref{fig:UDG_images}, identified and characterised in the g-band Subaru data with \textsc{SExtractor} and \textsc{galfit} respectively. These images highlight the wide variation in morphology of NUDGEs in this study. The properties of the redMaPPer clusters are given in Table~\ref{tab:cl_table1}, including the number of NUDGEs identified in each cluster (NUDGE$_{cl}$) and background annulus (NUDGE$_{bg}$) after applying a radius and surface brightness cut similar to \cite{vanderburg2016} (4579 NUDGEs, described below in section~\ref{result_char}) and the redshift dependent completeness percentage (described in section~\ref{sec:completenesscorrection}), resulting in a total weighted number of 5057 NUDGEs. In the following sections we discuss the radius vs surface brightness distribution, spatial distribution and the overall characteristics of the NUDGEs in our cluster sample.

\subsection{Radius vs Surface brightness}

\begin{figure}
\centering
\includegraphics[width=0.48\textwidth,trim={1.5cm 0.5cm 2.cm 1.9cm},clip]{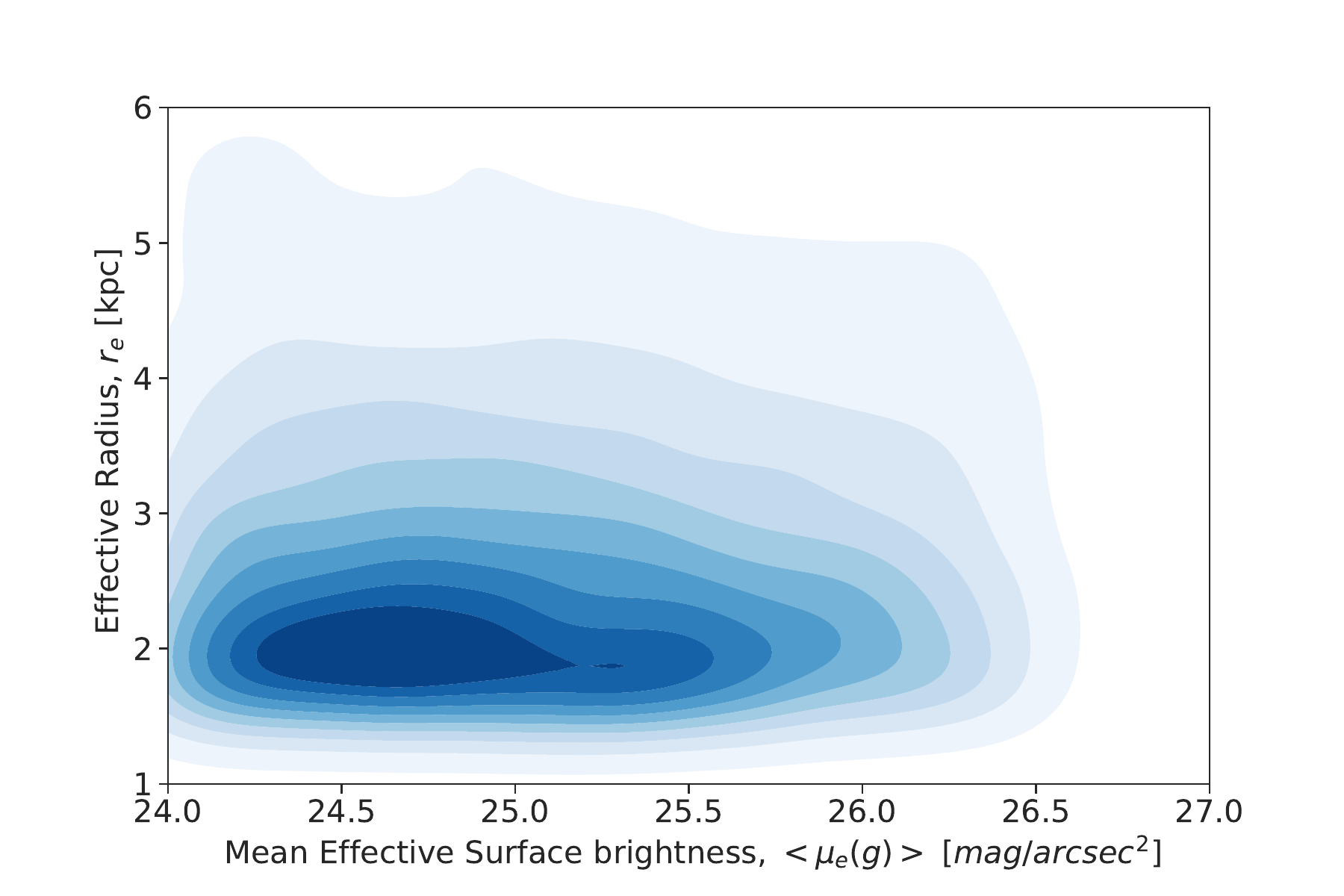}
\caption{\label{fig:udg_rad_Vs_sb2} Effective radius vs mean effective surface brightness density distribution for all the NUDGEs in our sample after applying the similar criteria to \protect\cite{vanderburg2016} (24.0~$\leq\, \langle\mu_e(g)\rangle\, \leq$~\surfbright{26.5} and 1.5~$\leq\,r_e\,\leq$~7.0$\,$kpc). The contours and colour gradient indicates the relative change in density.}
\end{figure}

Figure~\ref{fig:udg_rad_Vs_sb2} shows the distribution of NUDGEs after applying the similar criteria to \cite{vanderburg2016} (24.0~$\leq\, \langle\mu_e(g)\rangle\, \leq$~\surfbright{26.5} and 1.5~$\leq\,r_e\,\leq$~7.0$\,$kpc). The weighted mean effective radius of the UDG sample is  $r_e$~=~2.3~kpc and weighted mean of the mean surface brightness is $\langle\mu_e(g)\rangle$~=~25.1~mag/arcsec$^2$. This distribution of NUDGEs in the radius-surface brightness plane is comparable to other studies of UDGs in clusters \citep{vandokkum,vanderburg2016,mancera2019}, indicating a similar decrease in UDG counts towards larger radii and fainter surface brightnesses. Also, this distribution is affected by detection limitations, i.e. the smallest faint galaxies are difficult to observe and characterize.

\subsection{Spatial distribution}

The spatial distribution of UDGs in clusters provides information on where UDGs are abundant and where they are sparse. This may reflect the influence of the cluster environment on the existence of UDGs, i.e. whether UDGs preferentially form at the denser cores of clusters or in the lower density outskirts. Subsequently, this might give an insight into which formation mechanisms are likely to result in the abundances of UDGs in certain regions.\\

\begin{figure*}
\centering

\begin{minipage}{.365\textwidth}
  \centering
  \includegraphics[width=1.\linewidth,trim={0.0cm 5.05cm 3.45cm 0.3cm},clip]{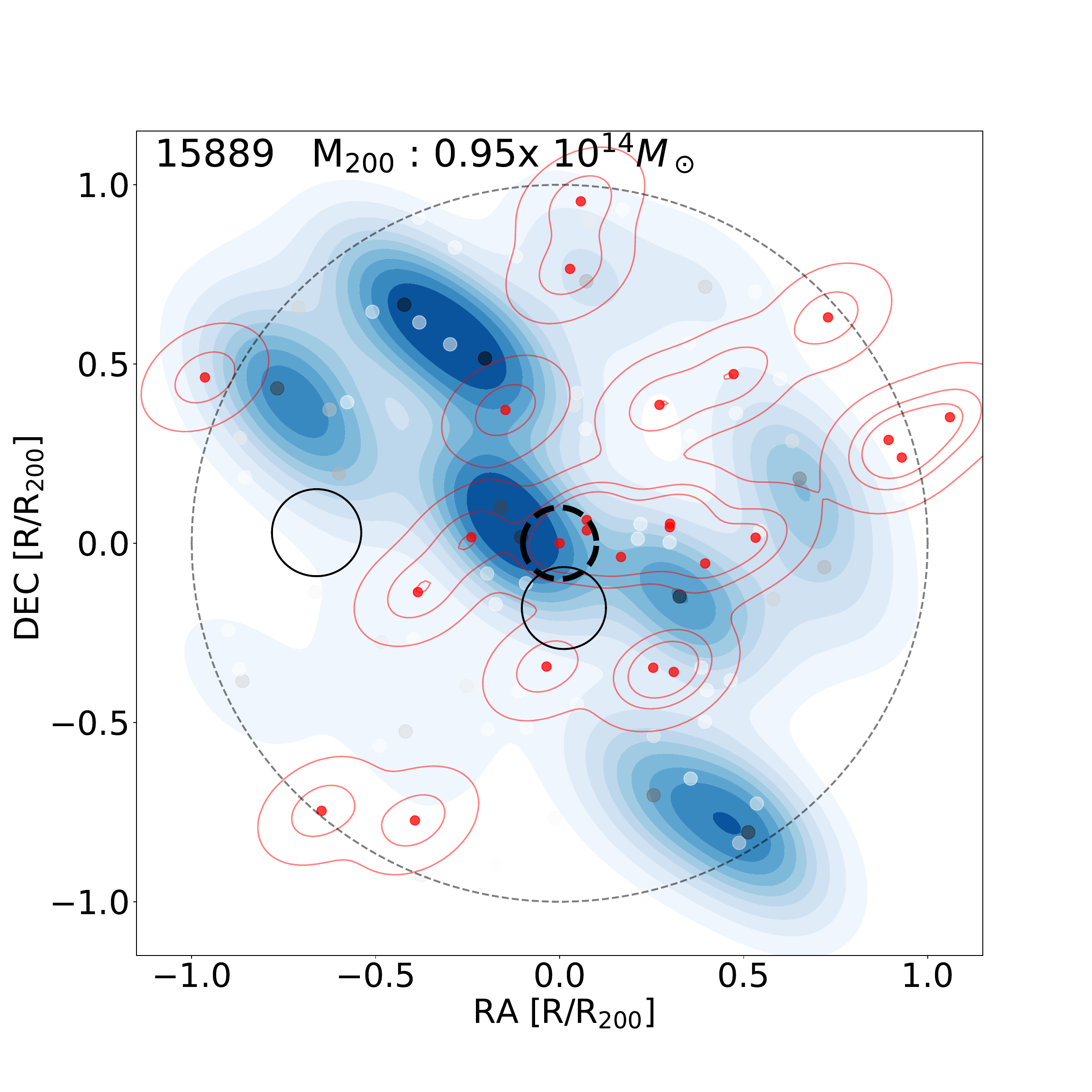}
  \label{fig:test1}
\end{minipage}%
\begin{minipage}{.315\textwidth}
  \centering
  \includegraphics[width=1\linewidth,trim={4.95cm 5.05cm 3.75cm 0.3cm},clip]{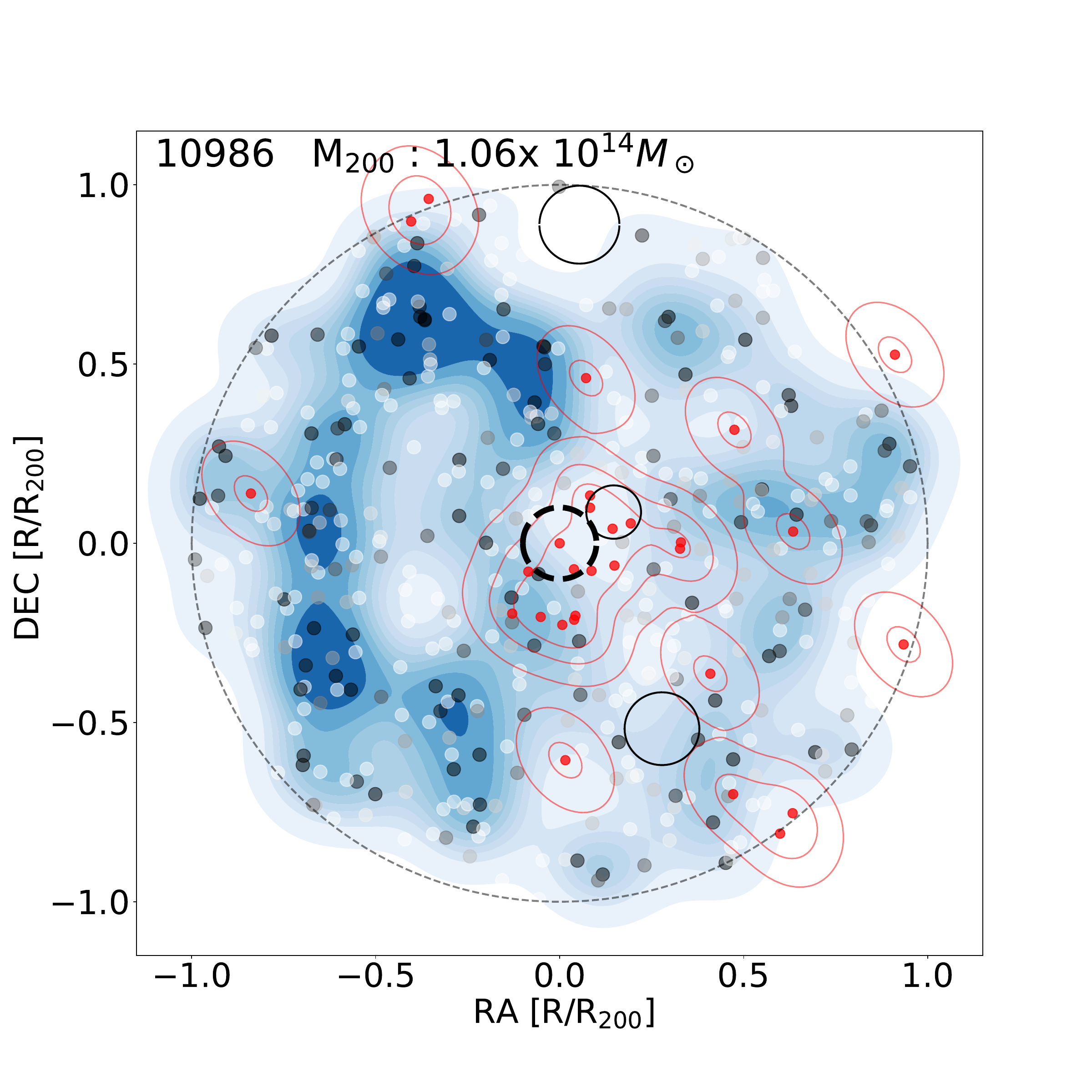}
  \label{fig:test2}
\end{minipage}%
\vspace{-2.5ex}
\begin{minipage}{.315\textwidth}
  \centering
  \includegraphics[width=1\linewidth,trim={4.95cm 5.05cm 3.75cm 0.3cm},clip]{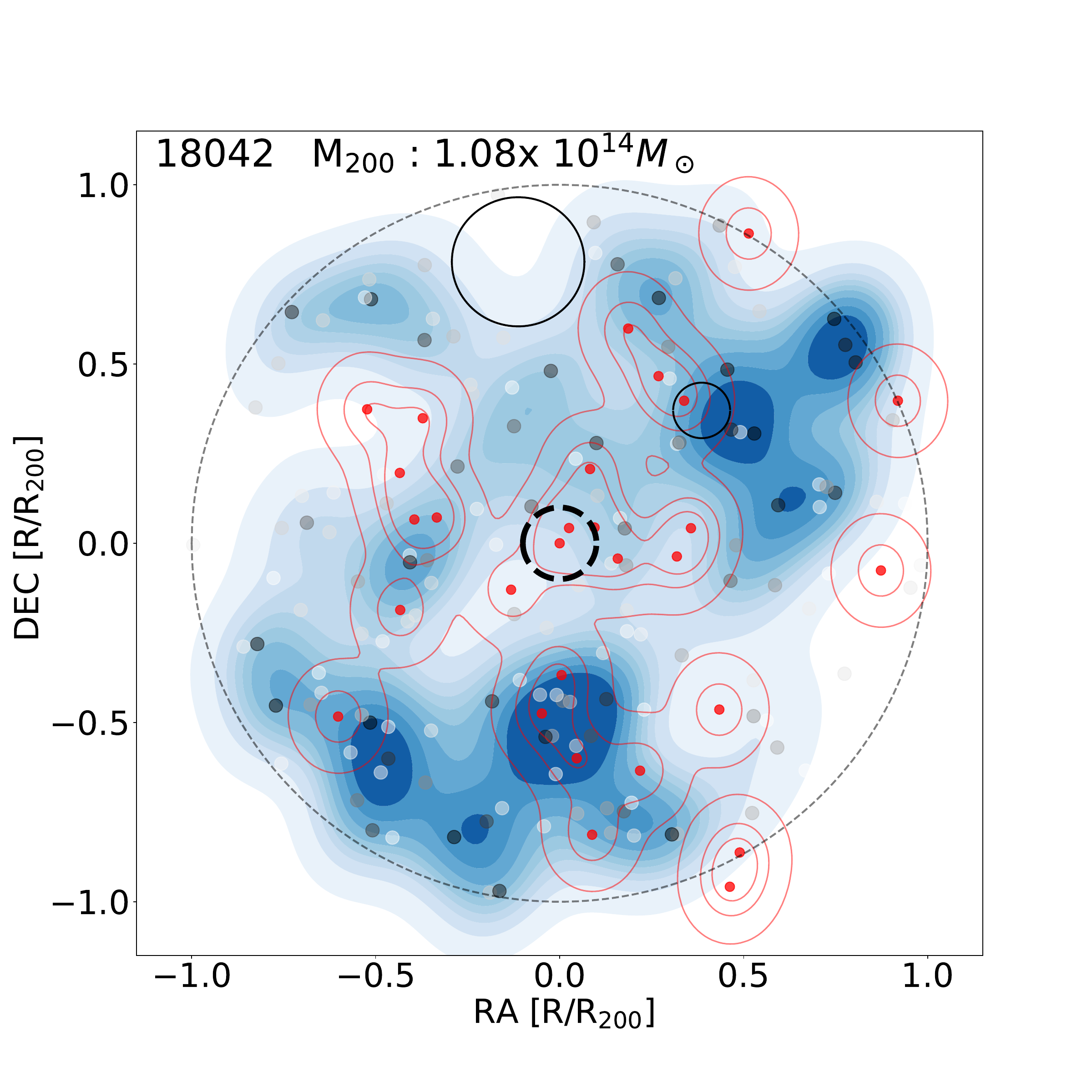}
  \label{fig:test3}
\end{minipage}
\begin{minipage}{.365\textwidth}
  \centering
  \includegraphics[width=1\linewidth,trim={0.0cm 5.05cm 3.45cm 4.8cm},clip]{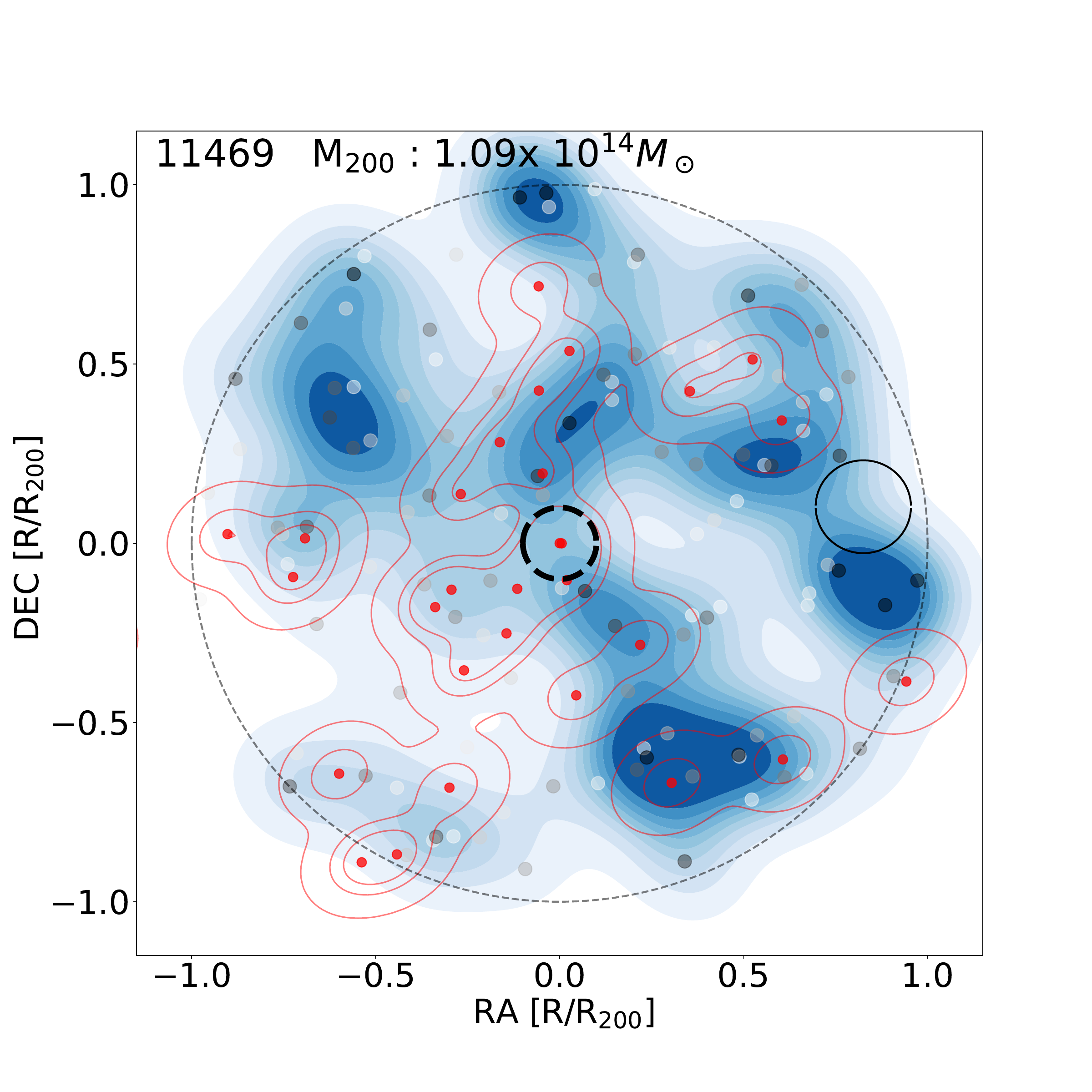}
  \label{fig:test4}
\end{minipage}%
\begin{minipage}{.315\textwidth}
  \centering
  \includegraphics[width=1\linewidth,trim={4.95cm 5.05cm 3.75cm 4.8cm},clip]{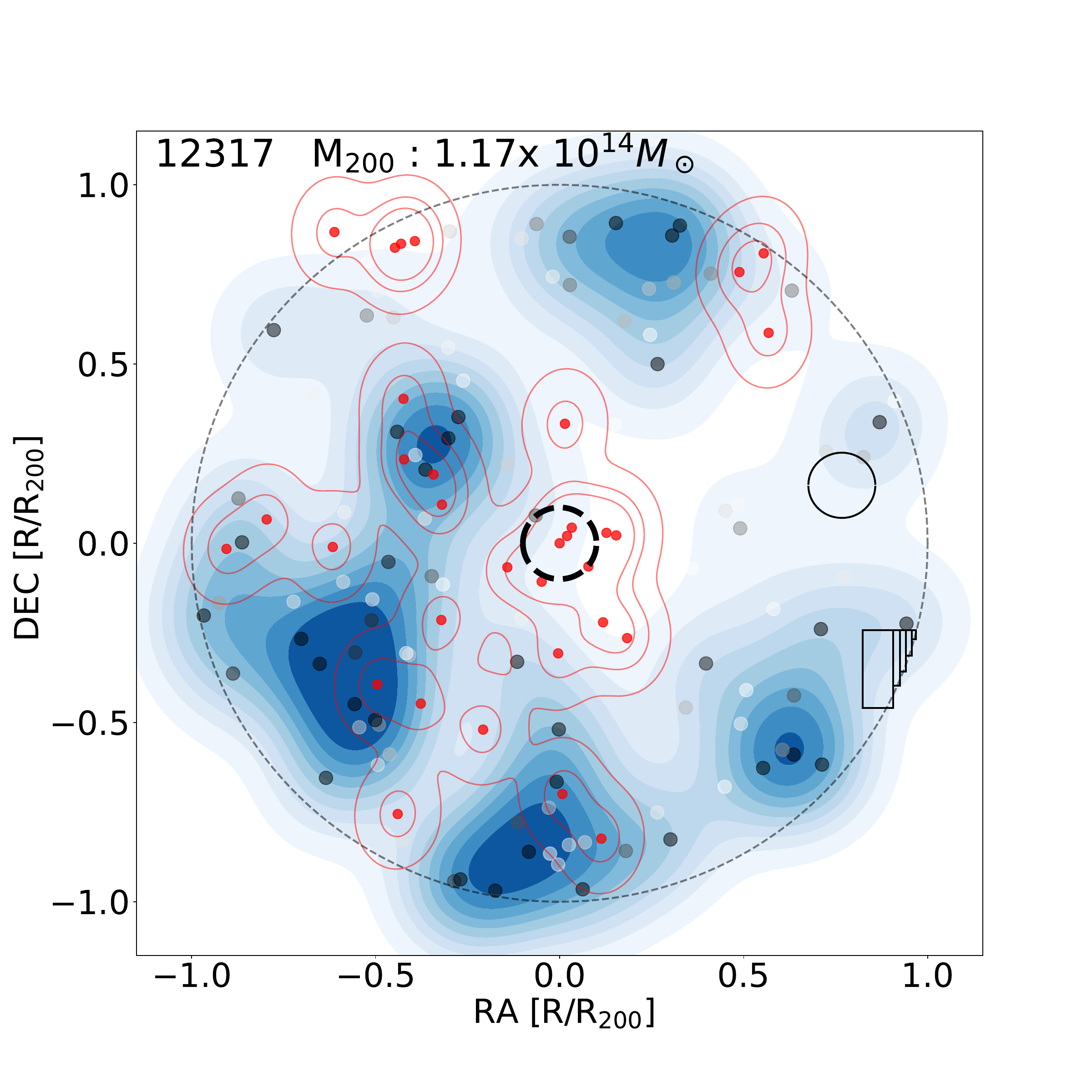}
  \label{fig:test5}
\end{minipage}%
\vspace{-2.5ex}
\begin{minipage}{.315\textwidth}
  \centering
  \includegraphics[width=1\linewidth,trim={5.0cm 5.05cm 3.75cm 4.8cm},clip]{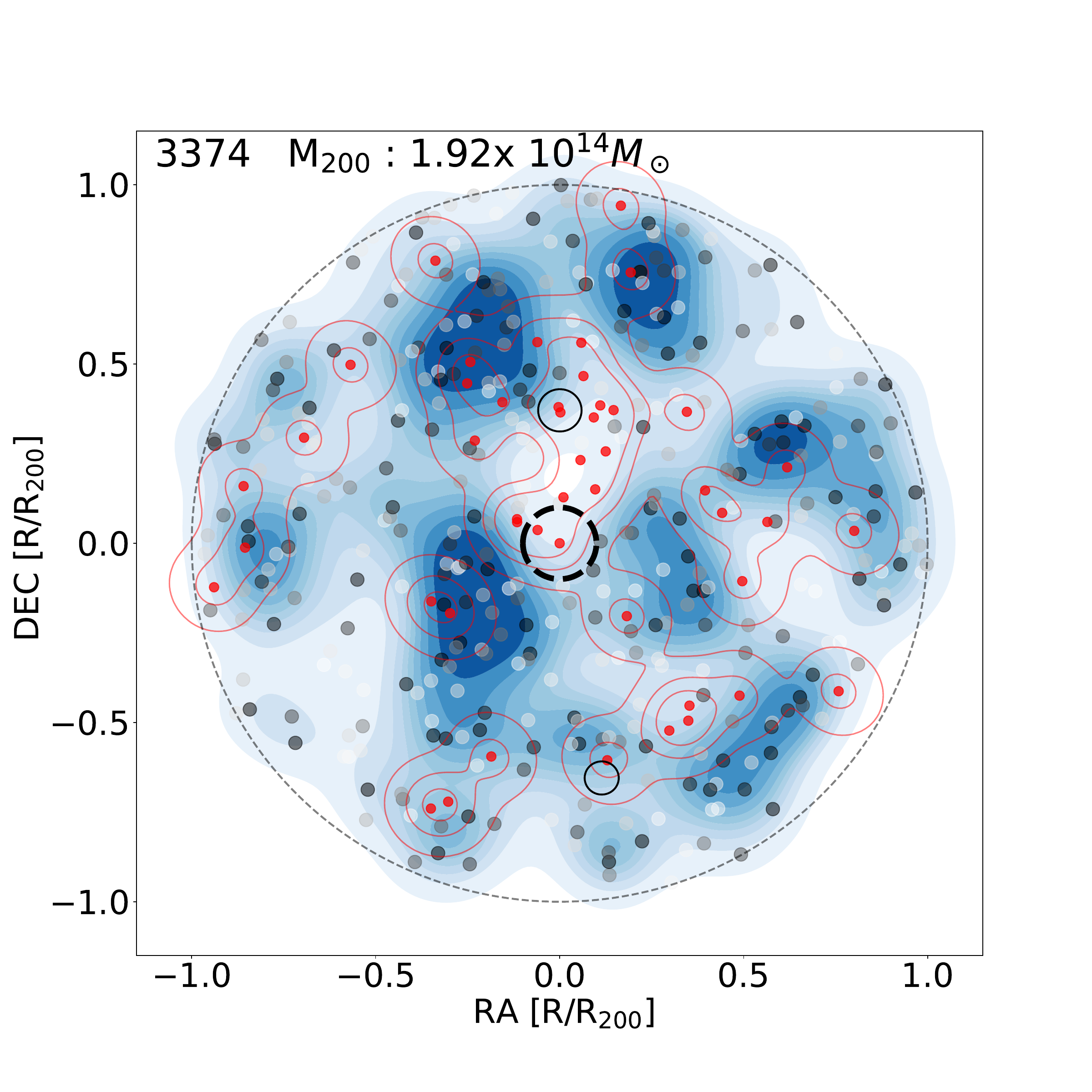}
  \label{fig:test6}
\end{minipage}
\begin{minipage}{.365\textwidth}
  \centering
  \includegraphics[width=1\linewidth,trim={0.0cm 2cm 3.45cm 4.8cm},clip]{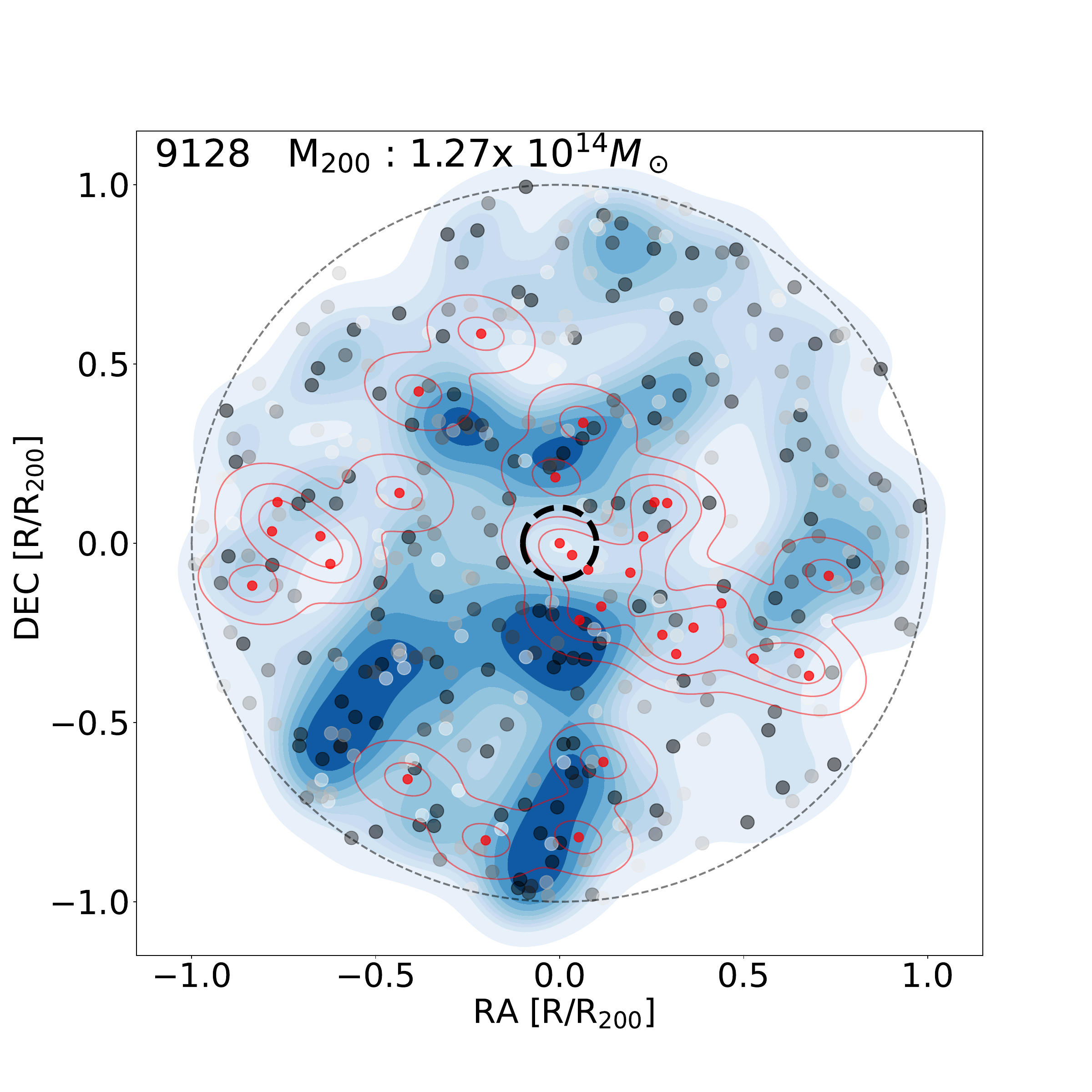}
  \label{fig:test7}
\end{minipage}%
\begin{minipage}{.315\textwidth}
  \centering
  \includegraphics[width=1\linewidth,trim={4.95cm 2cm 3.75cm 4.8cm},clip]{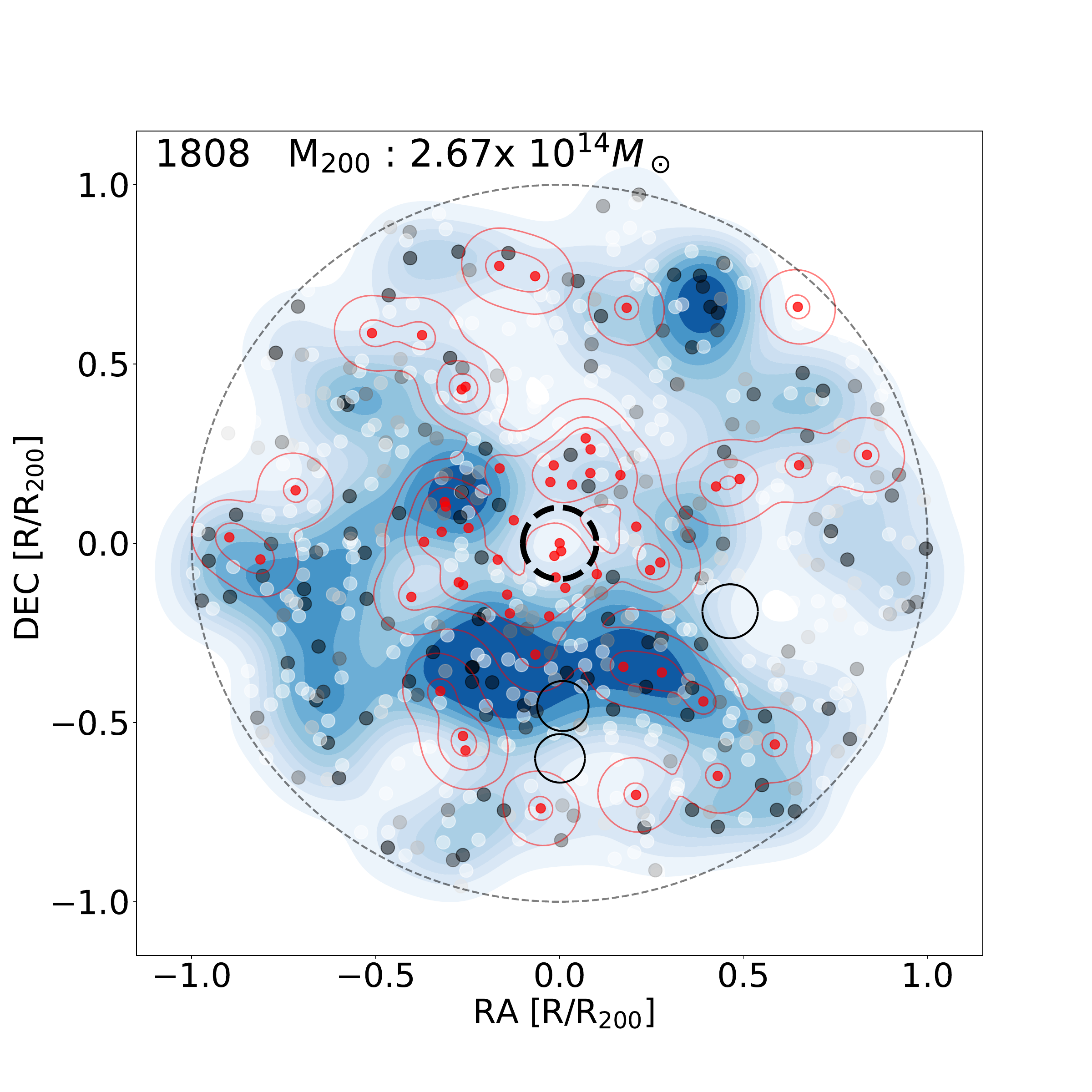}
  \label{fig:test8}
\end{minipage}%
\vspace{-2.5ex}
\begin{minipage}{.315\textwidth}
  \centering
  \includegraphics[width=1\linewidth,trim={5.0cm 2cm 3.75cm 4.8cm},clip]{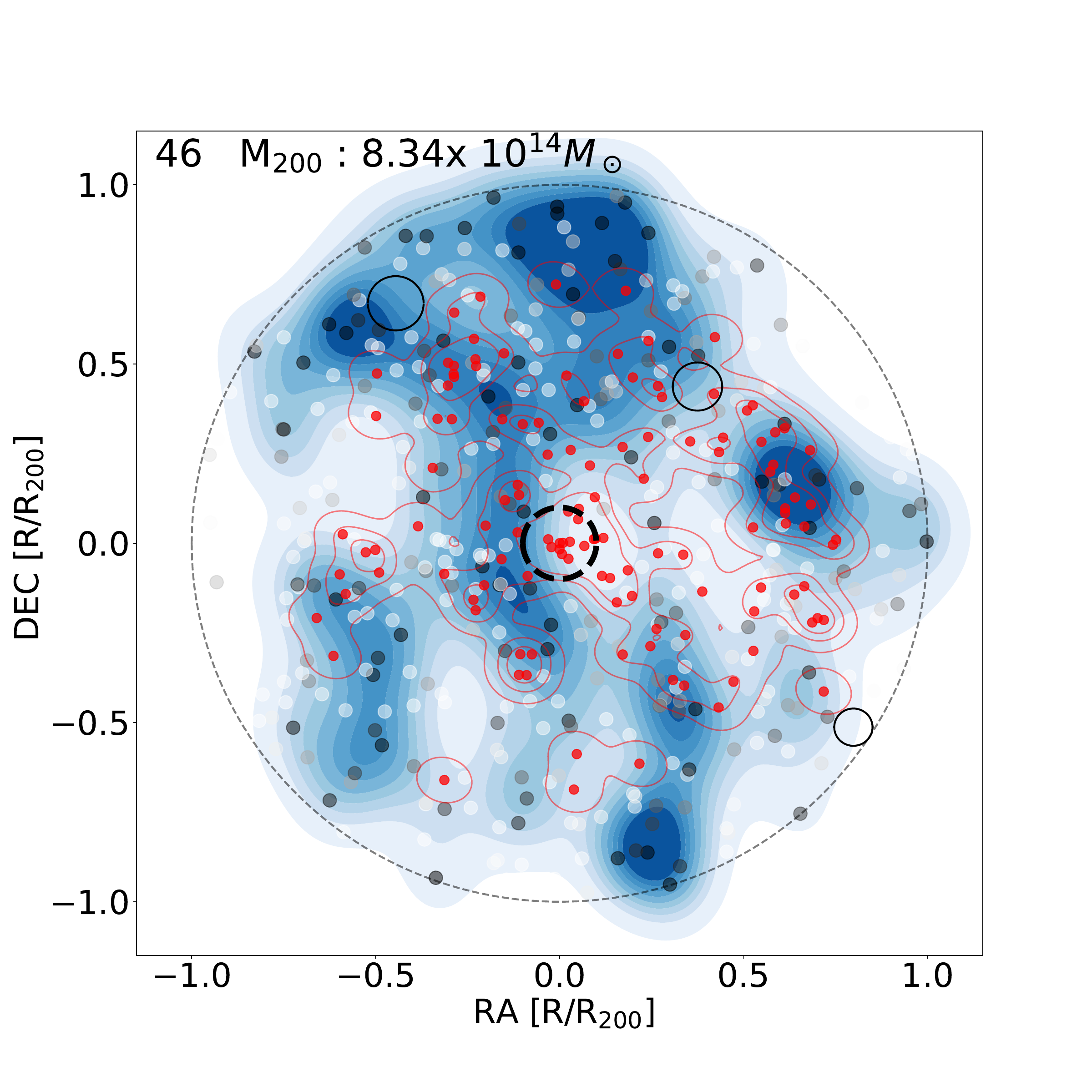}
  \label{fig:test9}
\end{minipage}
\caption{\label{fig:spatial}Spatial distribution of NUDGEs within the $R_{200}$ radius (indicated by outer dashed circle) of a sample of clusters, identified by their cluster ID number as referenced in \protect\cite{Redmapper} and cluster mass (M$_{200}$). The grey weighted dots indicate the positions of NUDGEs and their weightings after background subtraction, with the blue contour heat map showing their weighted densities. The red dots and contours indicate the positions of the redMaPPer galaxies and their densities. The selected sample of clusters demonstrates the wide variety of different cluster environments in our cluster sample. The thick dashed circle indicates the central 0.1R$_{200}$ masked region and the solid rectangles and circles indicate regions masked due to bright source over-subtraction and artefacts.}
\end{figure*}

The spatial distributions of NUDGEs and redMaPPer (bright) cluster members for a subset of our clusters are shown in Figure~\ref{fig:spatial}. The grey weighted dots indicate the spatial positions of NUDGEs and the blue shaded contours show their density distributions (determined using the \textsc{seaborn} statistical data visualization \textsc{python} package, \citealt{seaborn2021}). A caveat of using photometric data for distribution measurements is that we are looking at a projection. Without distance information, it could be that the projected spatial density overestimates the actual volumetric density. The red dots indicate the spatial positions of the redMaPPer galaxies and the red contours indicate their density distributions. The sample of clusters shown here displays the variation in spatial distributions of both the redMaPPer galaxies and NUDGEs. For example, in the top row of clusters, we observe that the over-dense regions of NUDGEs seem to be near dense regions of redMaPPer galaxies, often in between them. In the second row, we notice some of the dense NUDGE regions overlap with dense redMaPPer regions. In the bottom row we find some dense NUDGE regions are not near dense redMaPPer galaxy regions. We also observe variations of NUDGE density within clusters of similar masses. The clusters in the top row, the first and second column of the second row and the first column of the third row have similar masses ranging from $0.95\times10^{14}\text{M}_{\odot}$ to $1.27\times10^{14}\text{M}_{\odot}$, while their numbers of NUDGEs range from 8 to 178. These plots demonstrate the diversity in NUDGE density and spatial distribution in clusters.\\

The significance of the internal and external processes within clusters may vary depending on the spatial locations of the NUDGEs. The proximity between NUDGEs and the redMaPPer galaxies and dense cluster centre indicate that these NUDGEs likely experience strong tidal fields from massive galaxies and the cluster potential which may result in tidal stripping \citep{boselli2022}. NUDGEs found further from massive galaxies or in the outskirts of clusters do not experience such harsh tidal fields and therefore may form due to internal processes like strong gas outflows \citep{papstergis2017}.\\

Comparing the spatial distributions of NUDGEs to the distributions of redMaPPer cluster members we find no clear relation after visual inspection. We also compared the NUDGE density within each cluster to the proximity to redMaPPer galaxies. For each redMaPPer galaxy we measured the NUDGE density in radial annuli with the redMaPPer galaxy at the centre. The NUDGE density in our cluster sample remained constant with respect to the distance to redMaPPer galaxies. It is clear that NUDGEs are found in both high and low density regions of clusters, and therefore are likely to form via different mechanisms. 

\subsection{\label{result_char}Characteristics}
The formation of UDGs through different processes in different environments may result in contrasting UDG properties, which could shed light on their different formation processes. The number of UDGs - cluster halo mass relation determined by \cite{vanderburg2016} used a UDG sample with particular selection criteria (in the $r$-band). We applied similar cuts to our sources in the g-band:~24.0~$\leq\, \langle\mu_e(g)\rangle\, \leq$~\surfbright{26.5} and 1.5~$\leq\,r_e\,\leq$~7.0$\,$kpc, which makes our sample comparable to their work. Due to our criteria including slightly brighter sources than \cite{vandokkum}, we regard our sample as UDG-analogues (NUDGEs). We note some studies have implemented a red colour selection to their sample of UDGs (e.g. \citealt{roman2017a,lee2017,mancera2018}) to remove background and foreground contamination. The UDG criteria used may impact conclusions made on the potential formation pathways of UDGs as shown in \cite{vannest2022}, therefore, to avoid introducing biases in the measured properties we only apply selection cuts in radius and surface brightness. The properties of the NUDGEs in our cluster sample are shown in Figure~\ref{fig:hist}. The weighted mean of each property was calculated using $\Bar{x} = \Sigma[x_i \times w_i] / \Sigma [w_i]$ and the weighted standard deviation using $\sigma = \sqrt{\Sigma[w_i (x_i-\Bar{x})^2] \big/ \Sigma [w_i]}$, where $x_i$ and $w_i$ represent the values of properties and their weights respectively. \\

\begin{table}
    \centering
    \begin{tabular}{c|c|c|c}
        \hline
        Property & Weighted Mean & Weighted Std & Unit\\
        \hline
        Colour $(g-r)_0$ & 0.6 & 0.44 &  mag\\
        S\'ersic Index & 0.8 & 0.9 &  \\
        Axis Ratio & 0.6 & 0.2 &  \\
        Effective Radius & 2.7 & 1.1 & kpc \\
        Surface Brightness & 25.1 & 0.7 & mag/arcsec$^2$ \\
        \hline\hline
        
    \end{tabular}
    \caption{The weighted mean values and standard deviations of our NUDGE candidates after applying the criteria similar to \protect\cite{vanderburg2016}.}
    \label{tab:characteristicsummary}
\end{table}

\subsubsection{Colour}
The $(g\,-\,r)$ colour is an indicator of the metallicity of stellar populations and indicates the relative age of stellar populations. The clusters were identified in the redMaPPer cluster catalogue which in essence relies on the red sequence of galaxies in close proximity to each other. The magnitudes used were corrected for Galactic extinction \citep{schlafly2011}. The 1st row left panel of Figure~\ref{fig:hist} shows the~$(g\,-\,r)_0$ colour distribution and we observe a wide range of values with a weighted mean of~0.6~$\pm\,$0.44\,mag. The sample includes both blue and red NUDGEs, and we have not applied a cut or any limiting restriction on colour (see appendix \ref{app:colourUDGs} comparing properties of blue and red NUDGEs). The UDGs studied by \cite{vanderburg2016} have a mean~$(g\,-\,r)$ colour of 0.6, and the sample studied by \cite{mancera2019} show a similar mean~$(g\,-\,r)$ colour of 0.59. The colour we observe agrees with both of their results as well as their colour distributions, although \cite{mancera2019} applied a maximum colour cut of $(g\,-\,r)~\,<\,$1.2.\\      

\subsubsection{S\'ersic Index and Axis Ratio}
The S\'ersic index describes the light profile of galaxies and the relative distribution of the stellar light with respect to radius. The 1st row right panel of Figure~\ref{fig:hist} shows the distribution of S\'ersic indices. We observe a steep drop-off with increasing S\'ersic index and the distribution has a weighted mean of~0.8~$\pm\,$0.9. The large spread in S\'ersic indices implies a range of light profiles (see Figure \ref{fig:hist}). Note that we have applied a minimum S\'ersic index cut of $n\,>\,$0.05 to remove poor galaxy characterisations. Previous studies have shown UDGs to have exponential light profiles with mean S\'ersic indices, $n\,\sim\,$1 \citep{koda2015, vanderburg2016, mancera2019}. This distribution suggests a sizeable portion of our NUDGE sample have low S\'ersic indices, which indicates more gaussian-like profiles; flat cores with sharply truncated wings. These low S\'ersic index profiles are similar to some of the dwarf and low surface brightness galaxies studied in the Fornax \citep{eigenthaler2018, venhola2018} and Virgo clusters \citep{ferrarese2020} and may imply dwarfs as potential UDG progenitors. The tail of the S\'ersic index distribution shows a relatively small population of NUDGEs with S\'ersic indices $n\,>\,$2, suggesting a steep light profile typically found in elliptical galaxies \citep{kormendy2009}.\\

The axis ratio distribution of our sample is shown in the 2nd row left panel of Figure~\ref{fig:hist} and describes how elongated a galaxy appears. The axis ratio distribution of our NUDGE sample shows a large range of values from highly elliptical to round galaxies with a weighted mean axis ratio of~0.6~$\pm\,$0.2. This is comparable to the results by \cite{vandokkum}, who found a median axis ratio of 0.74 with a standard deviation of 0.16 in the Coma UDGs. \cite{mancera2019} found the mean of median axis ratios to be 0.72~$\pm\,$0.15 within the $R_{200}$ of the eight nearby clusters in their sample. The sample of dwarf and low surface brightness galaxies in the Fornax and Virgo clusters are more spherical with mean axis ratios of $b/a=0.72\pm$0.16 and $b/a=0.78\pm$0.16 respectively \citep{eigenthaler2018,ferrarese2020}.\\

\subsubsection{Effective Radius}

Generally small galaxies are more abundant in the universe than larger galaxies due to galaxy formation mechanisms and environment. This is also the case with UDGs. The 2nd row right panel of Figure~\ref{fig:hist} shows how the number of NUDGEs decreases with increasing effective radius and we observe that smaller NUDGEs are more common than larger NUDGEs, with a weighted mean effective radius of 2.7$\,\pm\,$1.1 kpc. The mean effective radius found by \cite{vandokkum} in the Coma cluster UDGs is 2.8 kpc (ranging from 1.5 to 4.5 kpc). The median effective radius within the $R_{200}$ of the eight nearby clusters in the \cite{mancera2019} sample ranges from 1.82 to 2.34 kpc with a mean of the median effective radius values of 1.95 kpc. Our weighted mean effective radius, $r_e$=~2.7$\,\pm\,$1.1~kpc is consistent with the values reported by \cite{vandokkum} and \cite{mancera2019}.

\subsubsection{Surface Brightness}

\begin{figure}
\centering
\includegraphics[width=0.48\textwidth,trim={0.4cm 0.3cm 0.0cm 0.25cm},clip]{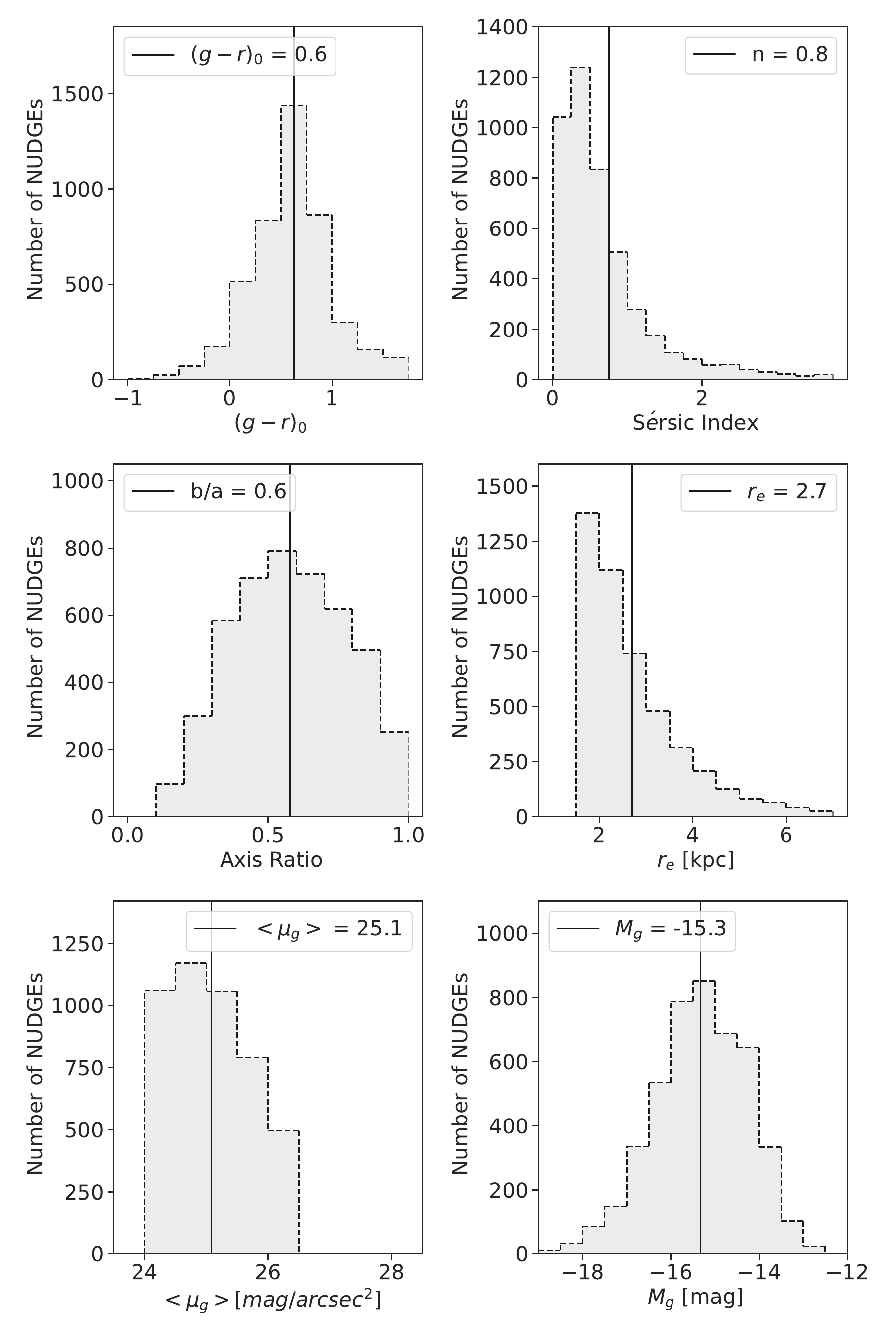}
\caption{\label{fig:hist} Histograms of the properties of NUDGEs in the Subaru wide field clusters before completeness correction. The grey histograms display the properties for all NUDGEs in our sample after applying cuts similar to \protect\cite{vanderburg2016} (4579 NUDGEs). The weighted mean values for each of the respective properties are indicated by the vertical lines.}
\end{figure}

The distribution in mean effective surface brightness is shown in the 3rd row left plot of Figure~\ref{fig:hist}. The g-band mean effective surface brightness of our sample of NUDGEs shows decreasing counts at fainter surface brightnesses. The weighted mean value of the mean effective surface brightness in our NUDGE sample is~25.1$\,\pm\,$0.7~mag/arcsec$^2$.\\

Comparing our sample to previous studies is slightly challenging due to the different measures of surface brightness and different photometric bands used. The original UDG criteria set by \cite{vandokkum} uses central surface brightness. To compare our NUDGEs to their sample we implement the following equation to recover a rough estimation of the central surface brightness from the weighted mean effective surface brightness ($\langle \mu_e \rangle\sim\,$25.1$\,\pm\,$0.7~mag/arcsec$^2$) and weighted mean S\'ersic index ($n\,\sim\,0.8\,\pm\,$0.9), obtained from \cite{graham2005} (equations 7,8 and 9) and S\'ersic coefficient from \cite{macarthur2003}(equation A1):
\begin{equation}
    \indent\mu_0 = \langle \mu_e \rangle - 1.392 .
\end{equation}

We calculate a median central surface brightness of~24.3\,mag/arcsec$^2$. \cite{vandokkum} found a median central surface brightness in the $g$-band of~25\,mag/arcsec$^2$ in their sample of UDGs in the Coma cluster. The overall distribution of surface brightness in our sample is similar to that found in \cite{vandokkum}. The $\sim\,0.7$ magnitude difference is due to the conservative surface brightness criteria we applied to directly compare to the \citealt{vanderburg2016} sample and their host clusters.\\

The absolute magnitudes of our NUDGE sample were calculated assuming distances given by the redshift of their respective host clusters. The 3rd row right panel displays the distribution of g-band absolute magnitudes. We measure a weighted mean absolute magnitude in the g-band of -15.3$\,\pm\,$1.1 mag. \cite{vandokkum} report a median $M_g$ = -14.3 mag. Considering the more inclusive criteria we use, the magnitude difference we observe is expected.

\section{Discussion : NUDGE Abundances}

\begin{figure}
\centering
\includegraphics[width=0.48\textwidth,trim={1.4cm 0.5cm 2.5cm 2.cm},clip]{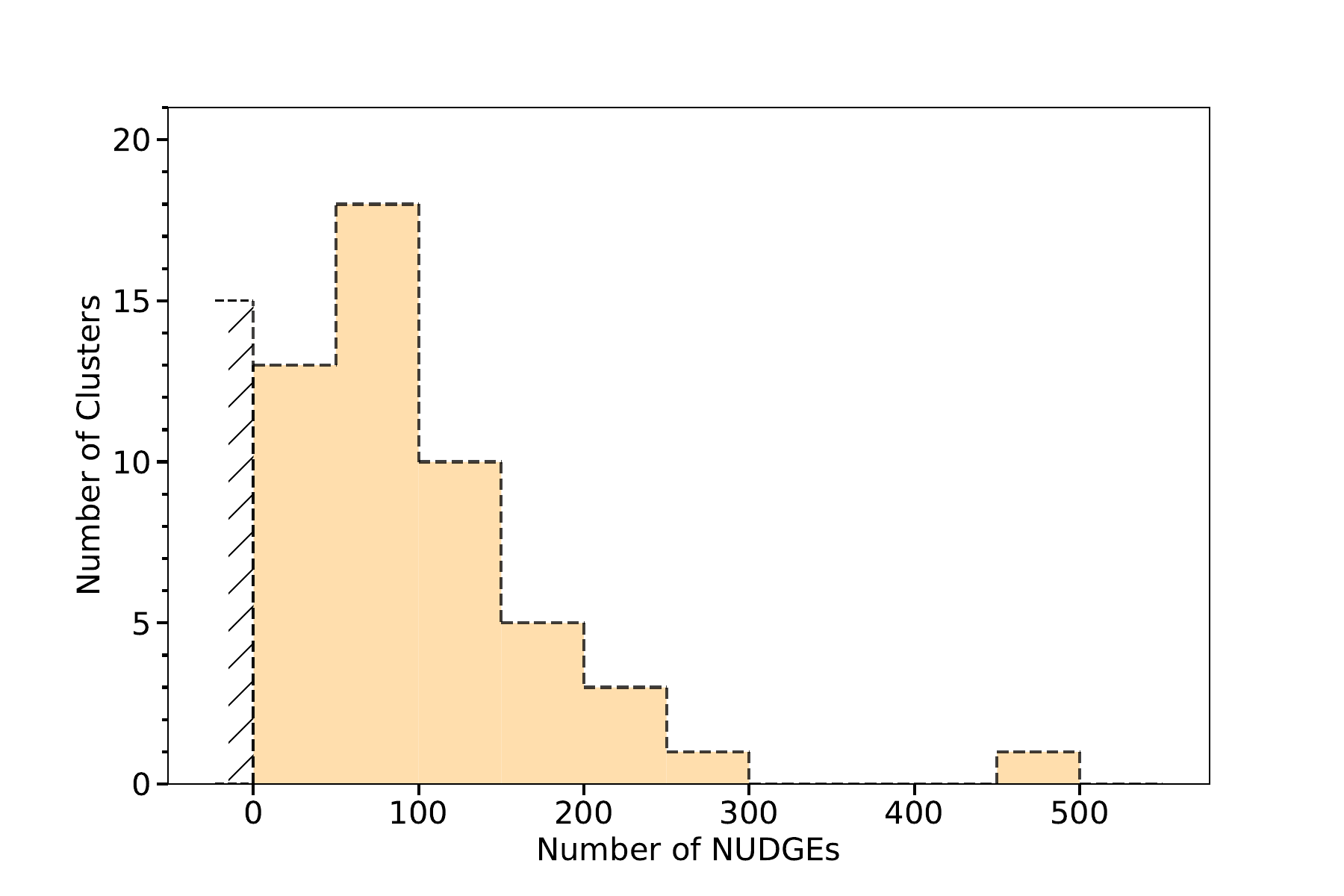}
\caption{\label{fig:hist_numUDGs_Vs_cl} Histogram comparing the number of NUDGEs in each cluster. Most clusters (41) with an abundance of NUDGEs have less than 150 NUDGEs and only 1 cluster has over 300 NUDGEs. The median number of NUDGE counts per cluster with an excess of NUDGEs is 71. We find 15 clusters with less NUDGEs than the background annulus.}
\end{figure}

Galaxy clusters with their high galaxy densities and hot intracluster medium are harsh environments which drive galaxy evolution processes. The clusters in our sample have been identified in the redMaPPer cluster catalogue and therefore have a significant population of red galaxies, however these clusters have a diverse range of properties as shown in the spatial distribution plots (Figure~\ref{fig:spatial}) and detailed in Table~\ref{tab:cl_table1}. We have found a large number of UDG-analogues (5057) throughout the sample, with an excess compared to the background in 51 of 66 clusters. Figure \ref{fig:hist_numUDGs_Vs_cl} displays the number distribution of NUDGEs in our clusters, showing that over 80\% of the clusters with an excess of NUDGEs have 150 NUDGEs or less, while one cluster has 453 NUDGEs (almost twice the second highest, 272). The median number of NUDGEs per cluster in our sample is 71 and we observe fewer clusters with higher numbers of NUDGEs. The cluster with 453 NUDGEs is distinct from the rest of the distribution and highlights the variability of cluster environments. The background counts and cluster counts (before subtraction) are always the same order of magnitude per cluster region but can vary quite widely from cluster to cluster (as shown in Table \ref{tab:cl_table1}). The sources observed in the clusters and background annuli have been studied with a systematic and uniform methodology. However, as observed in our study and mentioned in other studies (e.g. \citealt{vanderburg2017}), a statistical background subtraction may result in negative values for NUDGE abundances in some clusters.\\

In the following sections we discuss the relationship between the properties of NUDGEs and their host clusters in terms of mass and size. Specifically, we study the relationships between the number of NUDGEs and cluster halo mass, the NUDGE size distribution, and the NUDGE density with respect to cluster radius. 

\subsection{Number of NUDGEs vs cluster halo mass}

The cluster environment has a direct impact on the formation and evolution of cluster galaxies, where the densest cluster regions are expected to have the most significant effect. UDGs are a classification of the most diffuse galaxies, and therefore the cluster environment is expected to have a strong impact on their formation and evolution. An important indicator showing the effect of the cluster environment on UDGs is how the number of UDGs change with respect to the mass of the cluster. UDGs were not expected to be abundant in clusters, however, several studies have since shown that in fact the abundances of UDGs in clusters scale with cluster halo mass \citep{vanderburg2016,roman2017b,mancera2019}, i.e. more massive clusters are observed to have more UDGs. \\

\begin{figure*}
\centering
\includegraphics[width=0.95\textwidth,trim={1.0cm 2.0cm 0.25cm 1cm},clip]{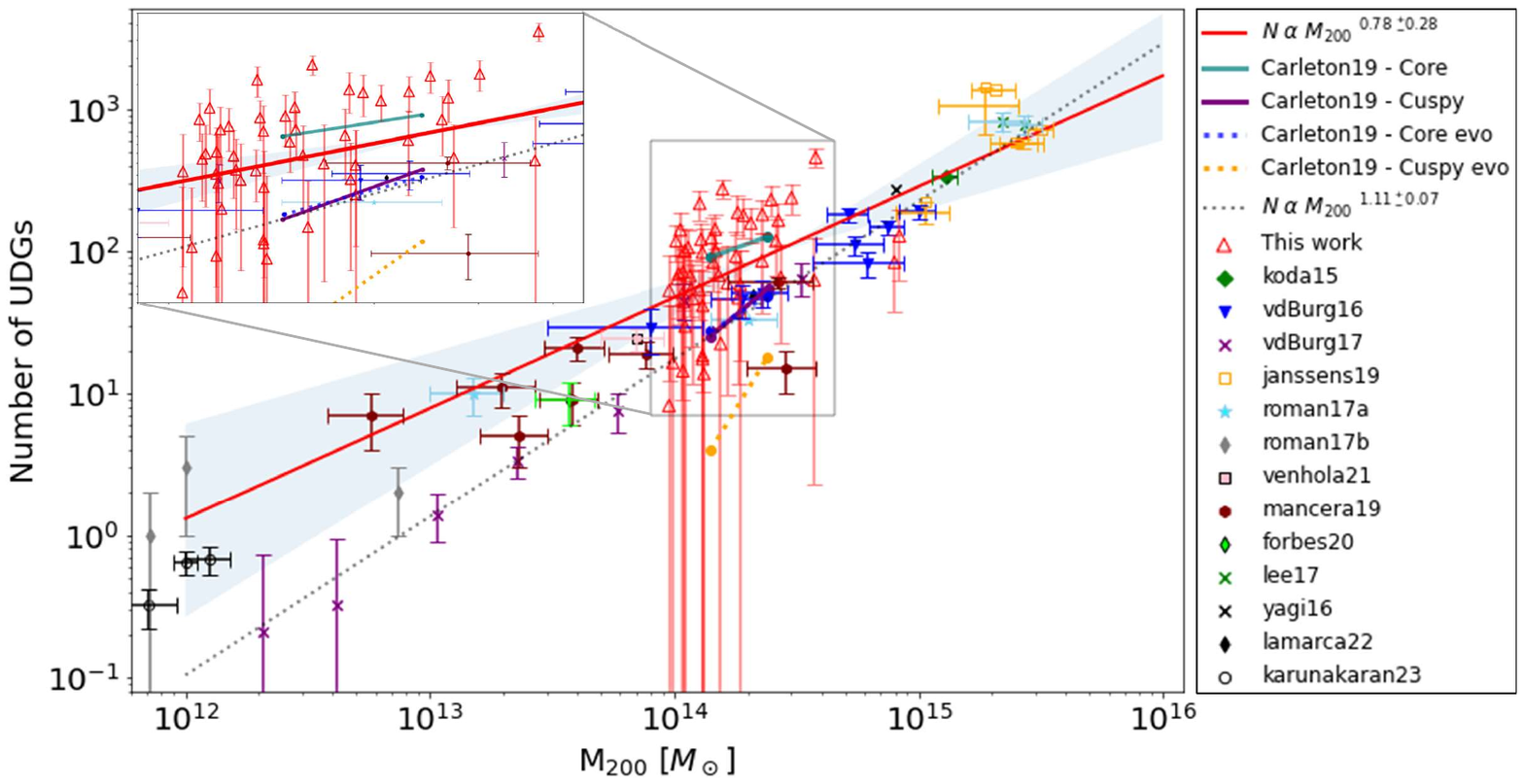}
\caption{\label{fig:udg_vs_m200_o} The number of UDGs vs cluster mass $M_{200}$, our clusters with an excess of NUDGEs are shown using red triangles compared to previous studies. We used our sample of NUDGEs in clusters (red triangles) to determine the best fit power law through bootstrap resample fitting of 10 000 power law fits with errors, indicated by the red line, $N\propto M_{200}^{0.78\pm\,0.28} $, and light blue shading. The power law obtained by \protect\cite{vanderburg2017} in their study of UDGs in nearby groups and clusters is shown by the black dotted line, $N\propto M_{200}^{1.11\pm\,0.07}$. The inset zooms in on the region studied in this work. The results of the \protect\cite{carleton2019} simulations are also shown, as indicated in the legend.}
\end{figure*}

The relationship between the number of UDGs and cluster halo mass for our sample of NUDGEs  is shown in Figure~\ref{fig:udg_vs_m200_o} (red triangles), with Poisson errors. The variation in the number of UDGs in clusters with similar masses can be seen, and demonstrates again the range of environments of the clusters in our sample, as previously illustrated in Figure~\ref{fig:spatial}. The numbers of NUDGEs in our cluster sample are comparable to previous studies, however on average we find more NUDGEs per cluster compared to previous work on UDGs. Our data is most comparable in terms of method and selection criteria to \cite{vanderburg2016} (blue triangles). The main difference between our sample and that of \cite{vanderburg2016} is that we use the $g$-band mean effective surface brightness (24.0~$\leq\, \langle\mu_e(g)\rangle\, \leq$~\surfbright{26.5}) whereas they have used the $r$-band. The median number of NUDGEs over the mass range $0.8\times 10^{14} - 4\times 10^{14} M_\odot$ (range of Figure \ref{fig:udg_vs_m200_o} inset) for our sample is 70 NUDGEs, approximately 1.5 times the median number (46) reported in \cite{vanderburg2016} over the same mass range. The higher numbers of NUDGEs vs M$_{200}$ relation (see Figure \ref{fig:udg_vs_m200_o}) compared to previous studies of UDGs may be attributed to several possible reasons (apart from the criteria difference): Firstly, our method for estimating foreground and background interlopers is quite different to previous studies. Previous works mostly use one or a few blank fields or averages of these to subtract from each of their individual clusters (e.g. \citealt{vanderburg2016, mancera2018}) whereas we use regions close to our clusters to better account for cosmic variance in different sky directions. Secondly, different studies use different methods for estimating cluster mass which can result in systematic offsets in the results for the different studies presented. \cite{mancera2018} illustrate this when they recalculate the cluster masses from \cite{vanderburg2016} and find a median factor of 3.2 difference in the cluster masses when using two different methods (both based on velocity dispersions). Furthermore, the samples of \cite{janssens2017, janssens2019} and \cite{lee2017} include a subsample of the same clusters, however they provide different M$_{200}$ values for those clusters (plotted in Figure \ref{fig:udg_vs_m200_o}). Given the different data types required for different mass estimate methods, and data availability, it's not currently possible to rescale all the cluster masses presented in Figure \ref{fig:udg_vs_m200_o} using a common estimation method, therefore we plot the values presented in the individual papers.\\

We used bootstrap resampling with replacement to determine a robust power law fit for our data only, by selecting the median best fit power law exponent after fitting 10 000 curves. The best fit power law relation is $N\propto M_{200}^{0.78\pm\,0.28}$ and is shown by the red line in Figure~\ref{fig:udg_vs_m200_o} with the 1$\sigma$ uncertainty region indicated by the light-blue shaded region. This relation agrees with many of the previous studies at both lower and higher halo masses \citep{koda2015,vanderburg2016,janssens2017,roman2017b,mancera2019}. Power law fits from the literature for UDGs in clusters are presented in Table~\ref{tab:powerlawsummary}. The relation determined by \cite{vanderburg2017} includes both UDGs in clusters as well as a sample of UDGs in groups binned by mass (blue triangles and purple crosses respectively). The power law relation determined by \cite{vanderburg2017}, $N\propto M^{1.11\pm\,0.07}$, is steeper than the relation we find, although many of the data points from their clusters agree within the 1$\sigma$ uncertainty of our power law fit. \cite{janssens2019} determined a best fit power law using their sample of six Frontier Field clusters as well as data from previous studies (including \citealt{vanderburg2016,vanderburg2017,roman2017a,roman2017b}), which is similar to the relation by \cite{vanderburg2017} and steeper than the relation we find. \cite{roman2017b} and \cite{mancera2019} study this relation for nearby groups and lower mass clusters, their fits are more similar to the relationship we find (only marginally steeper), and we observe that their data coincide within our power-law and 1$\sigma$ uncertainty.\\

\begin{table}
    \centering
    \begin{tabular}{c|c|c|c}
        \hline
        Study & Power law\\
        \hline
        This work & $N\propto M^{0.78\pm\,0.28}$ \\
        \cite{mancera2019} & $N\propto M^{0.81\pm\,0.17}$ \\
        \cite{janssens2019} & $N\propto M^{1.13\pm\,0.06}$ \\
        \cite{vanderburg2017} & $N\propto M^{1.11\pm\,0.07}$ \\
        \cite{roman2017b} & $N\propto M^{0.85\pm\,0.05}$ \\
        \cite{karunakaran2023} & $N\propto M^{0.89\pm\,0.04}$ \\
        \cite{carleton2019} cored& $N\propto M^{0.76}$ \\
        \cite{carleton2019} cuspy& $N\propto M^{1.24}$ \\
        \hline\hline
        
    \end{tabular}
    \caption{The power law relations for the different UDG samples, including: our work, \protect\cite{mancera2019}, \protect\cite{janssens2019}, \protect\cite{vanderburg2017}, \protect\cite{roman2017b}, \protect\cite{karunakaran2023} and the simulated UDGs forming in cored and cuspy dark matter haloes by \protect\cite{carleton2019}.}
    \label{tab:powerlawsummary}
    \vspace{-1ex}
\end{table}

\cite{carleton2019} simulated the formation of UDGs in cored and cuspy dark matter haloes in clusters with similar masses to our sample, shown in Figure \ref{fig:udg_vs_m200_o} (ranging from $1.40\times10^{14}\text{M}_{\odot}$ to $2.41\times10^{14}\text{M}_{\odot}$). A cored dark matter halo refers to a dark matter mass which is distributed with a shallow profile, whereas the cuspy dark matter haloes have mass profiles which are more concentrated at the centre. The simulated UDGs forming in cored dark matter haloes lie centrally within our cluster distribution and follow the number of UDGs to cluster halo mass relation of $N\propto M_{halo}^{0.76}$. This relation shows agreement with previous studies of NUDGEs in low and high halo mass clusters (including \citealt{koda2015,vanderburg2016,janssens2017,roman2017b,mancera2019}). This may suggest that our sample of NUDGEs might have formed in cored dark matter haloes and experienced significant tidal stripping by external mechanisms in their host clusters as shown by the simulated UDGs in \cite{carleton2019}. In contrast, \cite{carleton2019} show that UDGs forming in cuspy dark matter haloes follow a $N\propto M_{halo}^{1.24}$ relation, which agrees with many previous studies (including \citealt{koda2015,vanderburg2016,vanderburg2017,janssens2019,forbes2020} as shown in Figure \ref{fig:udg_vs_m200_o}). \cite{carleton2019} further propose a size–mass relation that evolves weakly with redshift, resulting in the abundance of cored UDGs in this evolving size regime overlapping with the abundance of cuspy UDGs, shown by the dashed blue line shown in Figure \ref{fig:udg_vs_m200_o}.

\subsection{\label{sec:func_rad}NUDGE density as a function of cluster radius}
The density of galaxies and the cluster potential in the central cluster region often results in the strongest tidal fields, whereas the outer regions are not as harshly impacted by these processes \citep{smith2016,vanderburg2016,wittman2017}. To investigate this further we studied the relationship between the number density of UDGs and the cluster radius.\\

\begin{figure}
\centering
\includegraphics[width=0.48\textwidth,trim={1.cm 0.5cm 3cm 1.0cm},clip]{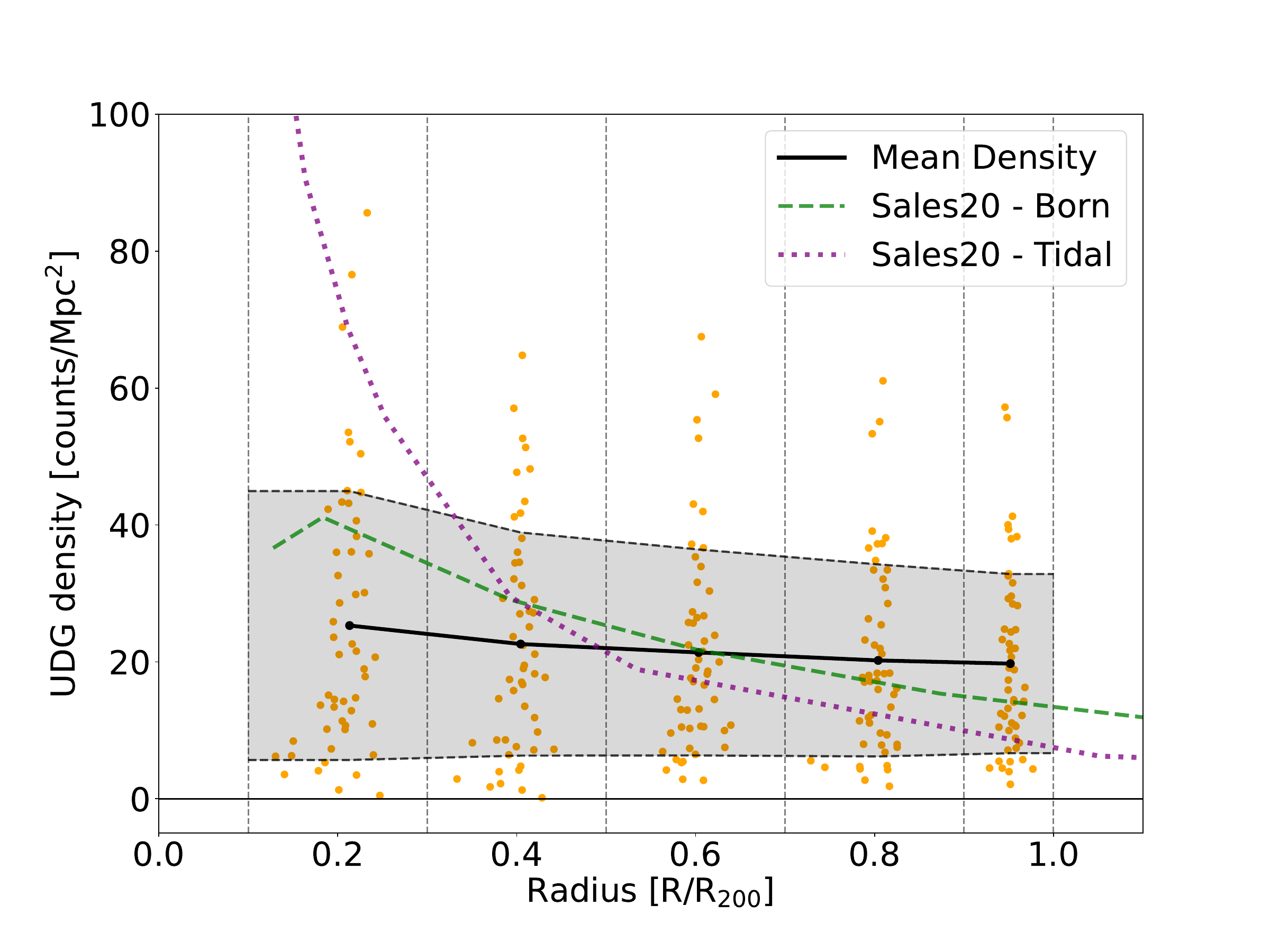}
\caption{\label{fig:udg_Density_counts1} The weighted mean NUDGE density as a function of clustercentric radius is shown by the black solid line. The orange dots indicate the NUDGE density per cluster vs their mean radial position in the bin. Black shading indicates standard deviation from the mean. The distribution of `born' and `tidal' UDGs determined from simulations by \protect\cite{sales2020} are shown with green dashed and purple dotted lines respectively, scaled (by a factor of 4) to compare the trends to our data.}
\end{figure}

We binned the clusters with respect to cluster radius in 5 bins, the first 4 ranging from $0.1 - 0.9\,R/R_{200}$ with an interval of 0.2 and the final bin ranging from $0.9 - 1.0\,R/R_{200}$. We excluded the central region ($<\,0.1\,R/R_{200}$) which was masked in most clusters due to saturation of the images caused by bright galaxies at the cluster centres. All regions with clear background over-subtraction and artefacts in the images were excluded and the annuli areas appropriately recalculated per cluster. Figure~\ref{fig:udg_Density_counts1} shows the NUDGE spatial number density (counts/Mpc$^2$) per radial bin for each cluster (orange dots). The mean trend is plotted with the black solid line and the standard deviation from the mean is shown in grey shading. The mean NUDGE density decreases slightly with respect to cluster radius. This shallow change in NUDGE density with respect to cluster radius, and therefore cluster density, may imply that NUDGEs are only very marginally affected by change in density within the cluster environment. Some studies, due to data coverage limitations, are only able to study the central regions of clusters and then apply an average geometrical correction (e.g.\citealt{janssens2017,janssens2019,lee2017}). However the wide coverage of the HSC-SSP data enable full coverage of the clusters in our sample and their background regions. We also plot the predicted spatial densities as simulated by \cite{sales2020} for two different UDG formation mechanisms in cuspy dark matter haloes, i.e. `born' and `tidal' UDGs. Compared to the simulated UDGs by \cite{sales2020} our NUDGE sample and errors are consistent with both simulated `born' and `tidal' UDGs. However, in the smallest radial bin, the `born' scenario matches our data better than the `tidal' formation scenario (we do not see the large increase in density at the centre expected from the tidal scenario). \\

\begin{figure}
\centering
\includegraphics[width=0.48\textwidth,trim={1.0cm 0.5cm 3cm 1cm},clip]{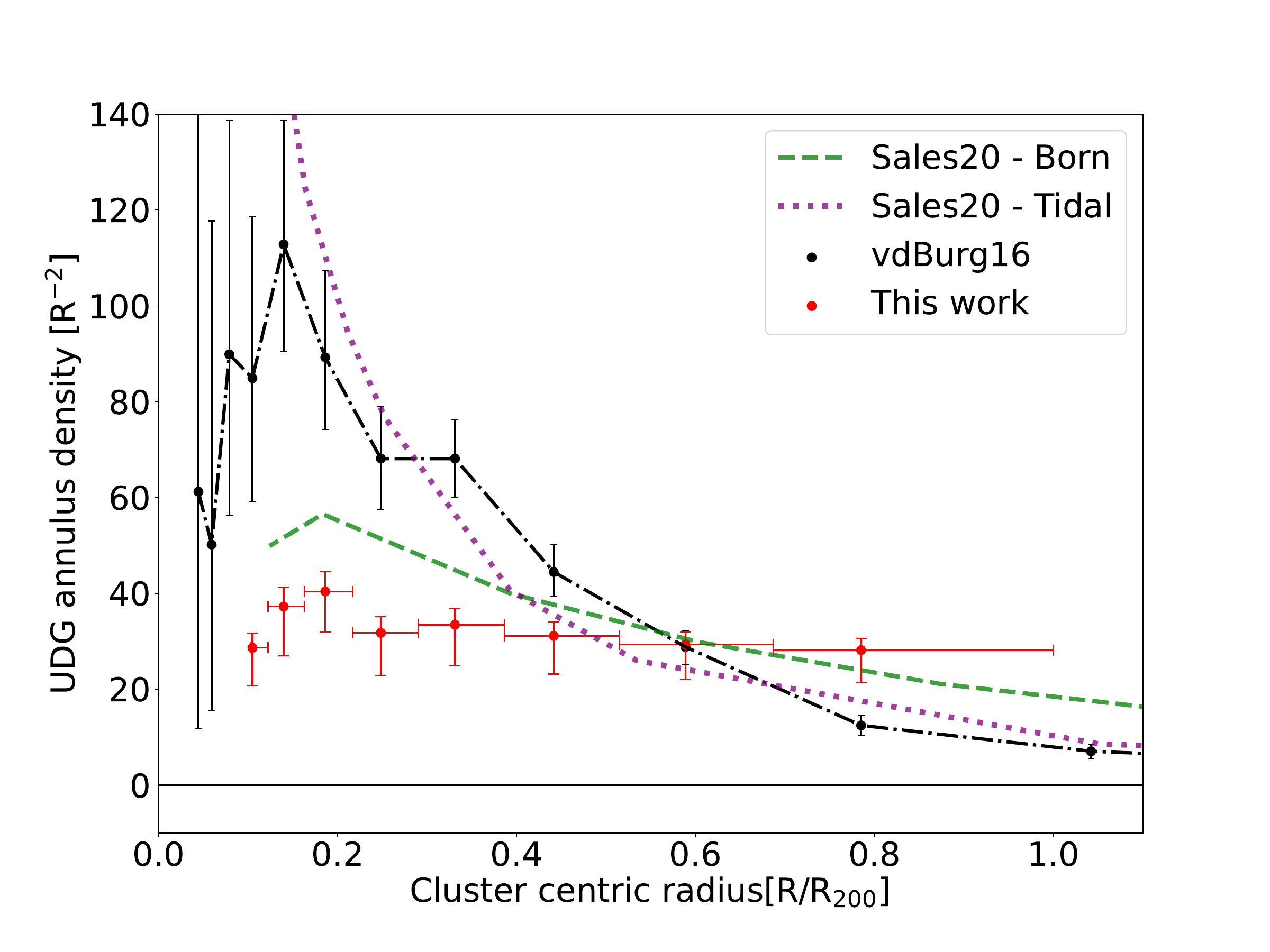}
\caption{\label{fig:udg_Density_counts2}Comparing the UDG density (per fraction of cluster area) vs fraction of cluster radius using similar binning to \protect\cite{vanderburg2016}. The errors are determined from 100 iterations of bootstrapping with replacement. The mean NUDGE density in each cluster radial bin is shown with the max and min density range shown as y error bars and x error bars indicate the cluster radius ranges. The simulated `born' and `tidal' UDGs by \protect\cite{sales2020} are shown with green and purple dashed lines respectively, scaled to compare the trends to our data.}
\end{figure}

Figure \ref{fig:udg_Density_counts2} shows the NUDGE spatial number density as a fraction of $R_{200}$ per radial bin. We used similar binning to \cite{vanderburg2016} to compare more closely to their work. We used bootstrapping with replacement (100 iterations), by selecting all NUDGEs in each cluster as a subset and bootstrapping with the 51 subsets. The mean NUDGE density per bin after 100 iterations is plotted and the range of density in the 100 iterations are shown as error bars. Compared to the \cite{vanderburg2016} distribution in black, our NUDGE distribution is significantly shallower; we find lower densities than they do near the cluster centre and higher densities in the outer regions. We note the same overall shape of the trends and similar radius at peak density. This small change (also shown in Figure \ref{fig:udg_Density_counts1}) in NUDGE density with respect to cluster radius indicates that the change in cluster density has a small impact on the number of NUDGEs. The Frontier Field clusters studied by \cite{janssens2019} show a significantly lower UDG density at all cluster radii compared to both our study and \cite{vanderburg2016}, however the shape of their density distributions show a similar peak at low cluster radii and decrease in UDG density with increasing cluster radii (figure 7 of \citealt{janssens2019}). The simulated UDGs by \cite{sales2020} are also shown in Figure \ref{fig:udg_Density_counts2}. The sample by \cite{vanderburg2016} is more similar to the `tidal' UDGs which indicate a strong influence of the cluster on their formation. \\

The density of NUDGEs in our sample shows a shallow decrease with respect to the cluster radius (as shown in Figures \ref{fig:udg_Density_counts1} and \ref{fig:udg_Density_counts2}) up to the cluster $R_{200}$. Within our sample some clusters show a NUDGE density that increases or remains constant with increasing cluster radius. This may be due to the varied properties of the clusters in our sample, shown in Figure~\ref{fig:spatial} and Table~\ref{tab:cl_table1}. We measured the NUDGE density distribution within each cluster's $R_{200}$ and a background annulus from $1.5\,\text{R}_{200}$ to $1.8\,\text{R}_{200}$, which is close enough to the cluster to ensure little cosmic variance. Overall, the densities of NUDGEs are higher than in the background annuli, however, in a minority of clusters (15 out of 66) we observe that the number of NUDGEs drops below the background counts. The ICL around bright central cluster galaxies (BCG) typically extends up to a radius of\,100\,-\,400\,kpc \citep{mihos2005,gonzalez2005,presotto2014,demaio2015} and could affect the detection and characterisation of faint sources in our cluster regions, which may result in decreased detections. We masked the central $0.1R_{200}$ cluster regions to account for ICL around the BCGs and we did not perform any systematic ICL correction beyond the $0.1R_{200}$ radius other than manual masking of bright sources with clear background over-subtraction and artefacts. The redMaPPer galaxy distributions in individual clusters do not decrease smoothly from the centre (as shown in Figure~\ref{fig:spatial}) which makes systematic corrections difficult. We avoid using an ICL correction with respect to cluster radius as we are investigating the density distribution of NUDGEs with respect to clustercentric radius and do not want to introduce corrections that may directly influence or add biases to our results.\\

\begin{figure}
    
  \centering
  \includegraphics[width=0.49\textwidth,trim={1cm 0.5cm 3cm 3cm},clip]{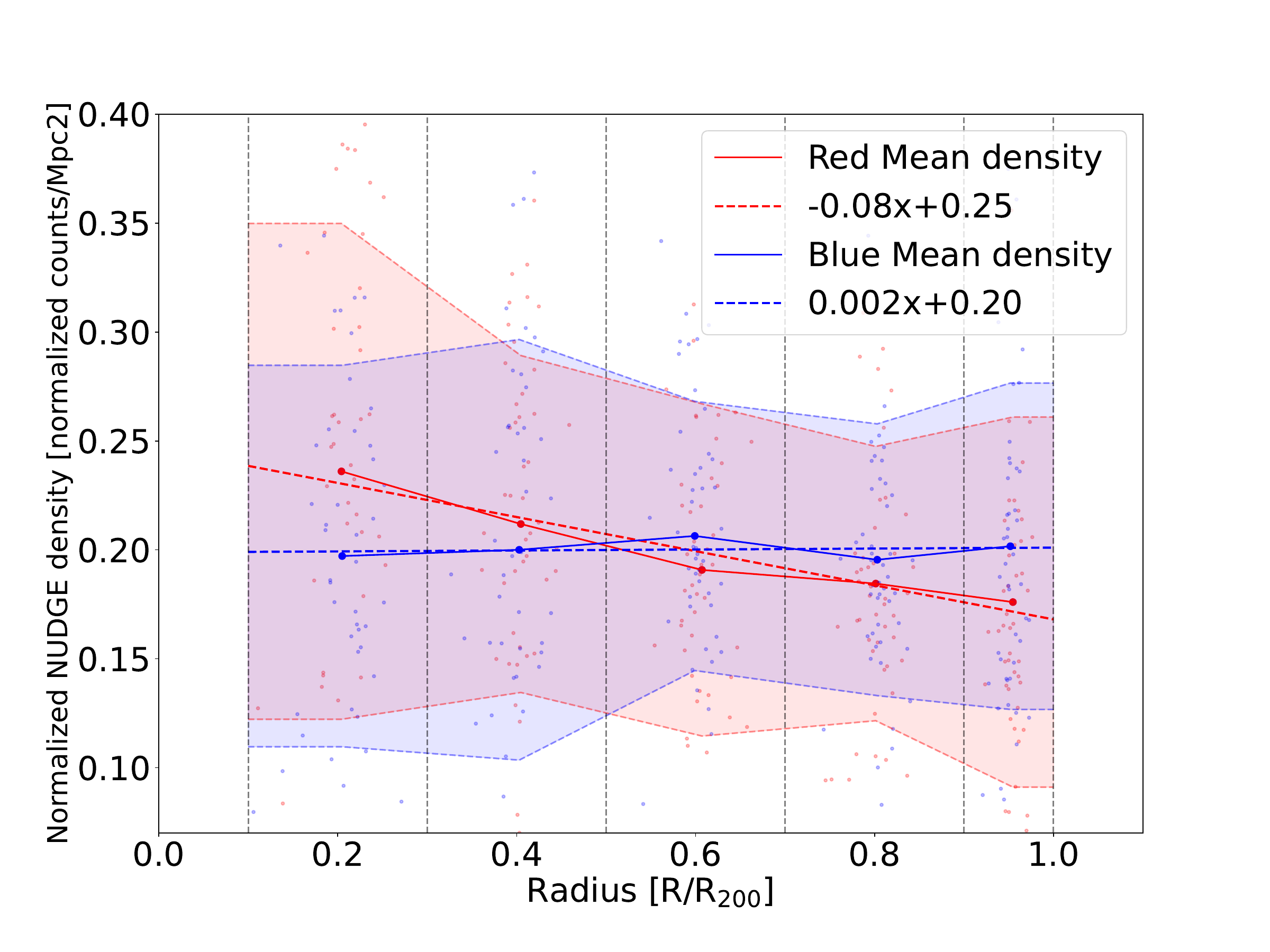}
  \label{fig:testf8}
\caption{Comparing the normalized mean NUDGE density of the red and blue NUDGEs as a function of clustercentric radius (normalized to the mean number of NUDGEs per cluster). The shading indicates standard deviation from the mean. The normalized NUDGE density per cluster for the red and blue NUDGE samples, shown with lighter shaded dots. The best fit lines showing the slope of the red and blue NUDGE samples are indicated by the dashed lines, and imply a statistically significant slope for the red NUDGEs and a slope consistent with zero for the blue sample. The best fit slope parameter for the red NUDGE sample is -0.08$\pm$0.01 $(\text{counts}/\text{Mpc}^2) / (\text{R}/\text{R}_{200})$ and for the blue NUDGE sample 0.002$\pm$0.008 $(\text{counts}/\text{Mpc}^2) / (\text{R}/\text{R}_{200})$.} 
\label{fig:udg_colour_Density3}
\end{figure}

The colour of galaxies is typically an indicator of the evolutionary stage of their stellar population. The morphology-density relation \citep{dressler1980} demonstrates how the density of the environment impacts the numbers of particular types of galaxies. For example, typically galaxies close to the centres of clusters tend to be redder than those on the edges. \cite{buzzo2024} studied the stellar population of UDGs in low-to-moderate density fields with spectral energy distribution fitting and determined two classes of UDGs. The first class were found to have a younger population and to exist in less dense environments, whereas the second class has an older population and were observed in more dense environments. We investigate the morphology-density relation for NUDGEs by studying their colours with respect to the host cluster radius, which may show how the cluster environment affects their stellar population. We used the red sequence of the redMaPPer cluster galaxies for each cluster to separate the red and blue NUDGEs. All NUDGEs with colours within 0.25 dex of the redMaPPer galaxies' red sequence relation and above were identified as red NUDGEs and all NUDGEs below were classified as blue. Due to the very faint sample selection, the separation into red sequence and blue cloud is not clear, however we can still use these samples to perform a statistical comparison of the bluer and redder NUDGEs (Figure~\ref{fig:hist_colour}). The properties of the blue and red NUDGEs overlap as expected with respect to most properties. The $g$-band mean effective surface brightness and $g$-band apparent magnitude show the median of the blue sample to be fainter than the red. Figure~\ref{fig:udg_colour_Density3} displays the average NUDGE densities (normalized by cluster counts) of the red and blue samples as a function of clustercentric radius. One might expect the dense central regions of clusters to have redder NUDGEs due to the strong tidal fields at the centres of clusters stripping galaxies of their gas and the outer cluster regions to have galaxies with relatively bluer populations. The scatter in the blue and red samples are consistent with each other. The slope of the red NUDGE sample is -0.08$\pm$0.01 $(\text{counts}/\text{Mpc}^2) / (\text{R}/\text{R}_{200})$ and the slope of the blue sample is 0.002$\pm$0.008 $(\text{counts}/\text{Mpc}^2) / (\text{R}/\text{R}_{200})$. This implies that the red NUDGEs decrease in density with respect to cluster radius while the clustercentric trend in the blue population is statistically consistent with zero. This may indicate that the change in cluster radius and thereby cluster density affects the red NUDGE sample to a greater extent than the blue sample. This agrees with \cite{buzzo2024}, showing that the redder, and by extension older population of NUDGEs, increase in density toward the more dense centres of clusters. \cite{ferremateu2023} studied the star formation histories of quiescent UDGs and similarly found that UDGs in dense environments were typically quenched at earlier times implying a redder population, whereas typical UDGS found in lower density environments only experienced quenching very recently. Note that in our sample of NUDGEs it is not possible to disentangle projection effects without spectroscopic data.

\subsection{\label{sec:size_dist}NUDGE size distribution}

\begin{figure}
\centering
\includegraphics[width=0.5\textwidth,trim={0.9cm 1.3cm 1cm 2.9cm},clip]{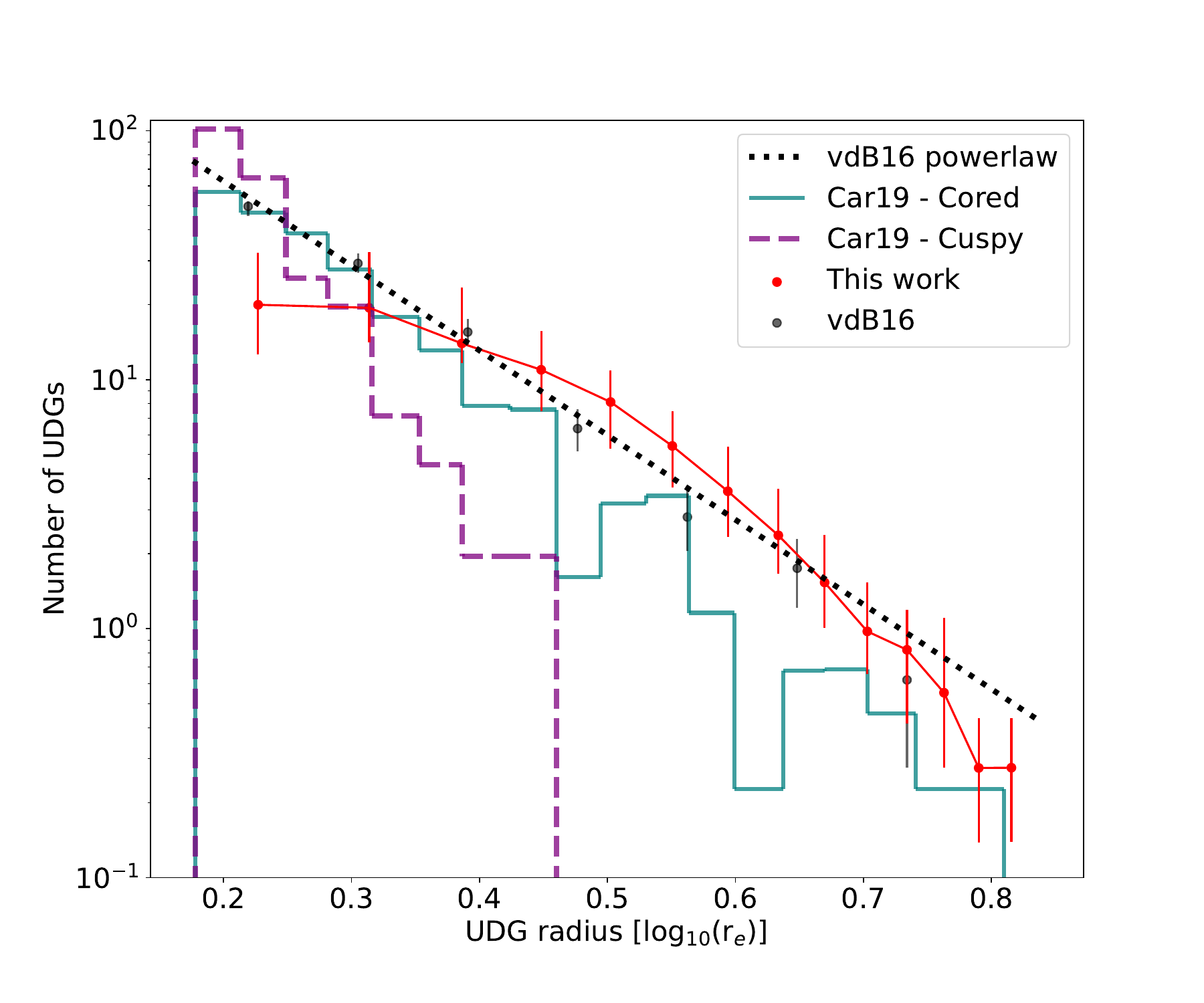}
\caption{\label{fig:udgdenvsudgrad} The size distribution of NUDGEs in this work compared to UDGs in simulations and previous work. We used a bin size of 0.37\,kpc and plot the median number of NUDGEs within each bin from 1.5 kpc to 6.5 kpc. Error bars indicate the interquartile range. The mean number of UDGs from the \protect\cite{vanderburg2016} study is shown with a power-law which describes their distribution, $n\propto r_{e}^{-3.4\pm\,0.19} $, and agrees with our sample distribution. The distribution of simulated UDGs by \protect\cite{carleton2019} are also shown, separated into two samples: UDGs formed in cored dark matter haloes and those formed in cuspy dark matter haloes.}
\end{figure}

\cite{vanderburg2016} studied the sizes of UDGs to determine whether they are a distinct category of galaxy, or whether they fall on the galaxy size continuum. We follow a similar approach, and Figure~\ref{fig:udgdenvsudgrad} shows the distribution of NUDGE sizes in our sample (median number of NUDGEs), compared to the sample studied by \cite{vanderburg2016} and the UDGs simulated by \citep{carleton2019}. The number of NUDGEs in our study decreases with increasing size, similar to \cite{vanderburg2016}. At large radii we get identical numbers, however at the smallest radii our median value and interquartile range falls below the \cite{vanderburg2016} value. The simulated UDGs are split into UDGs formed in cored dark matter haloes and UDGs formed in cuspy dark matter haloes. The simulated UDGs which form in cored dark matter haloes are shown to originate as satellite galaxies that undergo significant tidal stripping during accretion into the cluster \citep{carleton2019}. The UDGs which form in cuspy dark matter haloes are shown to experience stripping mainly in the outer regions of the galaxies \citep{carleton2019}. We find fewer NUDGEs than predicted by both the cuspy and cored dark matter halo models by \cite{carleton2019} for the smallest UDG radii, but thereafter (log$_{10}(r_e)\geq\,0.3$ ($\sim$2\,kpc)), our numbers match the cored dark matter halo formation model quite well and are also consistent with the results by \cite{vanderburg2016}. This suggests that the NUDGEs in our sample may have formed in cored dark matter haloes which experience significant tidal stripping as they are accreted into their host clusters. In the model the cuspy dark matter haloes are shown to not form UDGs beyond log$_{10}(r_e)\sim0.45$ ($\sim$2.8\,kpc), with a much steeper trend in their small range of radii than observed in our work and the \cite{vanderburg2016} sample. In their simulation the cuspy haloes do not experience sufficient mass-loss to form a significant number of large cluster UDGs with $r_e>$2.8\,kpc. \\

\section{Conclusion}
We studied the properties of NUDGEs in galaxy clusters identified in the Subaru wide field survey and the overall effect of the cluster environment on the density of NUDGEs. The clusters span the redshift range $0.08\,<\,z\,<\,0.15$ (350 - 630 Mpc) and they have a mass range of $0.95\times10^{14}\,\text{M}_\odot - 8.34\times10^{14}\,\text{M}_\odot$. Here, we summarize the main results of our study:

\begin{itemize}
    \item We observed an excess of NUDGEs in 51 of the 66 studied clusters.  Approximately 4579 NUDGEs satisfied our selection criteria (24.0~$\leq\, \langle\mu_e(g)\rangle\, \leq$~\surfbright{26.5} and 1.5~$\leq\,r_e\,\leq$~7.0$\,$kpc, adapted from \cite{vanderburg2016}) resulting in a total of 5057 NUDGEs after the completeness correction. 

    \item The overall spatial distribution of NUDGEs within in our cluster sample shows no clear relation with respect to cluster radius or the distribution of bright galaxies, represented by the redMaPPer galaxy density. Projection effects and limited distance information make it difficult to determine trends unambiguously.

    \item For our NUDGE sample the weighted mean $(g\,-\,r)_0$ colour is 0.6 mag, which agrees with previous studies \citep{vanderburg2016,mancera2019} and indicates the similarity between our samples. We find a weighted mean S\'ersic index $n\sim\,$0.8 indicating that our sample has a slightly flatter central surface brightness profile than compared to literature values (exponential profile) for UDGs in clusters. Our mean S\'ersic index is more similar to dwarf and low surface brightness galaxies in other nearby clusters (i.e. Fornax and Virgo; \citealt{eigenthaler2018,ferrarese2020}). The weighted mean axis ratio of our sample is 0.6, which is more elongated than previous UDG studies (for example \citealt{vandokkum,mancera2019}) and the dwarfs and low surface brightness galaxies studied in the Fornax and Virgo clusters \citep{eigenthaler2018,ferrarese2020}, which might indicate tidal effects in our cluster sample.

    \item The NUDGEs we identified have a similar radius and surface brightness distribution to previous studies - typically there are more small bright sources and few large faint NUDGEs (i.e. \citealt{vandokkum,vanderburg2016,mancera2019}). The weighted mean effective radius for our sample is 2.7 kpc and the weighted mean effective surface brightness is 25.1 mag/arcsec$^2$, similar to \cite{vanderburg2016}.  

    \item The relation between the number of NUDGEs and cluster halo mass for our sample of NUDGEs agrees with previous UDG studies. We determined a median best fit power law: $N\propto M_{200}^{0.78\pm\,0.28} $. This fit is shallower than the relation in \cite{vanderburg2016} ($N\propto M_{200}^{1.11\pm\,0.07} $), however it overlaps with other previous studies at both lower and higher halo masses \citep{koda2015,vanderburg2016,janssens2017,roman2017b,mancera2019}, as well as the simulated cored dark matter haloes by \cite{carleton2019}. This agreement with the simulated UDGs in cored dark matter haloes suggests that our sample of NUDGEs may have formed due to tidal stripping caused by external processes in the cluster. Although, we also note that many previous studies including \cite{koda2015,vanderburg2016,vanderburg2017,janssens2019,forbes2020} show agreement to the UDGs forming in cuspy dark matter haloes relation: $N\propto M_{halo}^{1.24}$, shown in Figure \ref{fig:udg_vs_m200_o}. Studies by \citep{mowla2017,forbes2020,buzzo2024} do not observe any visual signs of tidal interactions (i.e. elongated morphology). Additionally, the observed high numbers of globular clusters and metal poor stellar populations, indicate minimal disruptions and possibly imply a cuspy dark matter halo scenario.

    \item The density of NUDGEs with respect to clustercentric radius shows a large scatter across the different clusters in our sample, which is expected given their varied intrinsic properties. We observe a shallow decrease in density with respect to cluster radius, which is different to previous studies, for example, the UDGs in the \cite{vanderburg2016} sample show a steep drop with radius. However, we note the Abell frontier field clusters studied by \cite{janssens2019} show lower UDG densities over all cluster radii compared to both our work and \cite{vanderburg2016}. The shape of the density plots of this work as well as \cite{vanderburg2016} and \cite{janssens2019} are similar, showing distinct peaks at low cluster radii and slopes which decrease with increasing cluster radii. Our NUDGE distribution is consistent with both the simulated cuspy halo `born' UDGs and cuspy `tidal' UDGs \citep{sales2020} down to smaller clustercentric radii ( < 0.3 R/R$_{200}$) whereas the `born' model matches our data better and the `tidal' model diverges. Therefore we cannot rule out NUDGE formation through either internal processes or external cluster processes.

    \item The red NUDGEs in our sample show a slightly steeper relation between density and clustercentric radius with a slope of -0.08±0.01 $(\text{counts}/\text{Mpc}^2) / (\text{R}/\text{R}_{200})$ compared to the blue sample which shows a slope that is statistically consistent with zero (0.002±0.008 $(\text{counts}/\text{Mpc}^2) / (\text{R}/\text{R}_{200})$ ). This indicates that red UDGs are found preferentially in the denser regions of clusters. We note that the sample of blue and red NUDGEs have similar characteristics.      

    \item The distribution of NUDGE sizes in our work is consistent with the \cite{vanderburg2016} UDG sample and the simulated UDGs formed through tidal stripping in cored dark matter haloes by \cite{carleton2019}.
    
\end{itemize}

Finally, since our data are matched by models of UDGs with cored dark matter haloes and tidal formation mechanisms \citep{carleton2019} as well as models where UDGs are `born' and located in cuspy dark matter haloes \citep{sales2020}, we are not able at this time to differentiate between the formation mechanisms and await further development on the theoretical front.

\section*{Acknowledgements}

We thank the referee for the thorough report and useful feedback, which has improved the paper.

The Hyper Suprime-Cam (HSC) collaboration includes the astronomical communities of Japan and Taiwan, and Princeton University. The HSC instrumentation and software were developed by the National Astronomical Observatory of Japan (NAOJ), the Kavli Institute for the Physics and Mathematics of the Universe (Kavli IPMU), the University of Tokyo, the High Energy Accelerator Research Organization (KEK), the Academia Sinica Institute for Astronomy and Astrophysics in Taiwan (ASIAA), and Princeton University. Funding was contributed by the FIRST program from the Japanese Cabinet Office, the Ministry of Education, Culture, Sports, Science and Technology (MEXT), the Japan Society for the Promotion of Science (JSPS), Japan Science and Technology Agency (JST), the Toray Science Foundation, NAOJ, Kavli IPMU, KEK, ASIAA, and Princeton University. This paper makes use of software developed for Vera C. Rubin Observatory. We thank the Rubin Observatory for making their code available as free software at \url{http://pipelines.lsst.io/.} This paper is based on data collected at the Subaru Telescope and retrieved from the HSC data archive system, which is operated by the Subaru Telescope and Astronomy Data Center (ADC) at NAOJ. Data analysis was in part carried out with the cooperation of Center for Computational Astrophysics (CfCA), NAOJ. We are honored and grateful for the opportunity of observing the Universe from Maunakea, which has the cultural, historical and natural significance in Hawaii. The Pan-STARRS1 Surveys (PS1) and the PS1 public science archive have been made possible through contributions by the Institute for Astronomy, the University of Hawaii, the Pan-STARRS Project Office, the Max Planck Society and its participating institutes, the Max Planck Institute for Astronomy, Heidelberg, and the Max Planck Institute for Extraterrestrial Physics, Garching, The Johns Hopkins University, Durham University, the University of Edinburgh, the Queen’s University Belfast, the Harvard-Smithsonian Center for Astrophysics, the Las Cumbres Observatory Global Telescope Network Incorporated, the National Central University of Taiwan, the Space Telescope Science Institute, the National Aeronautics and Space Administration under grant No. NNX08AR22G issued through the Planetary Science Division of the NASA Science Mission Directorate, the National Science Foundation grant No. AST-1238877, the University of Maryland, Eotvos Lorand University (ELTE), the Los Alamos National Laboratory, and the Gordon and Betty Moore Foundation.

We acknowledge the use of the ilifu cloud computing facility – \url{www.ilifu.ac.za}, a partnership between the University of Cape Town, the University of the Western Cape, Stellenbosch University, Sol Plaatje University and the Cape Peninsula University of Technology. The ilifu facility is supported by contributions from the Inter-University Institute for Data Intensive Astronomy (IDIA – a partnership between the University of Cape Town, the University of Pretoria and the University of the Western Cape), the Computational Biology division at UCT and the Data Intensive Research Initiative of South Africa (DIRISA).

This research made use of Astropy,\url{http://www.astropy.org} a community-developed core Python package for Astronomy \citep{astropy2013, astropy2018}

NM acknowledges the bursary provided by National Research
Foundation through the South African Astronomical Observatory.

\section*{Data Availability}

The processed data and psf models are available from the HSC-SSP website through multiple tools which can be used to retrieve the HSC data, \url{https://hsc-release.mtk.nao.ac.jp/doc/index.php/data-access__pdr3/} and are described in \cite{aihara2022}.\\
The redMaPPer clusters catalogue and cluster member catalogue are available at \url{https://vizier.cds.unistra.fr/viz-bin/VizieR?-source=J/ApJ/785/104} and are described in \cite{Redmapper}.



\bibliographystyle{mnras}
\bibliography{example} 




\appendix

\section{Background contamination}\label{app:bg_cont}
\begin{figure}
\centering
\includegraphics[width=0.5\textwidth,trim={2.8cm 2.0cm 2.0cm 2.8cm},clip]{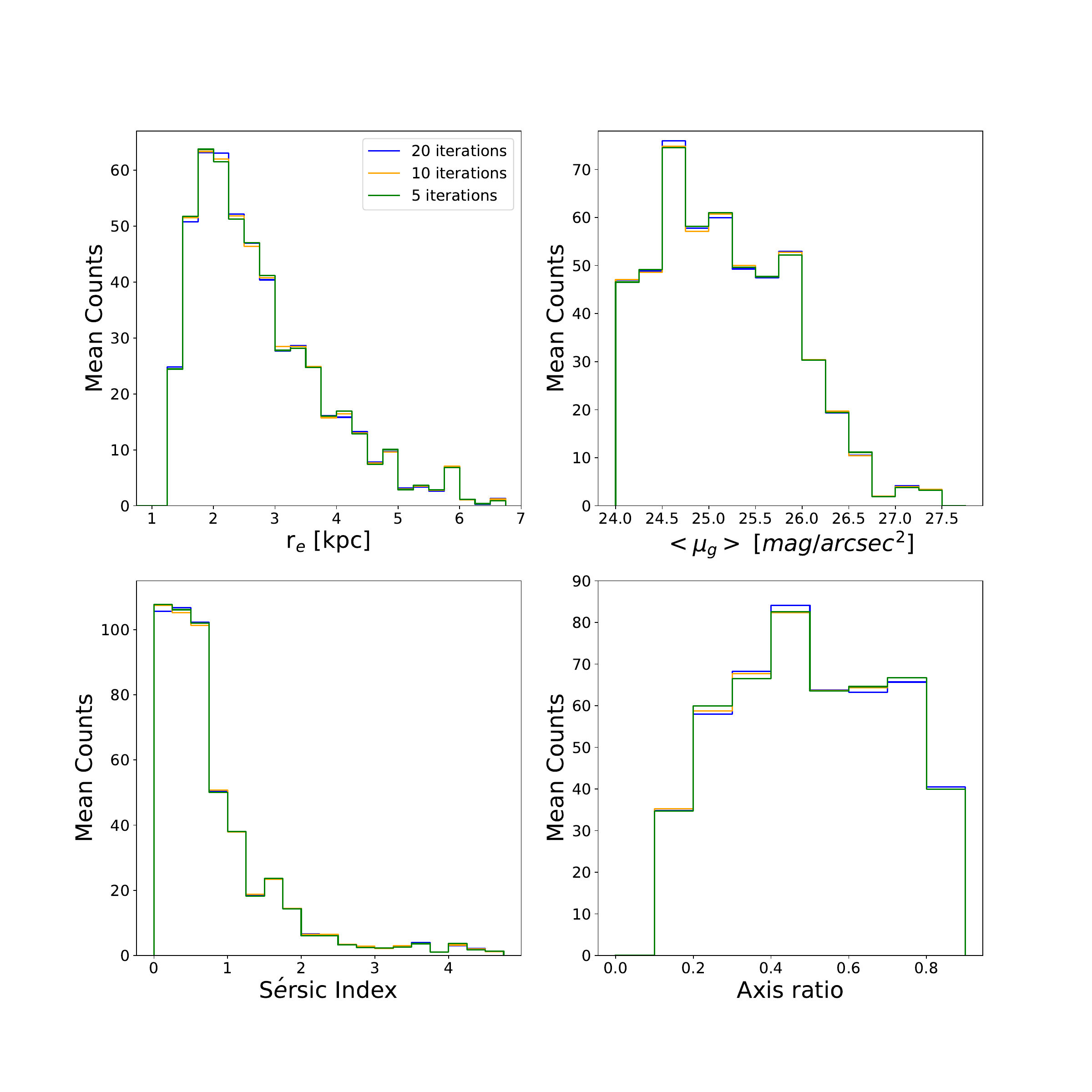}
\vspace{-4.5ex}
\caption{\label{fig:iteration} Histogram comparing the UDG properties after performing 5, 10 and 20 iterations of background subtraction while randomizing the order of background sources for sample cluster 730. The properties agree across the different numbers of iterations.}
\end{figure}

The background contamination is determined by removing cluster UDGs with a similar $(g\,-\,r)_0$ colour and $r$ magnitude to the background UDG-like sources. To avoid biases in which may result from the order of the subtracted sources we randomize the order of the background sources and repeat the background subtraction 5, 10, and 20 times. We used the average UDG weighting of all iterations to minimize any bias. Figure \ref{fig:iteration} shows how the distribution of properties change with the different number of iterations. The properties are found to be near identical at 5, 10 and 20 iterations, indicating the consistency using the colour-magnitude background subtraction.

\section{Red and Blue UDGs}\label{app:colourUDGs}

\begin{figure}
\centering
\includegraphics[width=0.5\textwidth,trim={0.4cm 0.3cm 0.0cm 0.25cm},clip]{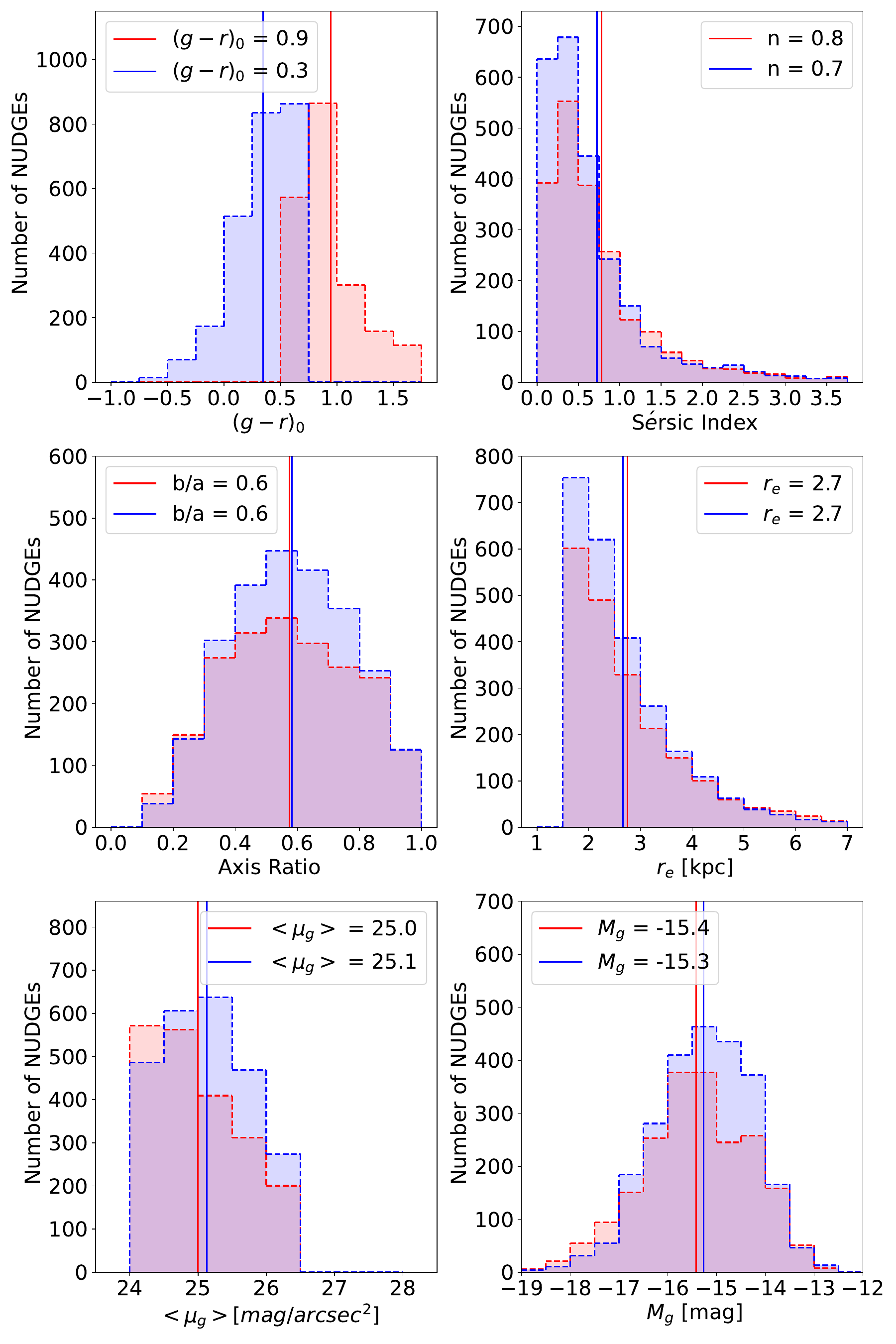}
\caption{\label{fig:hist_colour} Histograms of the properties of UDGs (after applying \protect\cite{vanderburg2016} cut) in the Subaru wide field clusters, separated by red sequence into blue and red subsets. The blue sample consists of  2560.3 UDGs and the red sample has 1961.7 UDGs. The weighted mean values for each of the respective properties are indicated by the vertical lines.}
\end{figure}

The red and blue samples of UDGs were separated using the red sequence of the redMaPPer galaxies. UDGs with $(g\,-\,r)_0$ colours within 0.25 dex and above were classed as red UDGs and sources below as blue UDGs. The distribution of properties in Figure \ref{fig:hist_colour} show that we have more blue UDGs in our sample than red and that their distribution in properties are quite similar.

\label{lastpage}
\end{document}